\documentclass[onecolumn,aps,prd,showpacs,nofootinbib,superscriptaddress,preprintnumbers,floatfix,11pt]{revtex4-2}

\usepackage[utf8]{inputenc}
\usepackage{graphicx}
\graphicspath{ {./Figures/}}
\usepackage{color}
\usepackage{multirow}
\usepackage{amsmath,amssymb}
\usepackage{hyperref}
\usepackage{bbm}
\usepackage{url}
\usepackage{slashed}
\usepackage{subfigure}
\usepackage[usenames,dvipsnames]{xcolor}
\usepackage{dsfont}
\usepackage{mathrsfs}
\usepackage{mathtools}
\usepackage{amsfonts}
\usepackage{booktabs}
\usepackage[mathscr]{euscript}
\usepackage{float} 
\usepackage{epsfig}
\usepackage{epstopdf}
\usepackage[normalem]{ulem}
\usepackage[sort&compress]{natbib}
\usepackage{MnSymbol}
\usepackage{physics}
\usepackage{pifont}

\newcommand{\cmark}{\ding{51}}
\newcommand{\xmark}{\ding{54}}

\usepackage{stackengine}
\newcommand\xrowht[2][0]{\addstackgap[.5\dimexpr#2\relax]{\vphantom{#1}}}

\def \nn{\nonumber}

\def \e{\textmd{e}}
\def \N{\textmd{N}}

\def \Tr{\textmd{Tr}}

\def \GNk{G_{\textmd{N},k}}
\def \pt{\partial}
\def \ptt{\partial_t}

\def \TT{\textmd{TT}} 
\def \T{\textmd{T}} 
\def \L{\textmd{L}}
\def \max{\textmd{max}}

\begin{document}
 
\title{Exploring new corners of asymptotically safe unimodular quantum gravity}

  \author{Gustavo P. de Brito}
   \email{gustavo@cp3.sdu.dk}
\affiliation{CP3-Origins, University of Southern Denmark, Campusvej 55, DK-5230 Odense M, Denmark}
  
\author{Antonio D.~Pereira}
   \email{adpjunior@id.uff.br}
\affiliation{Instituto de F\'isica, Universidade Federal Fluminense, Campus da Praia Vermelha, Av. Litor\^anea s/n, 24210-346, Niter\'oi, RJ, Brazil}
\affiliation{Institute for Mathematics, Astrophysics and Particle Physics (IMAPP),
Radboud University, Heyendaalseweg 135, 6525 AJ Nijmegen, The Netherlands}

\author{Arthur F.~Vieira}
   \email{arthurfv@id.uff.br}
\affiliation{Instituto de F\'isica, Universidade Federal Fluminense, Campus da Praia Vermelha, Av. Litor\^anea s/n, 24210-346, Niter\'oi, RJ, Brazil}

\begin{abstract}
The renormalization group flow of unimodular quantum gravity is investigated within two different classes of truncations of the flowing effective action. In particular, we search for non-trivial fixed-point solutions for polynomial expansions of the $f(R)$-type as well as of the $F(R_{\mu\nu}R^{\mu\nu})+R\,Z(R_{\mu\nu}R^{\mu\nu})$ family on a maximally symmetric background. We close the system of beta functions of the gravitational couplings with anomalous dimensions of the graviton and Faddeev-Popov ghosts treated according to two independent prescriptions: one based on the so-called background approximation and the other based on a hybrid approach which combines the background approximation with simultaneous vertex and derivative expansions. For consistency, in the background approximation, we employ a background-dependent correction to the flow equation which arises from the proper treatment of the functional measure of the unimodular path integral. We also investigate how different canonical choices of the endomorphism parameter in the regulator function affect the fixed-point structure. Although we have found evidence for the existence of a non-trivial fixed point for the two classes of polynomial projections, the $f(R)$ truncation exhibited better (apparent) convergence properties. Furthermore, we consider the inclusion of matter fields without self-interactions minimally coupled to the unimodular gravitational action and we find evidence for compatibility of asymptotically safe unimodular quantum gravity with the field content of the Standard Model and some of its common extensions.
\end{abstract}

\maketitle

\section{Introduction}
A fundamental quantum description of the gravitational field which is valid across arbitrarily short length scales remains unknown. Frequently, a theory of quantum gravity is claimed to necessitate tools which go beyond quantum-field theoretic tools due to the perturbative non-renormalizability of General Relativity (GR) \cite{tHooft:1974toh,Christensen:1979iy,Goroff:1985th}. However, perturbative renormalizability is neither necessary nor sufficient to define a fundamental quantum field theory. As a concrete example, a theory can be perturbatively renormalizable but not valid up to arbitrary scales due to a Landau pole \cite{PhysRev.95.1300,Gockeler:1997dn}, suffering from a triviality problem \cite{frohlich1982triviality,Gies:2004hy}. What suffices for a fundamental description is the finiteness of the running couplings of the underlying theory due to quantum fluctuations. One possible way of ensuring such a property is by the existence of a fixed point in the renormalization group flow. At such a point, the theory reaches a scale-invariant regime and a continuum limit, i.e., the removal of a ultraviolet (UV) cutoff, can be achieved \cite{Wetterich:2019qzx}. From this point of view, a theory of quantum gravity based on continuum quantum field theory techniques which features a fixed point can very well define a fundamental theory. If such a fixed point sits at vanishing value of the couplings, the theory is dubbed asymptotically free while for non-vanishing (interacting) fixed points, the theory is said to be asymptotically safe. In \cite{weinberg1979ultraviolet}, S. Weinberg conjectured that quantum gravity could be an asymptotically safe theory despite its perturbative non-renormalizability. A major obstacle in order to test the validity of such a conjecture lies on the fact that being an interacting fixed point, perturbation theory might not be applicable and non-perturbative tools are mandatory. 

Two main perspectives were taken over the past decades. On the one hand, as in lattice QCD, non-perturbative information can be extracted from a lattice formulation of quantum gravity. Such a field has developed into the (Causal) Dynamical Triangulations program and evidence for a well-defined continuum limit was collected in the recent years\footnote{Evidence for a suitable continuum limit in four dimensions was also obtained within quantum Regge calculus, a slightly different approach from Dynamical Triangulations, see, e.g., \cite{Hamber:2015jja}.} \cite{loll2019quantum}. Alternatively, on the continuum quantum field theory side, functional renormalization group (FRG) techniques were applied to gravity in the seminal paper \cite{Reuter:1996cp}, leading to the asymptotic safety program for quantum gravity, see, e.g., \cite{Percacci:2017fkn,Reuter:2019byg,Eichhorn:2018yfc,Pereira:2019dbn,Dupuis:2020fhh,Eichhorn:2020mte,Reichert:2020mja} for recent reviews on the topic. See also \cite{Donoghue:2019clr,Bonanno:2020bil} for critical discussions on this program for quantum gravity. The application of FRG techniques has enabled a systematic search for a non-trivial fixed point in the renormalization group (RG) flow of quantum gravity, as proposed in \cite{weinberg1979ultraviolet}. Recently, Dyson-Schwinger equations were also adapted to the context of quantum gravity in \cite{Hamber:2020isy}, opening up an alternative semi-analytical continuum field-theoretic path to probe the existence of a non-trivial fixed point. All those perspectives are anchored on the firm pillars of standard quantum field theory and have collected compelling evidence for the existence of a well defined continuum limit in quantum gravity. This seems to contradict the standard lore that there exists a riddle between quantum field theory and GR. Nevertheless, even if such proposals for a fundamental theory of quantum spacetime fail to describe our world, one can take an effective field theory perspective and still use quantum field theory techniques to compute quantum-gravitational corrections \cite{Donoghue:1993eb,Hamber:1995cq,Bjerrum-Bohr:2014zsa}, showing that GR and quantum field theory have no a priori incompatibility. In this work, we explore the asymptotic safety scenario for quantum gravity within the framework of the FRG.  

Yet, being able to define a fundamental theory of quantum gravity cannot be fully satisfactory. First, it must be compatible with our current observations and, second, it must be able to make predictions that will be eventually tested. However, at this point, another difficulty about quantum gravity becomes evident. Quantum gravity effects are usually suppressed by the Planck scale making the measurement of direct quantum-gravity effects extremely challenging. However, one promising path towards testing the consistency of quantum gravity that has been taken in asymptotically safe quantum gravity is the coupling with matter fields, see, e.g., \cite{Dou:1997fg,Percacci:2002ie,Percacci:2003jz,Shaposhnikov:2009pv,Narain:2009fy,Zanusso:2009bs,Eichhorn:2011pc,Eichhorn:2012va,Dona:2012am,Dona:2013qba,Dona:2014pla,Labus:2015ska,Oda:2015sma,Meibohm:2015twa,Dona:2015tnf,Meibohm:2016mkp,Eichhorn:2016esv,Eichhorn:2016vvy,Biemans:2017zca,Hamada:2017rvn,Christiansen:2017qca,Eichhorn:2017eht,Eichhorn:2017egq,Eichhorn:2017sok,Christiansen:2017cxa,Eichhorn:2017als,Christiansen:2017gtg,Christiansen:2017qca,Eichhorn:2017ylw,Eichhorn:2017lry,Alkofer:2018fxj,Eichhorn:2018akn,Eichhorn:2018ydy,Eichhorn:2018nda,Pawlowski:2018ixd,Eichhorn:2018whv,deBrito:2019epw,Wetterich:2019zdo,Wetterich:2019rsn,Wetterich:2019rsn,Reichert:2019car,Burger:2019upn,Domenech:2020yjf,Daas:2020dyo,Eichhorn:2020sbo,deBrito:2020dta}. Quantum fluctuations of matter degrees of freedom affect the running of gravitational couplings and might eventually change the fixed-point structure. Eventually, such fluctuations can be strong enough to destroy the existence of a scale-invariant regime, invalidating the proposal of asymptotic safety. Conversely, quantum-gravitational fluctuations contribute to the running of matter couplings and can therefore affect their behavior at high energies. So far, there are several hints that the asymptotic safety scenario for quantum gravity based on metric theories of gravity which are invariant under the full diffeomorphism group is compatible with the matter content of the Standard Model of particle physics plus minor extensions such as right-handed neutrinos and some dark matter candidates. See \cite{Eichhorn:2018yfc} for more details. 

From the FRG perspective, we do not declare a classical (or microscopic) action from which we derive the quantum dynamics. Rather, one starts with a set of symmetries which should be respected by the underlying quantum theory. Such symmetries can be deformed by the introduction of regulator terms which play the role of effectively suppressing the functional integration of ``slow-modes" in the Wilsonian sense \cite{Morris:1998da,Berges:2000ew,Pawlowski:2005xe,Gies:2006wv,Rosten:2010vm,Dupuis:2020fhh}. They act as momentum-dependent mass-like terms for the elementary fields of the theory. Thus, a flowing action $\Gamma_k$, with $k$ a momentum scale, which obeys an exact flow equation \cite{Wetterich:1992yh,Morris:1993qb,Ellwanger:1993mw} is constructed upon such deformed symmetries and interpolates between the full quantum action\footnote{The generating functional of one-particle-irreducible diagrams.} $\Gamma$ and the microscopic action $S$ which enters the Boltzmann factor\footnote{We restrict the discussion to an Euclidean setting from now on.} of the path integral. The infinitely-many terms that define $\Gamma_k$ are parameterized by an infinite set of couplings which, in their dimensionless versions, are coordinates of the so-called theory space. It is thus expected that different symmetries define different (and inequivalent) theory spaces. From a quantum-gravity perspective, alternative theories of the gravitational field based on different symmetry principles will, very likely, define different quantum theories. However, there is a situation where this issue becomes subtle: there are theories which are based on different symmetries but feature the same classical dynamics. A famous example is GR and Unimodular Gravity (UG) \cite{Anderson:1971pn,vanderBij:1981ym,Buchmuller:1988yn,Buchmuller:1988wx,Unruh:1988in,Henneaux:1989zc,Unruh:1989db,ellis2011trace}. Hence, it is an immediate question whether dynamical equivalence remains true in the quantum realm.

In UG, the determinant of the metric $g_{\mu\nu}$ is fixed (non-dynamical) to a specific scalar density $\omega^2 (x)$, i.e., 
\begin{equation}
\mathrm{det}\,g_{\mu\nu} = \omega^2 (x)\,.
\label{in1}
\end{equation}
The symmetry group is reduced from the group of diffeomorphisms (\textit{Diff}) to transverse diffeomorphisms (\textit{TDiff}). Such a group is generated by transverse vectors $\epsilon^{\mathrm{T}\alpha}$, which satisfy $\nabla_\alpha \epsilon^{\mathrm{T}\alpha} = 0$, where the covariant derivative is defined with respect to the unimodular metric $g_{\mu\nu}$. The equivalence between GR and UG at the classical level is established by the use of the Bianchi identities, see, e.g., \cite{Percacci:2017fsy}. Quantum mechanically, however, the situation is much more subtle. In particular, there is a long-standing debate in the literature, see, e.g., \cite{Ardon:2017atk,Percacci:2017fsy,Fiol:2008vk,Saltas:2014cta,Padilla:2014yea,Smolin:2009ti,Smolin:2010iq,Alvarez:2015sba,Bufalo:2015wda,Upadhyay:2015fna,Eichhorn:2013xr,Eichhorn:2015bna,Benedetti:2015zsw,deBrito:2020rwu} for some recent references, if equivalence remains when quantum fluctuations are taken into account. Naively, however, one would expect that they are not equivalent at the quantum level since the nature of their quantum fluctuations is very different, i.e., the sum over histories is performed in very different configuration spaces. In particular, with a view towards asymptotic safety, the theory space defined by \textit{Diff}-invariant operators is different from the one associated with \textit{TDiff}. 

An important and very subtle difference in GR and UG lies on the treatment of the cosmological constant. In GR, it corresponds to a parameter which is fixed from the beginning and added by hand, as a coupling constant. It is, generically, subject to quantum corrections. In UG, the cosmological constant arises as an integration constant and, therefore, must be fixed by initial conditions. However, as such, it does not enter the classical action of the theory invariant under \textit{TDiff} since the measure term is just a fixed scalar density. Thus, in the former case, the cosmological constant defines a direction in the theory space while in the latter, it does not. This can indicate that, if asymptotically safe, \textit{Diff}- and \textit{TDiff}-invariant theories will not be equivalent. However, as pointed out in \cite{Ardon:2017atk,Percacci:2017fsy,deBrito:2020rwu}, this discussion is more subtle than it sounds and this is still not completely understood. 

In this paper, we leave aside this discussion and take \textit{TDiff} as the fundamental symmetry of the would-be theory of quantum gravity and look for further hints for the existence of a non-trivial UV fixed point. For simplicity, we call the hypothetical asymptotically safe theory as Unimodular Quantum Gravity (UQG), although it does not mean that our starting point for the quantization is the unimodular version of the Einstein-Hilbert action. Earlier results on asymptotic safety and unimodular gravity can be found in \cite{Eichhorn:2013xr,Saltas:2014cta,Eichhorn:2015bna,Benedetti:2015zsw}, where a non-trivial fixed point was obtained within truncations of the flowing action $\Gamma_k$. We provide a systematic analysis of fixed points within certain classes of truncations of $\Gamma_k$ by taking into account the following refinements: We take $\Gamma_k$ to be a function of the Ricci scalar and the quadratic contraction of Ricci tensors, i.e., 
\begin{equation}
\Gamma_k = \int \mathrm{d}^dx\,\omega\,f_k (R,R^{\mu\nu}R_{\mu\nu})\,.
\label{in2}
\end{equation}
For concreteness, subdivide the analysis in two classes, the first being the so-called $f(R)$-truncations, i.e., $f_k (R,R^{\mu\nu}R_{\mu\nu}) = f_k (R)$ and the second is defined by $f_k (R,R^{\mu\nu}R_{\mu\nu}) = F_k(R^2_{\mu\nu})+R Z_k(R^{2}_{\mu\nu})$, where $F_k(R^2_{\mu\nu})$ and $Z_k(R^2_{\mu\nu})$ are arbitrary functions. The second class of truncations was introduced in \cite{Falls:2017lst}. For both classes, we restrict the analysis to polynomial expansions of the curvatures on a spherical background, for technical simplicity. The other improvement that we implement in this work is that we employ the modified flow equation for UQG introduced in \cite{deBrito:2020rwu} due to the properties of the functional measure of UQG discussed in \cite{Ardon:2017atk,Percacci:2017fsy}. Moreover, we treat the anomalous dimensions of the elementary fields in different approximations as discussed in the context of UG in \cite{deBrito:2020rwu}. Finally, we minimally couple matter fields and analyse the impact they play on the gravitational couplings. 

The paper is presented as follows: in Sect. \ref{MethodModel} we provide a brief discussion of the background-field method, Faddeev-Popov quantization and FRG techniques for UQG and describe the model that we investigate in this work. The flow equation is set up in Sect. \ref{FlowEqSec}. In Sect. \ref{Projectionsf} we discuss the two classes of polynomial projections of $f(R,R^{\mu\nu}R_{\mu\nu})$ and the extraction of beta functions. Results for the interacting gravitational fixed-point structure both for pure gravity and gravity-matter systems are collected in Sect. \ref{OverallResults}. Finally, we draw our conclusions in Sec. \ref{Conclusions}. Technical details and expressions for the anomalous dimensions used in this work are presented in the appendices.

\section{Method and Model}\label{MethodModel}
\subsection{UG and the background-field method}\label{ugbfm}
One of the challenges in the application of coarse-graining techniques to quantum gravity lies on the lack of a notion of external scale which tells what is coarser or finer. In order to define such a structure, the background-field method \cite{Abbott:1981ke} is employed, but see \cite{Falls:2020tmj}. The metric $g_{\mu\nu}$ is split as a fixed background metric $\bar{g}_{\mu\nu}$ and a fluctuating part $h_{\mu\nu}$, i.e., $g_{\mu\nu} = f(\bar{g},h)_{\mu\nu}$, where $f$ is an arbitrary function. For the purposes of this work, it is highly convenient to choose the so-called exponential parameterization or split of the metric. It is defined by,
\begin{equation}
g_{\mu\nu}=\bar{g}_{\mu\alpha}[\exp(\kappa\,{h^{.}}_{.})]^{\alpha}_{\,\,\,\nu}=\bar{g}_{\mu\nu}+\kappa\,h_{\mu\nu}+\sum_{n=2}^{\infty}\frac{\kappa^n}{n!}h_{\mu\alpha_1}\cdots\, h^{\alpha_{n-1}}_{\,\,\,\nu},
\label{ugbfm1}
\end{equation}
with $\kappa = (32\pi G_N)^{1/2}$, with $G_N$ being the Newton constant. Systematic studies employing more general parameterizations in \textit{Diff}-invariant theories can be found in, e.g., \cite{Kalmykov:1995fd,Kalmykov:1998cv,Ohta:2016npm,Ohta:2016jvw,Goncalves:2017jxq,Ohta:2018sze} for perturbative quantum gravity and in \cite{Nink:2014yya,Gies:2015tca,Falls:2015qga,deBrito:2018jxt} in the context of asymptotic safety.

The unimodularity condition $\mathrm{det}\,g_{\mu\nu} = \omega^2 (x)$ can be easily implemented in Eq.~\eqref{ugbfm1} by choosing a unimodular background  $\mathrm{det}\,\bar{g}_{\mu\nu} = \omega^2 (x)$ and traceless fluctuations $\bar{g}^{\mu\nu}h_{\mu\nu} \equiv h^{\mathrm{tr}} = 0$. In a path-integral formulation, the restriction to traceless fluctuations around a unimodular background in the exponential parameterization automatically restricts the configuration space to unimodular metrics. It must be emphasized that the tracelessness of the fluctuation field is not taken as a gauge condition but rather as a constraint from the very definition of such a field. Such a formulation of the unimodularity condition is called as ``minimal version" and was put forward in \cite{Eichhorn:2013xr,Eichhorn:2015bna,Benedetti:2015zsw,Ardon:2017atk,Percacci:2017fsy}. This is one particular way to implement the unimodularity condition in the path integral and it is the one we adopt in this work. Different strategies to implement such a condition may bring different conclusions with respect to those reported here. See, e.g., \cite{Saltas:2014cta,Alvarez:2015sba,Herrero-Valea:2020xaq} for different perspectives on how to implement the unimodularity condition in the path integral.

\subsection{Digression on the Faddeev-Popov quantization in UG}
The Euclidean path integral of UQG is performed over trace-free fluctuations in the exponential parameterization as discussed in Subsect.~\ref{ugbfm} and can be written formally as
\begin{equation}
\mathcal{Z}_{\textrm{UQG}}=\int\frac{{\mathcal{D}h^{\textrm{tr-free}}_{\mu\nu}}}{V_{\textrm{TDiff}}}\,\e^{-S_{\textrm{UG}}[h;\bar{g}]},
\end{equation}
where $V_{\textrm{TDiff}}$ stands for the volume of the \textit{TDiff} group and the classical unimodular action $S_{\text{UG}}[h;\bar{g}]$ is invariant under \textit{TDiff} - but does not need to coincide with the unimodular version of the Einstein-Hilbert action. Applying the standard Faddeev-Popov procedure\footnote{See \cite{Ardon:2017atk,Percacci:2017fsy,deBrito:2020rwu} for further details about the Faddeev-Popov procedure in UG in its minimal version.}, one inserts in the functional integral a formal identity as
\begin{equation}\label{partfunction1}
\mathcal{Z}_{\textrm{UQG}}=\int\frac{{\mathcal{D}h^{\textrm{tr-free}}_{\mu\nu}}}{V_{\textrm{TDiff}}}\,\bigg(\int\mathcal{D}\epsilon^{\textrm{T}}\Delta_{\text{FP}}\,\delta(F^{\T})\bigg)\,\e^{-S_{\textrm{UG}}[h;\bar{g}]},
\end{equation}
where $\Delta_{\text{FP}}$ corresponds to the Faddeev-Popov determinant and $F^{\T}_ {\mu}[h;\bar{g}]=0$ corresponds to a transverse gauge-fixing condition. The Faddeev-Popov unity is obtained by the integration over transverse contravariant vectors $\epsilon^{\mathrm{T}\mu}$, which are the generators of \textit{TDiff}. In addition, we assume that the integration measures  are invariant under \textit{TDiff}.

Differently from the standard situation in the Faddeev-Popov prescription, one cannot factor out the integral over the transverse vectors $\epsilon^{\mathrm{T}\mu}$ and associate it with the $V_{\textrm{TDiff}}$. The main reason is that the transverse vector is metric-dependent. Following \cite{Ardon:2017atk,Percacci:2017fsy}, the volume of \textit{TDiff} is defined as
\begin{equation}
V_{\textrm{TDiff}}=\int\mathcal{D}\epsilon\,\delta(\bar{\nabla}_{\mu}\epsilon^{\mu})\,,
\end{equation}
where it is used that for unimodular metrics, $\nabla_{\mu}\epsilon^{\mu}= \bar{\nabla}_{\mu}\epsilon^{\mu}$. Decomposing $\epsilon^{\mu}$ in terms of transverse and longitudinal parts, i.e., $\epsilon^{\mu}=\epsilon^{\textrm{T}\mu}+\bar{\nabla}^{\mu}\varphi,$ it is straightforward to find that
\begin{equation}
V_{\textrm{TDiff}}=\textrm{Det}^{-1/2}(-\bar{\nabla}^2)\int\mathcal{D}\epsilon^{\textrm{T}}\,,
\end{equation}
as a proper representation of the volume of the \textit{TDiff} group. Therefore, the final expression of the path integral of unimodular quantum gravity is represented as
\begin{equation}\label{partfunction2}
\mathcal{Z}_{\textrm{UQG}}=\int\mathcal{D}h^{\textrm{tr-free}}_{\mu\nu}\mathcal{D}\bar{C}_{\alpha}\mathcal{D}C^{\beta}\,\textrm{Det}^{1/2}(-\bar{\nabla}^2)\,\e^{-S_{\text{UG}}[h;\bar{g}]-S_{\text{g.f.}+\text{gh.}}[h,\bar{C},C;\bar{g}]},
\end{equation}
where $S_{\text{g.f.}+\text{gh.}}[h,\bar{C},C;\bar{g}]$ corresponds to a gauge-fixing action along with Faddeev-Popov ghosts $\bar{C}_{\alpha}$ and $C^{\beta}$ term. In summary, Eq. (\ref{partfunction2}) is the proper formal definition of the Euclidean functional integral of UG (in its minimal version) and the starting point for applying functional renormalization group techniques.

\subsection{Functional Renormalization Group for UG}
In order to search for a fixed point in the renormalization group flow, we employ functional renormalization techniques. They are based on the Wilsonian perspective of momentum shell-wise integration of modes in the path integral. It can be performed in a smooth fashion through the introduction of a regulator term $\Delta S_k[\phi]$ in the action appearing in the Boltzmann factor of the Euclidean path integral. It implements a suppression of all field modes associated with momentum lower than an infrared scale $k$. Hence, the scale-dependent path integral is written as
\begin{equation}\label{partfunctionreg}
\mathcal{Z}_{k}[J]=\int\mathcal{D}\phi_{\Lambda_{\textrm{UV}}}\,\e^{-S[\phi]-\Delta S_k[\phi]+\int \mathrm{d}^dx\,J(x)\phi(x)}\,,
\end{equation}
where $\phi(x)$ represents a generic field content of the theory and  $J$ denotes its corresponding external source. The UV cutoff $\Lambda_{\textrm{UV}}$ is placed in order to make the measure well-defined. The regulator term is quadratic in the fields with a kernel function $\mathbf{R}_k(\Delta)$ as
\begin{equation}
\Delta S_k=\frac{1}{2}\int \mathrm{d}^dx\,\phi(x)\,\mathbf{R}_k(\Delta)\,\phi(x)\,.
\end{equation}
The suppression of field modes is achieved according to the spectrum of the Laplacian operator in $\mathbf{R}_k(\Delta)$, i.e., field configurations associated with eigenvalues $p^2$ such that $p^2<k^2$ will be suppressed in the functional integration. In this sense, the regulator is such that the path integral is evaluated over a shell from $\Lambda_{\textrm{UV}}$ to $k$, where $k$, therefore, acts as an infrared cutoff scale. Its introduction allows us to define the \textit{effective average action}, or the flowing action $\Gamma_k$, which is a scale-dependent functional and contains the effect of large quantum fluctuations. The flowing action interpolates between the full quantum effective action ($\Gamma_{k\rightarrow 0}=\Gamma$) and the classical/microscopic UV action $(\Gamma_{k\rightarrow \Lambda_{\textrm{UV}}}=S_{\Lambda})$. The flow of $\Gamma_k$ with $k$ is described by the Wetterich equation~\cite{Wetterich:1992yh,Morris:1993qb,Ellwanger:1993mw},
\begin{equation}
\partial_t\Gamma_k=\frac{1}{2}\textrm{STr}\bigg[\big(\Gamma^{(2)}_k+\textbf{R}_k\big)^{-1}\partial_t\textbf{R}_k\bigg]\,.
\end{equation}
where $\partial_t\equiv k\partial_k$, $\Gamma_k^{(2)}=\delta^2\Gamma_k/\delta\Phi\delta\Phi$ is the Hessian and $\textrm{STr}$  denotes the supertrace which contains a negative sign for Grassmann-valued fields and a factor of 2 for complex fields. The Wetterich equation receives an extra contribution coming from the extra determinant in (\ref{partfunction2}). Regularizing the extra determinant as $\textrm{Det}(-\bar{\nabla}^2)\mapsto \textrm{Det}(P_k(-\bar{\nabla}^2))$, where $P_k(z)=z + R_k(z)$, the flow equation for $\Gamma_k$ becomes
\begin{equation}\label{RegularizedFlowEq}
\partial_t\Gamma_k=\frac{1}{2}\textrm{STr}\bigg[\big(\Gamma^{(2)}_k+\textbf{R}_k\big)^{-1}\partial_t\textbf{R}_k\bigg]-\frac{1}{2}\textrm{Tr}\bigg(\frac{\partial_tR_k(-\bar{\nabla}^2)}{P_k(-\bar{\nabla}^2)}\bigg)\,.
\end{equation}
Note that, according to the procedure adopted earlier, the extra determinant does not generate additional fluctuation vertices and it arises from a proper application of the Faddeev-Popov procedure in UG. As a consequence, it contributes only to the ``background flow'' $\partial_t\Gamma_k[0;\bar{g}]$.

\subsection{Setting the truncation for unimodular gravity-matter systems}
In this work the key strategy to obtain information about the fixed-point structure is based on a mixed approach which combines the background-field approximation, vertex, and derivative expansion, similarly to what has been done in \cite{Eichhorn:2009ah,Eichhorn:2010tb,Christiansen:2012rx,Codello:2013fpa,Dona:2013qba,Dona:2014pla,Dona:2015tnf,deBrito:2020rwu}. On the one hand, in the background approximation the extraction of the beta functions of the dimensionless gravitational background couplings is obtained from the flow equation by turning off all the fluctuating fields after the computation of the Hessian. Moreover, the anomalous dimension of the graviton is identified with the running of the background Newton coupling. The ghost and matter anomalous dimensions are set to zero. On the other hand, a simultaneous vertex and derivative expansion generate  one-loop-structure diagrams as corrections to the flow of the two-point function of the fields and allow to unambiguously derive independent anomalous dimensions of all fluctuating fields. In this sense, the extra functional trace associated with the path integral measure only contributes to the background flow, since it only depends on the background metric. Furthermore, as an approximation, the different avatars of the Newton coupling, see, e.g., \cite{Eichhorn:2018akn,Eichhorn:2018ydy} in the vertices and graviton propagator are identified with its background value. 

We consider a truncation for the flowing action $\Gamma_k$ in the unimodular setting containing an arbitrary number of massless Gaussian matter fields\footnote{Meaning that we do not consider matter self-interactions.}, namely scalar, Abelian vector and Dirac fields, minimally coupled to gravity in four-dimensional Euclidean spacetime. Throughout the work we investigate a truncation of the form
\begin{equation}\label{workingtruncation}
\Gamma_{k}=\Gamma_k^{\textrm{gravity}}+\Gamma_{k}^{\textrm{matter}}+\Gamma_k^{\textrm{g.f.}}+\Gamma_k^{\textrm{gh.}}\,,
\end{equation}
where we follow~\cite{Falls:2017lst,Ohta:2018sze} and write the gravitational sector as\footnote{Herein we use the shorthand notation $\int_x\equiv\int \mathrm{d}^4x$.}
\begin{equation}
\Gamma_k^{\textrm{gravity}}[g_{\mu\nu}]=\frac{1}{16\pi \GNk}\int_x\,\omega\,f_{k}(R,R_{\mu\nu}^2)\,,
\end{equation}
where $\GNk$ is the dimensionful Newton coupling, $f_k$ is an arbitrary function of the Ricci scalar and the square of the Ricci tensor, $R_{\mu\nu}^2=R_{\mu\nu}R^{\mu\nu}$. The $k$-dependence comes from the scale-dependent renormalization factors and couplings of curvature invariants. The matter sector of the effective average action is composed of $N_{\phi}$ scalar fields, $N_{A}$ Abelian vector fields and $N_{\psi}$ Dirac spinors. Its complete action is given by
\begin{align}
\Gamma_k^{\textrm{matter}}[g,\phi,\bar{\psi},\psi,A]&=\frac{1}{2}\sum_{i=1}^{N_{\phi}}\int_x\,\omega\,g^{\mu\nu}\partial_{\mu}\phi_i\partial_{\nu}\phi_i\,+\,\sum_{i=1}^{N_{\psi}}\int_x\,\omega\,i\bar{\psi}_i\slashed{\nabla}\psi_i \nn\\
&+\frac{1}{4}\sum_{i=1}^{N_{A}}\int_x\,\omega\,g^{\mu\alpha}g^{\nu\beta}F_{i,\mu\nu}F_{i,\alpha\beta}\,,
\end{align}
where the summation index $i$ runs over the particle species and $F_{i,\mu\nu}$ is the field-strength of the Abelian gauge field $A_{i,\mu}$. We do not consider the running of wave-function renormalization factors of the matter fields as they do not lead to self-consistent results within the hybrid approximation adopted in this work (see Subsect. \ref{Results_PureGravity} for details).
The covariant Dirac operator $\slashed{\nabla}=e_{a}^{\,\,\,\mu}\gamma^a \nabla_{\mu}$ satisfies the Lichnerowicz relation
\begin{equation}
\Delta_{\textmd{L}_{\frac{1}{2}}}\psi_i=-\slashed{\nabla}^2\psi_i=\bigg(-\nabla^2+\frac{R}{4}\bigg)\psi_i\,.
\end{equation}
The fermion-gravity interaction is achieved through the vierbein and spin connection. In a spacetime manifold with vanishing torsion, these are not independent fields and can both be expressed in terms of $h_{\mu\nu}$ adapted to the exponential decomposition once the local $O(4)$ gauge invariance is gauge-fixed by a Lorentz symmetric condition (see App. B in Ref. \cite{deBrito:2019umw})\footnote{Alternatively, for the Dirac covariant operator $\slashed{\nabla}$, one could use the spin-base formalism \cite{Gies:2013noa,Gies:2015cka,Lippoldt:2015cea} expressed in accordance with the exponential parameterization. Both prescriptions are equivalent at the level of our computations.}. Moreover, due to the relation $g_{\mu\nu}={e^{a}}_{\mu}{e^{b}}_{\nu}\eta_{ab}$, the vierbein also obeys the unimodularity condition, i.e., $\det e^{a}_{\,\,\mu}=\omega$. Besides featuring a $\mathbb{Z}_2$ symmetry for the scalar sector under which $\phi_i \mapsto -\phi_i$, this toy model also features a shift symmetry $\phi_i\mapsto\phi_i+\text{const.}$, which prevents a scalar mass term. Additionally, an axial U(1) symmetry, i.e., $\psi_i \rightarrow e^{i\alpha\gamma_5}\psi_i$, $\bar{\psi_i}\rightarrow \bar{\psi_i}e^{i\alpha\gamma_5}$, prohibits a Dirac mass term. In this model the scalars and ``chiral'' fermions  are uncharged, not leading to gauge interactions. 

The gauge-fixing action for the \textit{TDiff} and the Abelian gauge symmetry is given by~\cite{Eichhorn:2013xr,Eichhorn:2015bna,Ardon:2017atk,Percacci:2017fsy,deBrito:2019umw}
\begin{equation}
\Gamma_k^{\textrm{g.f.}}[h,A;\bar{g}]
=\frac{1}{2a}\int_x\,\omega\,\bar{g}^{\mu\nu}F_{\mu}^{\text{T}}[h;\bar{g}]F_{\nu}^{\text{T}}[h;\bar{g}]+\frac{1}{2\zeta}\sum_{i=1}^{N_A}\int_x\,\omega\,(\bar{g}^{\mu\nu}\bar{\nabla}_{\mu}A_{i,\nu})^2,
\end{equation}
where $a$ and $\zeta$ represent gauge parameters for the gravitational and Abelian sectors, respectively. Using the transverse projector with respect to the background metric $(\mathscr{P}_{\text{T}})_{\mu\nu}=\bar{g}_{\mu\nu}-\bar{\nabla}_{\mu}(\bar{\nabla}^2)^{-1}\bar{\nabla}_{\nu}$, we define the transverse gauge-fixing function  as $F_{\mu}^{\text{T}}[h;\bar{g}]=\sqrt{2}\,(\mathscr{P}_{\text{T}})_{\mu}^{\,\,\,\nu}\,\bar{\nabla}^{\alpha}h_{\nu\alpha}$. This particular prescription of the gauge-fixing function is necessary since only the transverse diffeomorphism sector should be fixed. In this work we adopt the Landau gauge for both gravitational and Abelian sectors, i.e., $a\rightarrow 0$ and $\zeta\rightarrow 0$. The introduction of the transverse projector makes the gauge fixing for \textit{TDiff} a non-local functional. This could be avoided by allowing a higher-derivative operator in the gauge-fixing, see, e.g., \cite{Benedetti:2015zsw}.

Accompanying the gauge-fixing term there is the action for the Faddeev-Popov ghosts\footnote{Alternatively, the gauge-fixing and ghost terms for different formulations of unimodular gravity can be derived through BRST transformations, see \cite{Upadhyay:2015fna,Alvarez:2015sba}. See also \cite{Baulieu:2020obv,Baulieu:2020rpv} for a discussion of the BRST implementation of the unimodular gauge.} which reads
\begin{equation}
\Gamma_k^{\text{gh.}}[h,\bar{C},C,\bar{c},c;\bar{g}]=\int_x\,\omega\,\bar{C}_{\mu}\,\bar{g}^{\mu\nu}\frac{\delta F_{\nu}^{\T}[h;\bar{g}]}{\delta h_{\alpha\beta}}\delta^{Q}_{C}h_{\alpha\beta}+\sum_{i=1}^{N_A}\int_x\,\omega\,\bar{c}_i(-\bar{\nabla}^2)c_i,
\end{equation}
where $C^{\mu}\, (\bar{C}_{\mu})$ and $c_i\, (\bar{c}_i)$ are ghost and anti-ghost for the gravitational and Abelian sectors, respectively. In the unimodular setting, the Faddeev-Popov ghost for the gravitational sector is constrained by the transversality condition $\nabla_{\mu}C^{\mu}=\bar{\nabla}_{\mu}(g^{\mu\nu}C_{\nu})=0$ as a consequence of the transverse nature of the \textit{TDiff} generator. Furthermore, $\delta^{Q}_{C}h_{\alpha\beta}$ corresponds to the ``quantum'' transformation of the fluctuation field with generator being the ghost field $C^{\mu}$. The explicit implementation of the gravitational ghost sector suitable for the exponential split of the metric is discussed in \cite{Eichhorn:2017sok,deBrito:2020rwu}.

A minimal and diagonal Hessian together with an exact inversion of the kinetic operators can be achieved in a spherical background  
and by decomposing the fluctuation field $h_{\mu\nu}$ into the York basis~\cite{York:1973ia}, namely,
\begin{equation}
h_{\mu\nu}=h_{\mu\nu}^{\textrm{TT}}+2\bar{\nabla}_{(\mu}\xi_{\nu)}+\Big(\bar{\nabla}_{\mu}\bar{\nabla}_{\nu}-\frac{1}{4}\bar{g}_{\mu\nu}\bar{\nabla}^2\Big)\sigma.
\end{equation}
We emphasize the absence of the trace mode in the decomposition due to the unimodularity condition. No non-local field redefinitions are performed and, as a consequence, the Jacobians arising from the change of variables are taken into account in the flow equation by a suitable regularization of the resulting determinants. Furthermore, appropriate wave-function renormalization factors are introduced for the gravitational ghost fields and for each spin sector of the gravitational fluctuation according to
\begin{subequations}
\begin{align}\label{renormfactors}
h_{\mu\nu}^{\text{TT}}\mapsto \mathcal{Z}_{k,\text{TT}}^{1/2}\,h_{\mu\nu}^{\text{TT}}, \qquad \xi_{\mu}\mapsto \mathcal{Z}_{k,\xi}^{1/2}\,\xi_{\mu}, \qquad \sigma\mapsto \mathcal{Z}_{k,\sigma}^{1/2}\,\sigma,
\end{align}
\begin{align}\label{renormghost}
C^{\mu}\mapsto \mathcal{Z}_{k,C}^{1/2}\,C^{\mu},\qquad \bar{C}_{\mu}\mapsto \mathcal{Z}_{k,C}^{1/2}\,\bar{C}_{\mu}.
\end{align}
\end{subequations}
The wave-function renormalization factors $\mathcal{Z}_{k,\Phi}$ with $\Phi=(\TT,\xi,\sigma,C,\bar{C})$ generate anomalous dimensions $\eta_{\Phi}=-\partial_t\ln \mathcal{Z}_{k,\Phi}$
and contribute to the system of beta functions of Newton and higher curvature couplings.

The Abelian gauge potentials are also decomposed into its transverse and longitudinal parts,
\begin{equation}\label{vectordecomposition}
A_{i,\mu}=A_{i,\mu}^{\textrm{T}}+\bar{\nabla}_{\mu}\big[(-\bar{\nabla}^2)^{-1/2}A_{i}^{\textrm{L}}\big], \qquad \bar{\nabla}_{\mu}A_{i}^{\textrm{T}\mu}=0.
\end{equation}
Contrary to the fluctuation field decomposition, herein we chose to insert an inverse square root of the Bochner Laplacian, $-\bar{\nabla}^2$, so that the Jacobian associated with this field redefinition is a simple identity. 
 
\section{Setting the flow equation}\label{FlowEqSec}
 
At the practical level, the right-hand side of the flow equation is expanded on the same basis as the one chosen for the truncation such that a suitable projection rule selects the beta functions associated to each coupling. The beta functions of the background gravitational couplings can be read off at zeroth order in the fluctuating fields and the elements of the Hessian employed in such a computation are listed in the App. \ref{hessianelements}. The entries of the regulator function $\mathbf{R}_k$ are built from the following prescription~\cite{Codello:2008vh}
\begin{equation}
\mathbf{R}_{k,\varphi_i\varphi_j}(\Delta)=\Gamma^{(2)}_{k,\varphi_i\varphi_j}(\Delta)\bigg|_{\Delta\mapsto\Delta+R_k(\Delta)}-\Gamma^{(2)}_{k,\varphi_i\varphi_j}(\Delta)\,,
\end{equation}
where $\Delta$ is an appropriate coarse-graining operator and $\Gamma^{(2)}_{k,\varphi_i\varphi_j}$ denotes the second functional derivative of $\Gamma_k$ with respect to the fields $\varphi_i$ and $\varphi_j$. For the regulator kernel (i.e., for the shape function that enters the regulator) we choose the Litim-type cutoff~\cite{Litim:2001up}
\begin{equation}\label{litimcutoff}
R_k(\Delta)=(k^2-\Delta)\theta(k^2-\Delta)\,.
\end{equation}
 In particular, we adopt two types of regularization schemes distinguished by two common choices of coarse-graining operators~\cite{Codello:2008vh} namely, the Bochner-Laplacian, $-\bar{\nabla}^2$ (Type I), and the Lichnerowicz-Laplacians, $\Delta_{\L s}$ (Type II), which are connected by the Lichnerowicz relations on a four-dimensional maximally symmetric Euclidean background
\begin{align}
\Delta_{\textmd{L}_2} = -\bar{\nabla}^2 + \frac{2}{3}\bar{R} \,,\quad\qquad
\Delta_{\textmd{L}_1} = -\bar{\nabla}^2 + \frac{1}{4}\bar{R} \,,\quad\qquad
\Delta_{\textmd{L}_0} = -\bar{\nabla}^2 \,. \qquad
\end{align}

Inspired by \cite{Alkofer:2018fxj,Ohta:2015fcu,Demmel:2015oqa,Dona:2013qba}, in order to accommodate both regularization prescriptions, we define the ``interpolating'' coarse-graining operator for each spin-$s$ sector as $\Delta_s=\Delta_{\L s}-\gamma_s\bar{R}$, where the endomorphism parameters were introduced such that the choice $\gamma_0 = \gamma_{\frac{1}{2}} = \gamma_1 = \gamma_2=0$ implements the Lichnerowicz-Laplacians and $\gamma_0=0$, $\gamma_{\frac{1}{2}}=1/4$, $\gamma_1=1/4$ and $\gamma_2=2/3$ result in the Bochner-Laplacian.
According to~\cite{Dona:2012am}, in order to account for a correct sign for the fermionic contributions to the Newton coupling constant, a Type II regularization must be adopted. The fermionic regulator function then is written as
\begin{equation}
\mathbf{R}_{k,\psi\psi}(z)=i\big[\sqrt{1+(k^2/z-1)\theta(k^2/z-1)}-1\big]\slashed{\nabla},
\end{equation}
where $z=\Delta_{\L_{\frac{1}{2}}}$. Furthermore, since for massless scalar fields both types of regularizations are equal, we adopt for simplicity the Type II regularization prescription for the gauge fields as well~\cite{Dona:2013qba}. Henceforth, we explore in this work both types of coarse-graining operators only in the gravitational sector.

For the truncation (\ref{workingtruncation}), the running of the dimensionless gravitational couplings can be read off, at zeroth order in the fields, from the following flow equation written in the Landau gauge, i.e., $a=0,$
\begin{align}\label{Flow_Eq_York_Decomposed}
\pt_t \Gamma_{k} &=\frac{1}{2}\Tr_{(2)}\Big[\textbf{G}_{\TT} \,
\pt_t\textbf{R}_{k,\TT} \Big] +
\frac{1}{2}\Tr_{(1)}^{\prime}\Big[\textbf{G}_{\xi\xi} \,
\pt_t\textbf{R}_{k,\xi\xi} \Big] +
\frac{1}{2}\Tr_{(0)}^{\prime\prime}\Big[\textbf{G}_{\sigma\sigma} \,
\pt_t\textbf{R}_{k,\sigma\sigma} \Big] -
\Tr_{(1)}\Big[\textbf{G}_{C\bar{C}} \,
\pt_t\textbf{R}_{k,C\bar{C}} \Big]   \nn \\
&-
\frac{1}{2}\Tr_{(0)}^\prime\bigg[\frac{}{}\!\!\left( \Delta_0 + R_{k}(\Delta_0)\right)^{-1} \!
\pt_t R_{k}(\Delta_0) \bigg]\,+\,\frac{N_{\phi}}{2}\Tr_{(0)}\Big[\textbf{G}_{\phi\phi}\,\pt_t\textbf{R}_{k,\phi\phi}\Big]-N_{\psi}\Tr_{(1/2)}\Big[\textbf{G}_{\psi\psi}\,\pt_t\textbf{R}_{k,\psi\psi}\Big] \nn \\
&+\frac{N_A}{2}\Tr_{(1)}\Big[\textbf{G}_{A^{\T}A^{\T}}\,\pt_t\textbf{R}_{k,A^{\text{T}}A^{\text{T}}}\Big]+\frac{N_A}{2}\Tr_{(0)}^{\prime}\Big[\textbf{G}_{A^{\L}A^{\L}}\,\pt_t\textbf{R}_{k,A^{\text{L}}A^{\text{L}}}\Big]-N_{A}\Tr_{(0)}\Big[\textbf{G}_{c\bar{c}}\,\pt_t\textbf{R}_{k,c\bar{c}}\Big]\nn \\
&+\,\mathcal{T}^{\textmd{Jacob.}}_{(1)} \,+\, \mathcal{T}^{\textmd{Jacob.}}_{(0)}\,,
\end{align}
with $\textbf{G}_{ij}=\Big[\big(\Gamma_{k}^{(2)}+\textbf{R}_{k}\big)^{-1}\Big]_{ij}$ for every pair $(i,j)$. The first term in the second line corresponds to the extra determinant accounting for an appropriate treatment of the volume of the \textit{TDiff} group. The last two terms in the fourth line denote additional contributions coming from the Jacobian associated with the change of variables $h_{\mu\nu}\mapsto \{h^{\TT}_{\mu\nu},\xi_{\mu},\sigma\}$. Upon regularization $\Delta_s\mapsto \Delta_s+R_k(\Delta_s)$, these contributions manifest themselves as the following additional traces
\begin{subequations}
	\begin{align}
	\mathcal{T}^{\textmd{Jacob.}}_{(1)} = 
	-\frac{1}{2}\Tr^\prime\left[ \left(\Delta_1 +  R_k(\Delta_1)+\frac{2\gamma_1-1}{2}\bar{R} \right)^{-1} \pt_t R_k(\Delta_1)\right] \,,
	\end{align}
	\begin{align}
	\mathcal{T}^{\textmd{Jacob.}}_{(0)} = 
	&-\frac{1}{2}\Tr^{\prime\prime}\left[\left(\Delta_0 + R_k(\Delta_0) - \frac{1}{3}\bar{R} \right)^{-1} \pt_t R_k(\Delta_0)\right] \nn\\
	&-\frac{1}{2}\Tr^{\prime\prime}\left[ \!\frac{}{}\left(\Delta_0 +  R_k(\Delta_0) \right)^{-1} \pt_t R_k(\Delta_0)\right] \,.
	\end{align}
\end{subequations}

The computation of the traces in the FRG equation is performed with standard heat kernel techniques. All the necessary technical tools and notation are collected in App. \ref{heatkernel}. In general, the result of the trace computation leads to very long expressions and, therefore, we shall not report explicit results here. The anomalous dimensions can be computed by acting with two functional derivatives with respect to the fields on the flow equation (\ref{Flow_Eq_York_Decomposed}) and expanding the full scale-dependent effective action in powers of the fields on a flat background. Their extraction is then obtained by means of suitable projection rules.  We follow the same strategy as in \cite{deBrito:2019umw,deBrito:2020rwu}. The explicit expressions for the anomalous dimensions used in this work are reported in App.~\ref{etas}.

\section{$f(R,R_{\mu\nu}R^{\mu\nu})$ projections and extraction of beta functions}\label{Projectionsf}

In this section we discuss the extraction of beta functions from two different types of polynomial projections of the $f_k(R,R_{\mu\nu}R^{\mu\nu})$-truncation minimally coupled with Gaussian matter degrees of freedom in the unimodular setting.

To extract the beta functions of the background gravitational couplings from the FRG equation, we can adopt a projection which consists in setting to zero all fluctuation fields. Within the background approximation, the truncation (\ref{workingtruncation}), inserted in the left-hand side of (\ref{Flow_Eq_York_Decomposed}), leads to a flow equation of the form
\begin{equation}\label{General_Form_Flow_Eq}
\frac{1}{16\pi\GNk}\bigg[-\eta_{\N}f_{k}\left(\bar{R},\bar{R}_{\mu\nu}^2\right)+\ptt f_k\left(\bar{R},\bar{R}_{\mu\nu}^2\right)\bigg]=\mathscr{F}\left(f_k,f_k^{(m,n)},\eta_i,\ptt f_k, \ptt f_k^{(m,n)},N_{\Psi}\right),
\end{equation}
where the left-hand side of (\ref{General_Form_Flow_Eq}) features the ``background anomalous dimension'' $\eta_\N=-\ptt\ln\mathcal{Z}_\N$ with $\mathcal{Z}_\N=(16\pi\GNk)^{-1}$ and $N_{\Psi}=(N_{\phi},N_{A},N_{\psi})$. The dependence on $\vec{\eta}=(\eta_{\TT},\eta_{\xi},\eta_{\sigma},\eta_{c})$ on the right-hand side of (\ref{General_Form_Flow_Eq}) comes from the regulator insertion $\ptt\mathbf{R}_k$ associated with each field sector. Moreover, we adopt the compact notation
\begin{align}
f_k^{(m,n)} = \frac{\pt^{m+n} f_k(\bar{R},\bar{X})}{\pt \bar{R}^m \pt \bar{X}^n} \,,
\end{align}
with $\bar{X}=\bar{R}_{\mu\nu}^2$. In order to obtain concrete results, we resort to polynomial truncations.  In principle, had we performed all calculations in a generic background, the most general polynomial expansion (within the class of the $f_k(R,R_{\mu\nu}^2)$-truncation) would be of the form
\begin{equation}
f_k(R,R_{\mu\nu}^2)=\sum_{n_1,n_2}\bar{\alpha}_k^{(n_1,n_2)}R^{n_1}\,(R_{\mu\nu}R^{\mu\nu})^{n_2}
\end{equation}
where $\bar{\alpha}_k^{(n_1,n_2)}$ denotes the scale-dependent couplings. The running of the couplings $\bar{\alpha}_k^{(n_1,n_2)}$ can be extracted by expanding both sides of the flow equation (\ref{General_Form_Flow_Eq}) in powers of $\bar{R}$ and $\bar{R}_{\mu\nu}$ and comparing the coefficients of the same curvature invariants on both sides order by order. Unfortunately, this procedure carries an ambiguity for a spherical background as the invariant $\bar{R}_{\mu\nu}^2$ collapses to $\frac{1}{4}\bar{R^2}$. As a consequence, the running of any two couplings $\bar{\alpha}_k^{(p_1,p_2)}$ and $\bar{\alpha}_k^{(q_1,q_2)}$ can no longer be disentangled for all pairs $(p_1,p_2)$ and $(q_1,q_2)$ satisfying the relation $p_1+2p_2=q_1+2q_2$. A way to bypass this ambiguity, without appealing to a generic background, is to impose some restriction on the function $f_k(R,R_{\mu\nu}^2)$.

\subsection{$f(R)$ polynomial projection}

In this subsection, we consider the particular case corresponding to the $f(R)$-approximation, which can be directly obtained by neglecting the $R_{\mu\nu}^2$-dependence in our truncation. For practical computations, we focus on the polynomial approximation
\begin{equation}\label{fR_Truncation}
f_k(R)=-R+\sum_{n=2}^{N}k^{2-2n}\alpha_{k,n}R^n\,,
\end{equation}
where $\alpha_{k,n}$ corresponds to scale-dependent dimensionless couplings and the parameter $N$ stands for a positive integer number that fixes the maximal degree of the polynomial truncation. This truncation was largely explored in the context of ``standard" asymptotic safety. See, e.g., \cite{Codello:2007bd,Machado:2007ea,Codello:2008vh,Demmel:2012ub,Dietz:2012ic,Dietz:2013sba,Demmel:2013myx,Demmel:2014sga,Falls:2013bv,Falls:2014tra,Demmel:2014hla,Demmel:2015oqa,Ohta:2015efa,Ohta:2015fcu,Gonzalez-Martin:2017gza,Christiansen:2017bsy,Falls:2017lst,Alkofer:2018fxj,Alkofer:2018baq,deBrito:2018jxt,Falls:2018ylp,Burger:2019upn}. The coefficient of the first term is normalized to $-1$ in order to recover the unimodular Einstein-Hilbert truncation once higher-order powers of the curvature scalar are neglected. Furthermore, the zeroth-order term, which would be proportional to the cosmological constant, is absent since we are dealing with a unimodular theory space\footnote{It is important to emphasize that, since the introduction of the regulator breaks BRST invariance, mass-like terms for the graviton can be generated and mimick the effect of the cosmological constant~\cite{deBrito:2020rwu}. Nevertheless, such terms arise as a symmetry-breaking effect due to the regulator and are not present in the background approximation.}.

We extract the system of beta functions associated with the dimensionless Newton coupling $G_k$ and the set of dimensionless couplings $\{\alpha_{k,n}\}_{n=2,\cdots, N}$ by plugging Eq.~(\ref{fR_Truncation}) into the flow equation~(\ref{General_Form_Flow_Eq}) and expanding both sides of it up to order $\bar{R}^{N}$. In this case, the flow equation leads to the following structure
\begin{align}
&\frac{\eta_\N}{16\pi G_k}k^2\bar{R}+\frac{1}{16 \pi G_k}\sum_{n=2}^{N}\bigg((2-2n-\eta_\N)\alpha_{k,n}+\beta_{\alpha}^{(n)}\bigg)k^{4-2n}\bar{R}^n \equiv\sum_{n=1}^N\mathscr{H}_{n}\left(\alpha_k,N_{\Psi},\vec{\eta},\beta_{\alpha}^{(m)}\right)\,,
\end{align}
where $G_k=k^{2}\,G_{\N,k}$ is the dimensionless Newton coupling and we have defined $\beta_{\alpha}^{(n)}=\ptt \alpha_{k,n}$. The function $\mathscr{H}_n$ has the general schematic form
\begin{align}
&\mathscr{H}_{n}\left(\alpha_k,N_{\Psi},\vec{\eta},\beta_{\alpha}^{(m)}\right)\equiv\mathscr{A}_n(\alpha_k)+\tilde{\mathscr{A}}_n(N_{\Psi})+\sum_{j=1}^{4}\mathscr{B}_{n,j}(\alpha_k)\eta_{j}+\sum_{m=2}^{N}\mathscr{M}_{n,m}(\alpha_k)\beta_\alpha^{(m)}\,.
\end{align}
The coefficients $\mathscr{A}_n$, $\tilde{\mathscr{A}}_n$, $\mathscr{B}_{n,j}$ and $\mathscr{M}_{n,m}$ are scheme-dependent quantities and can be computed analytically for the Litim's cutoff. By matching contributions according to the power of scalar curvature, we arrive at the RG equations
\begin{subequations}
\begin{align}
\beta_G=2G_k\bigg[1+8\pi G_k\,\mathscr{H}_{1}\left(\alpha_k,N_{\Psi},\vec{\eta},\beta_{\alpha}^{(m)}\right)\bigg]\,,\label{betaGfR}
\end{align}
\begin{align}
\beta_\alpha^{(n)}=(\eta_\N+2n-2)\alpha_{k,n}+16\pi G_k\,\mathscr{H}_{n}\left(\alpha_k,N_{\Psi},\vec{\eta},\beta_{\alpha}^{(m)}\right)\,,\label{betaAlphafR}
\end{align}
\end{subequations}
with $n=2,\,\ldots,\,N$. In Eq.~(\ref{betaGfR}) we have used $\eta_\N=G_k^{-1}\beta_G-2$. We highlight that the system of RG equations defined by (\ref{betaGfR}) and (\ref{betaAlphafR}) provides only implicit results for the beta functions $\beta_G$ and $\beta_\alpha^{(n)}$. Furthermore, the system is not closed because of the presence of the anomalous dimensions $(\eta_{\TT},\eta_{\xi},\eta_{\sigma},\eta_{c})$. In principle, this system can be solved analytically in order to extract explicit results for $\beta_G$ and $\beta_\alpha^{(n)}$ once a prescription to obtain the anomalous dimensions is adopted. In Sect. \ref{Results_PureGravity}, we will consider two types of prescriptions: the standard ``RG-improvement'' approximation\footnote{In Appendix C, the reader can find more details about the identification of the background anomalous dimension with the one derived from the second derivative of the flowing action with respect to fluctuations.} and a hybrid semi-perturbative approximation based on an independent calculation of the anomalous dimensions using the derivative expansion around a flat background. We emphasize that the latter prescription is somewhat not self-consistent since it glues together results obtained under different schemes and backgrounds. Nonetheless, we take this tentative choice to obtain results beyond the background approximation. As usual in functional methods, the use of hybrid schemes might be justified \textit{a posteriori} if the underlying results find good convergence properties. Nevertheless, the final expressions for the system of RG equations are very lengthy and not worth being reported here.

The so-called non-trivial or non-Gaussian fixed-point (NGFP) solutions (denoted as $G^*$ and $\alpha^*_n$) may be obtained in terms of the following equations
\begin{subequations}
\begin{align}
2G^*\bigg[1+8\pi G^*\,\mathscr{H}_1\left(\alpha^*,N_{\Psi},\vec{\eta}\,\big|_*,0\right)\bigg]=0\label{FP_EQ1_fR},
\end{align}
\begin{align}
(2n-4)\alpha_n^*+16\pi G^*\,\mathscr{H}_n\left(\alpha^*,N_{\Psi},\vec{\eta}\,\big|_*,0\right)=0\label{FP_EQ2_fR}.
\end{align}
\end{subequations}
 The notation $(\cdots)\big|_*$ indicates that the quantity in parenthesis is evaluated at the fixed-point solution. In Sect. \ref{Results_PureGravity}, we report numerical evidence for interacting fixed-point solutions associated with various choices of $N$.

\subsection{\texorpdfstring{$F(R_{\mu\nu}R^{\mu\nu})+R\,Z(R_{\mu\nu}R^{\mu\nu})$}{TEXT} polynomial projection}
Another way of bypassing the technical problem of distinguishing the invariants $R^2$ and $R_{\mu\nu}^2$ on a spherical background is to consider an alternative class of truncation, which is characterized by the following decomposition\footnote{Hereafter we refer to such decomposition as FZ-truncation.}
\begin{equation}
f_k(R,R_{\mu\nu}^2)=F_k(R_{\mu\nu}^2)+R\,Z_k(R_{\mu\nu}^2),
\end{equation}
where $F_k(R_{\mu\nu}^2)$ and $Z_k(R_{\mu\nu}^2)$ denote scale-dependent arbitrary functions of the invariant $R_{\mu\nu}^2$. This class of truncation was first investigated in \cite{Falls:2017lst} as an approach to include effects beyond the tensor structure of the Ricci scalar. For practical calculations, we restrict our analysis to polynomial truncations defined by
\begin{subequations}
\begin{align}
F_k(R_{\mu\nu}R^{\mu\nu})=\sum_{n=1}^{N_{F}}k^{2-4n}\rho_{k,2n}(R_{\mu\nu}R^{\mu\nu})^n\label{FfunctionFZ}\,,
\end{align}
\begin{align}
Z_k(R_{\mu\nu}R^{\mu\nu})=-1+\sum_{n=1}^{N_Z}k^{-4n}\rho_{k,2n+1}(R_{\mu\nu}R^{\mu\nu})^n\label{ZfunctionFZ}\,,
\end{align}
\end{subequations}
where $N_F=\lfloor N/2\rfloor$ and $N_Z=\lfloor (N-1)/2\rfloor$, with $\lfloor \cdots \rfloor$ representing the floor function. We denote as $\{\rho_{k,n}\}_{n=2,\,\cdots,\,N}$ the set of scale-dependent dimensionless couplings. This particular decomposition allows us to unambiguously extract the beta functions associated with the set of higher-curvature couplings $\{\rho_{k,n}\}_{n=2,\,\cdots,\,N}$ even in a spherical background. This follows from the fact that $F_k(R_{\mu\nu}R^{\mu\nu})$ and $R \,Z_k(R_{\mu\nu}R^{\mu\nu})$ contribute to the left-hand side of the flow equation with even and odd powers of $\bar{R}$, respectively, when projected onto a spherical background.
Following the same procedure outlined in the $f(R)$-approximation, we extract the system of RG equations associated with the dimensionless couplings $G_k$ and $\{\rho_{k,n}\}_{n=2,\,\cdots,\,N}$ by plugging (\ref{FfunctionFZ}) and (\ref{ZfunctionFZ}) into both left-hand side and right-hand side of Eq. (\ref{General_Form_Flow_Eq}) to obtain the following expressions
\begin{subequations}
\begin{align}
\frac{1}{16\pi\GNk}&\bigg[-\eta_{\N}f_{k}\left(\bar{R},\bar{R}_{\mu\nu}^2\right)+\ptt f_k\left(\bar{R},\bar{R}_{\mu\nu}^2\right)\bigg]\bigg|_{S^4}=\nn\\
&=\frac{\eta_\N}{16 \pi G_k}k^2\bar{R}+\frac{1}{16\pi G_k}\sum_{n=1}^{N_F}\frac{k^{4-4n}}{4^n}\left(\beta_{\rho}^{(2n)}+(2-4n-\eta_\N)\rho_{k,2n}\right)\bar{R}^{2n}\nn\\
&+\frac{1}{16\pi G_k}\sum_{n=1}^{N_Z}\frac{k^{2-4n}}{4^n}\left(\beta_{\rho}^{(2n+1)}-(4n+\eta_\N)\rho_{k,2n+1}\right)\bar{R}^{2n+1}\,,
\end{align}
for the right-hand side of the flow equation and
\begin{align}
&\mathscr{F}\left(f_k,f_k^{(m,n)},\eta_i,\ptt f_k,\ptt f_k^{(m,n)},N_{\Psi}\right)\bigg|_{S^4}=\nn\\
&=\sum_{n=1}^{N}\bigg(\mathscr{A}_n(\rho_k)+\tilde{\mathscr{A}}_n(N_{\Psi})+\sum_{j=1}^{4}\mathscr{B}_{n,j}(\rho_k)\eta_{j}+\sum_{m=2}^{N}\mathscr{M}_{n,m}(\rho_k)\beta_{\rho}^{(m)}\bigg)\,,
\end{align}
\end{subequations}
for the left-hand side of the flow equation. The notation $(\cdots)\big|_{S^4}$ denotes the projection on the spherical background. Performing an order by order comparison in the curvature scalar $\bar{R}$, one easily obtains the system of RG equations for the FZ-truncation and the equations for the fixed-point solutions $G^*$ and $\{\rho^*_{n}\}_{n=2\,\cdots\,N}$. The final expressions are quite similar to the condensed expressions in (\ref{betaGfR}) and (\ref{betaAlphafR}) for the flow equations and (\ref{FP_EQ1_fR}) and (\ref{FP_EQ2_fR}) for the fixed-point equations, respectively. Nevertheless, the explicit form of the coefficients $\mathscr{A}_n$, $\tilde{\mathscr{A}}_n$, $\mathscr{B}_{n,j}$ and $\mathscr{M}_{n,m}$ obtained within the FZ-truncation differs considerably from the ones extracted via $f(R)$-approximation.

\section{Results for the interacting gravitational fixed-point structure}\label{OverallResults}

\subsection{Pure gravity systems}\label{Results_PureGravity}
In the following, we present our results regarding the fixed-point structure extracted within the two previously defined polynomial truncations, focusing on the case without matter fields, i.e., by setting $N_{\phi}=N_A=N_{\psi}=0$. The analysis including matter contributions is reported in Sect. \ref{Impact_Matter}.

The fixed-point equations for both truncations are considerably complicated so that we resort to a numerical recursive solution of the fixed-point equations for the higher-order couplings in terms of the two lowest ones and adopt a bootstrap search strategy~\cite{Falls:2013bv,Falls:2014tra} to select suitable fixed-point solutions and critical exponents. Within the background approximation, a canonical choice of closure for the system of RG equations is obtained with the RG-improved anomalous dimensions (see, e.g., \cite{Codello:2013fpa}) $\eta_{\TT}=\eta_{\sigma}=G_k^{-1}\ptt G_k-2$ and $\eta_{\xi}=\eta_c=0$. Alternatively, a hybrid closure is obtained by improving the background approximation with anomalous dimensions computed in an independent way via the vertex expansion (see, e.g., \cite{Christiansen:2012rx,Codello:2013fpa,deBrito:2020rwu}).

For the $f(R)$-approximation within the RG-improved closure, we have performed the search for fixed-point candidates at each order of the approximation from $N=1$ to $N=20$. It is worth mentioning that, in the case of standard ASQG, i.e., where the theory space is defined by all \textit{Diff}-invariant operators, the fixed-point analysis has been performed within polynomial expansions involving terms up to $R^{70}$~\cite{Falls:2018ylp}.

\begin{figure}[!t]
   \begin{center}
        \includegraphics[height=5.0cm]{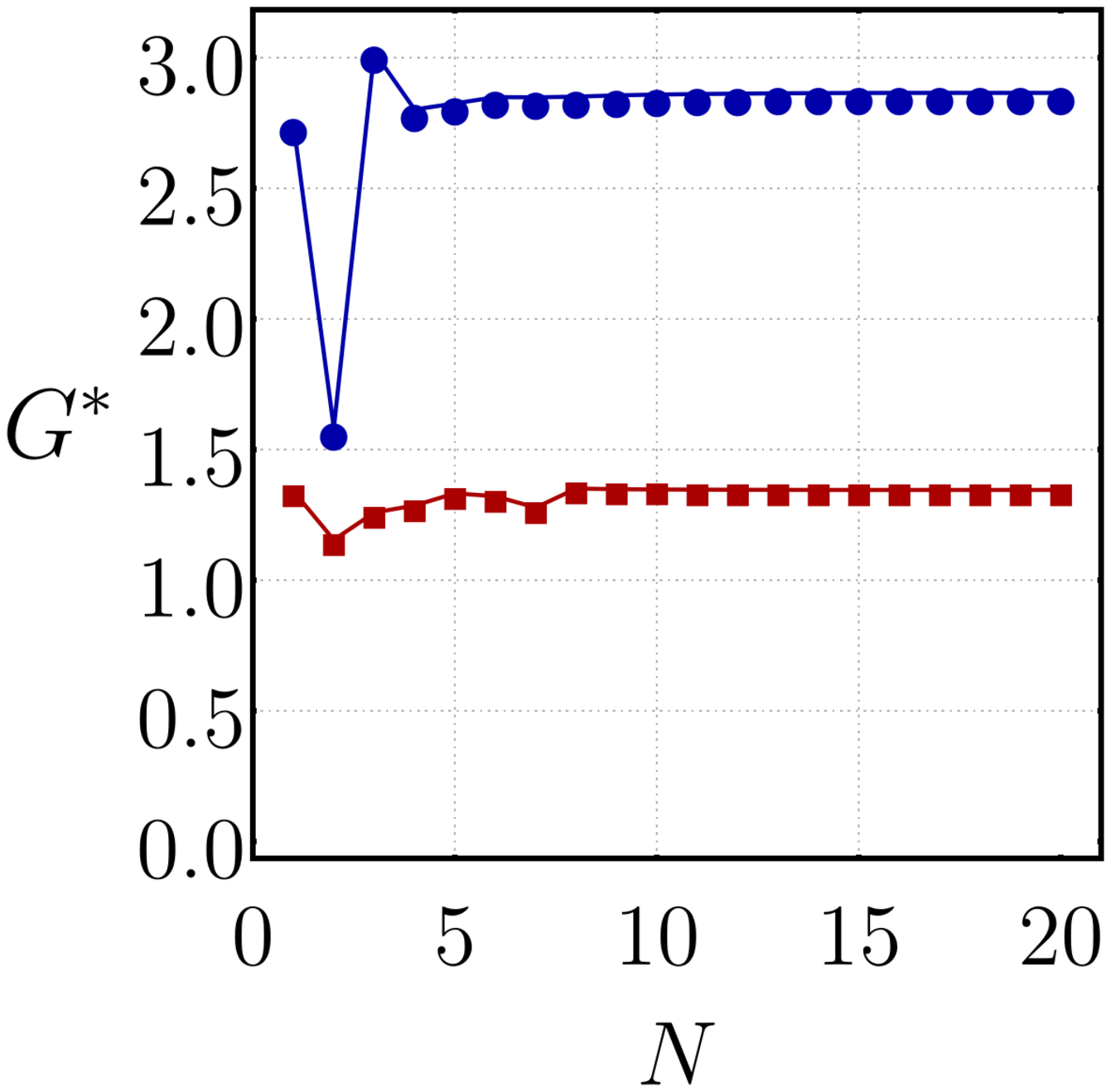} \quad
        \includegraphics[height=5.0cm]{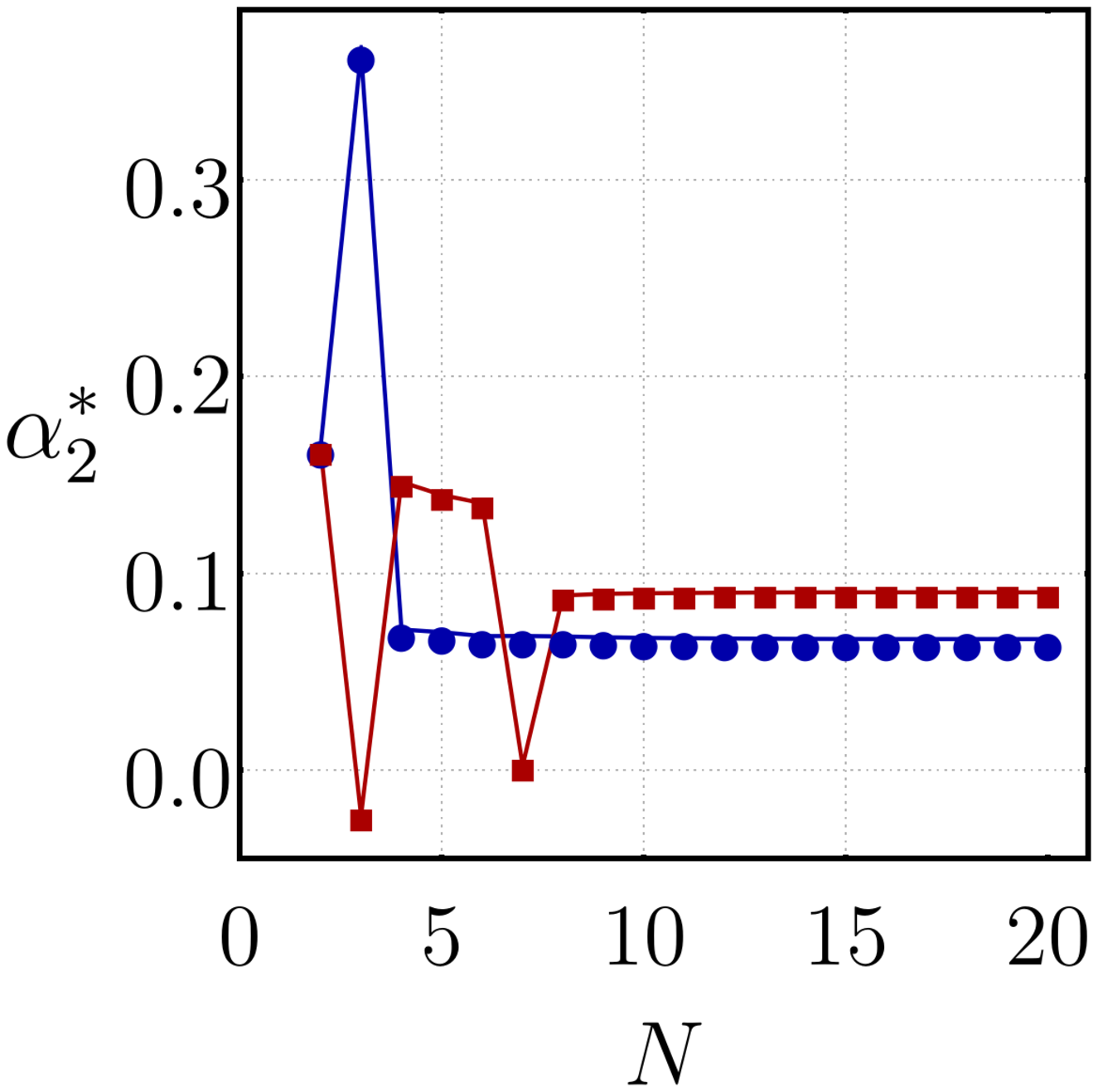} \quad
        \includegraphics[height=5.0cm]{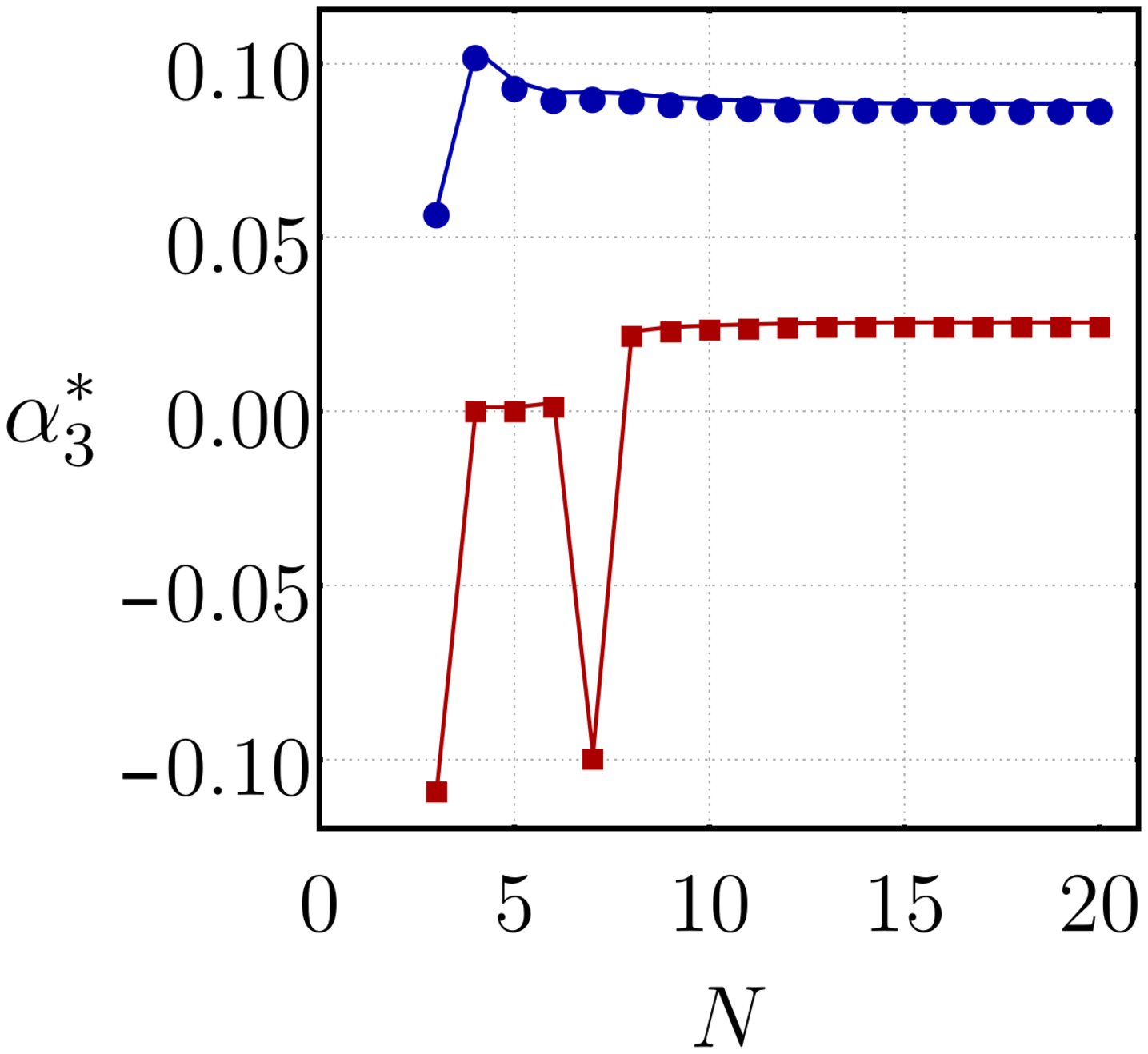} \quad
        \includegraphics[height=5.0cm]{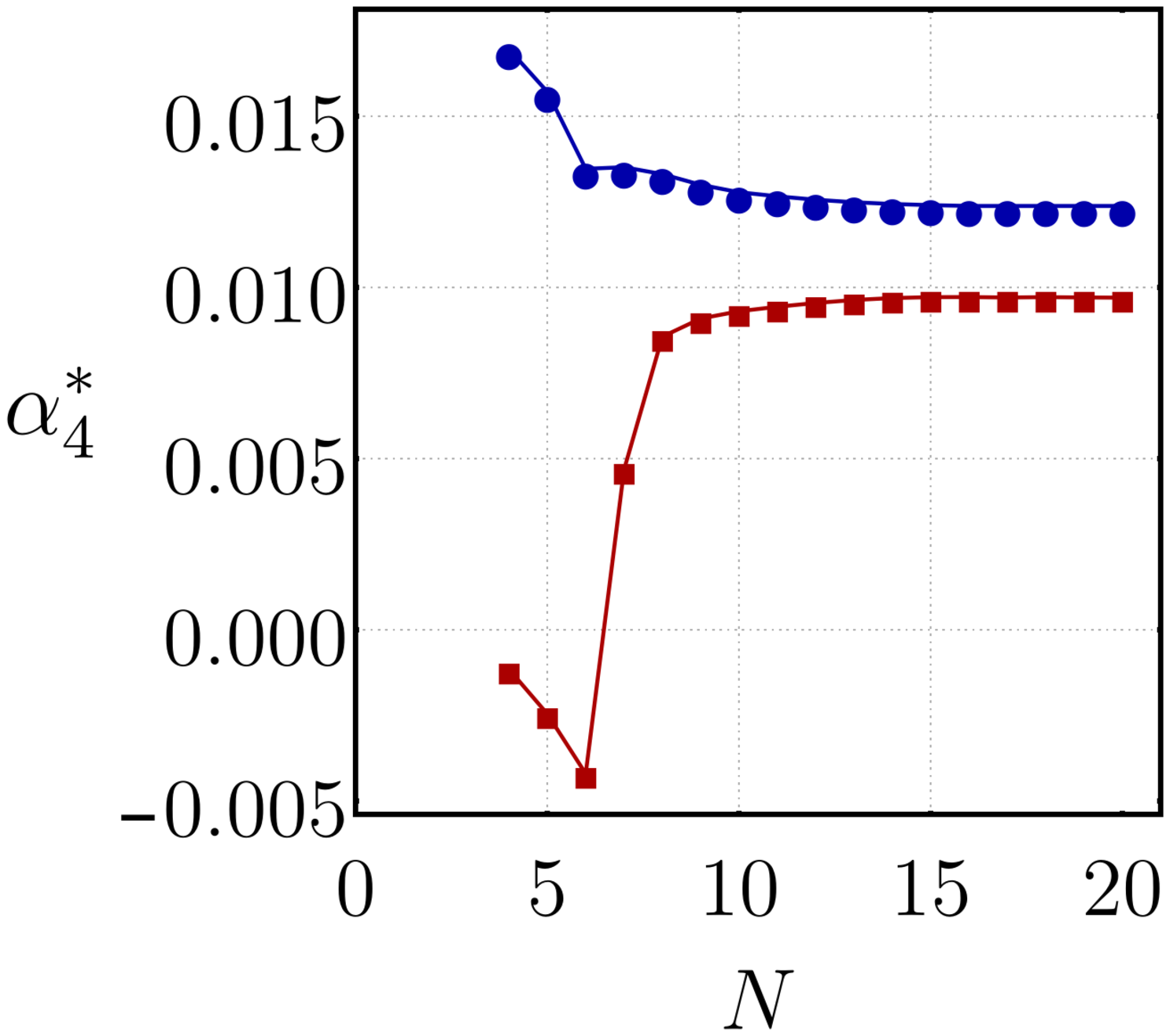} \quad
        \includegraphics[height=5.0cm]{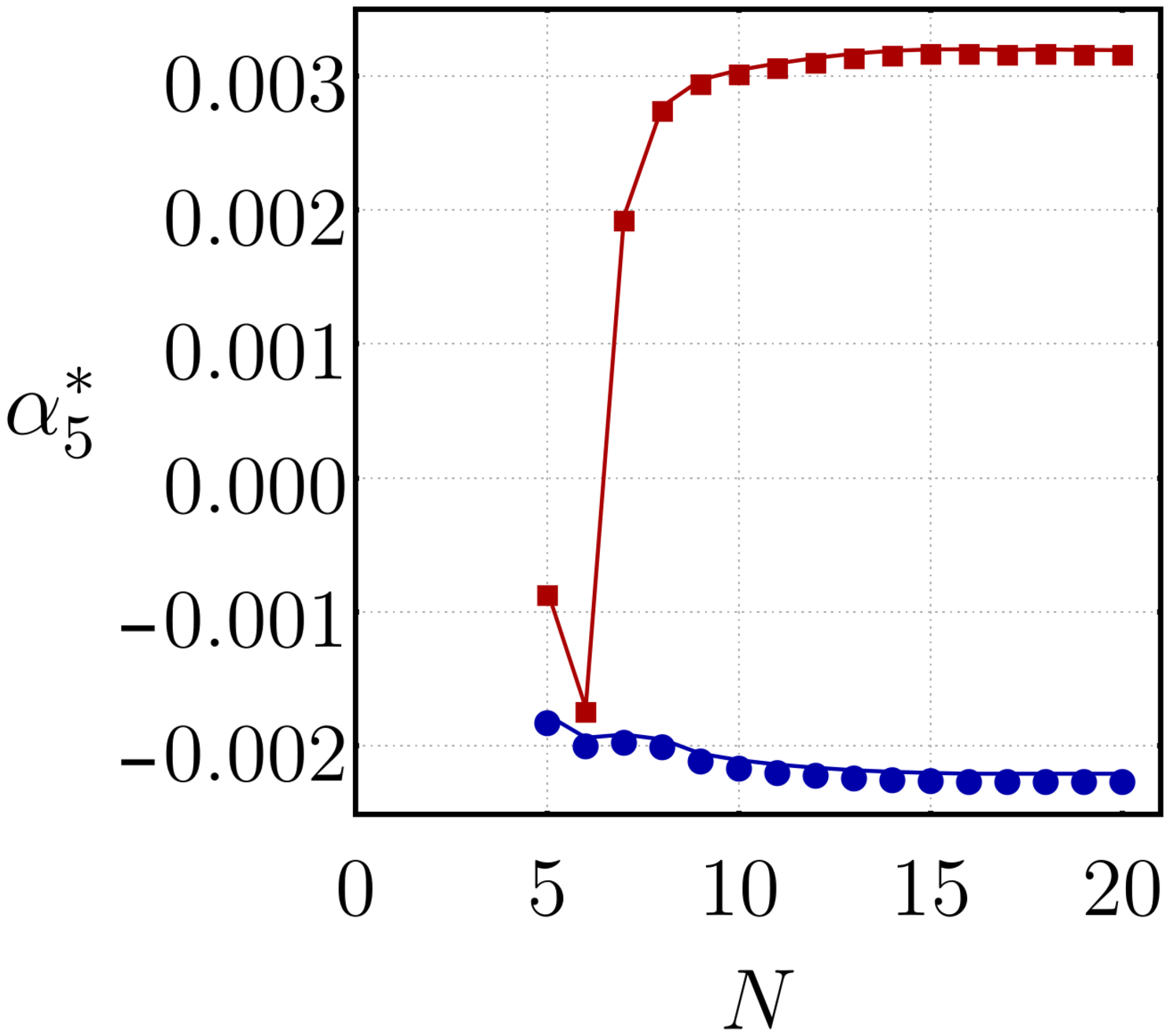} \quad
        \includegraphics[height=5.0cm]{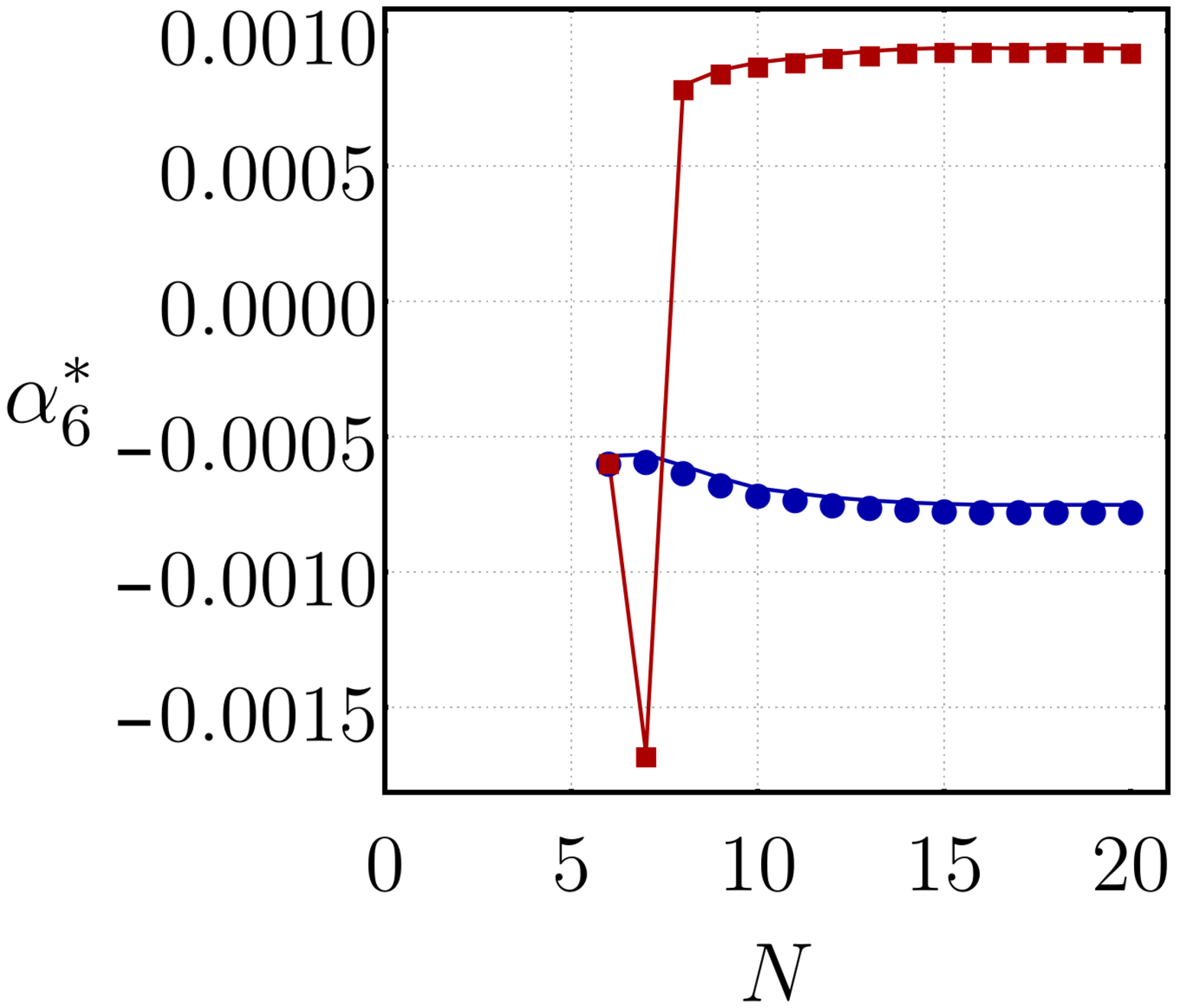}
        \caption{\footnotesize{Fixed-point values of the couplings $G_k$, $\alpha_{k,2}$, $\alpha_{k,3}$, $\alpha_{k,4}$, $\alpha_{k,5}$ and $\alpha_{k,6}$ in the $f(R)$-truncation. The blue circle indicates the Type I regularization (Bochner-Laplacian), whereas the red square indicates the Type II regularization (Lichnerowicz-Laplacians). All plots are computed within the RG-improved prescription.}}
  \label{fig:FPs_fR_RGImprov}
  \end{center}
\end{figure}

In Fig.~\textbf{\ref{fig:FPs_fR_RGImprov}}, we show the results of the fixed-point values for some of the dimensionless couplings (up to $\alpha^{*}_6$), defined in the polynomial $f(R)$-decomposition - see Eq. (\ref{fR_Truncation}), as functions of the order of approximation $N$. For higher-order couplings ($\alpha_7^*,\,\cdots,\,\alpha_{20}^*$), the fixed-point coordinates are of order $|\alpha_n^*|<10^{-4}$. In each plot, the results computed with the Bochner-Laplacian (Type I) as a coarse-graining operator are represented by a blue circle, whereas the ones computed employing the Lichnerowicz-Laplacians (Type II) are distinguished by a red square. We observe that the fixed-point values stabilize against the inclusion of higher-order operators. Albeit quantitatively different, the fixed-point structure is similar for both coarse-graining operators and, in particular, it displays an apparent stabilization for sufficiently large truncation.

\begin{figure}[!t]
  \begin{center}
    \includegraphics[height=5.8cm]{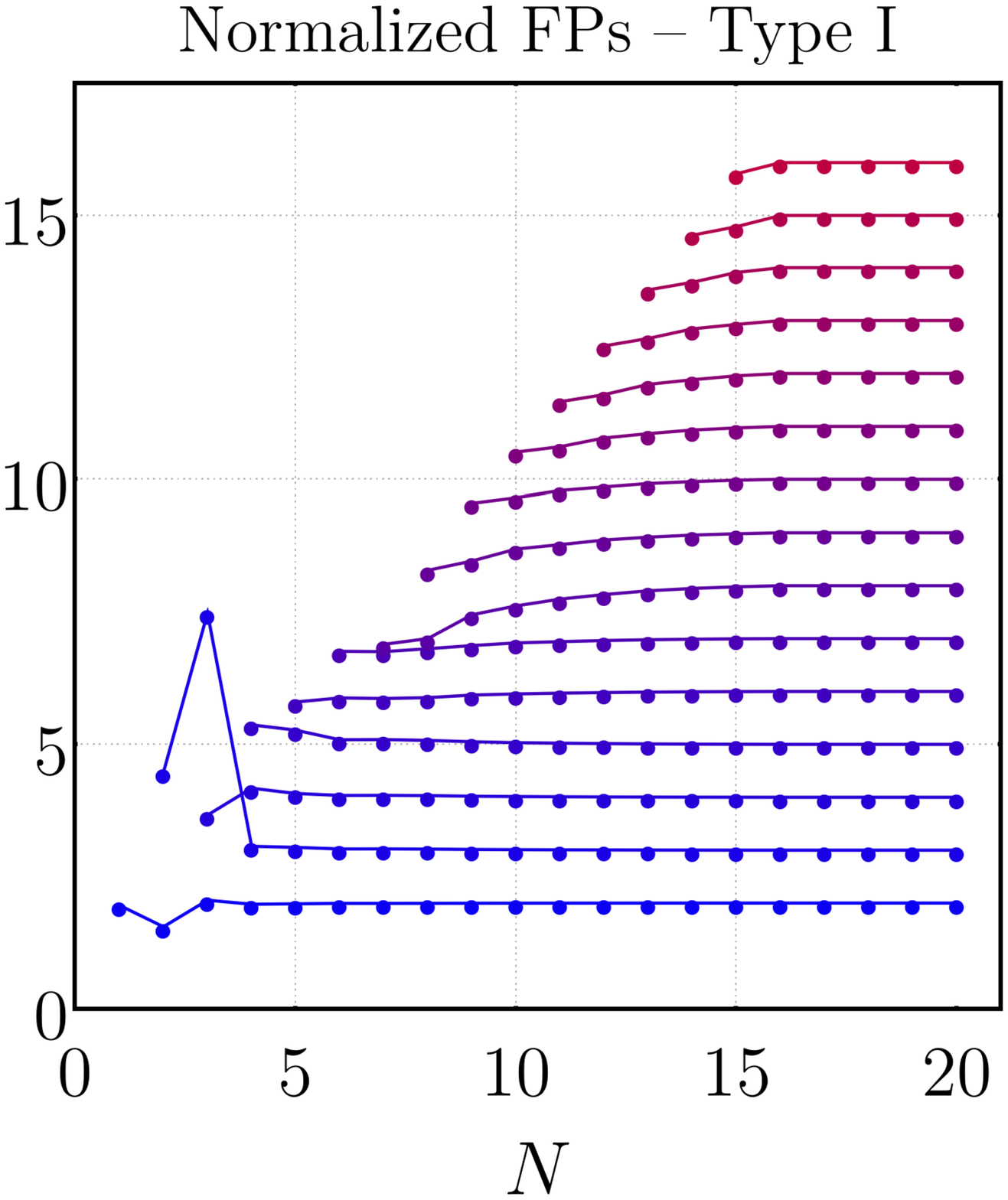} \quad
    \hspace{3em}
    \includegraphics[height=5.8cm]{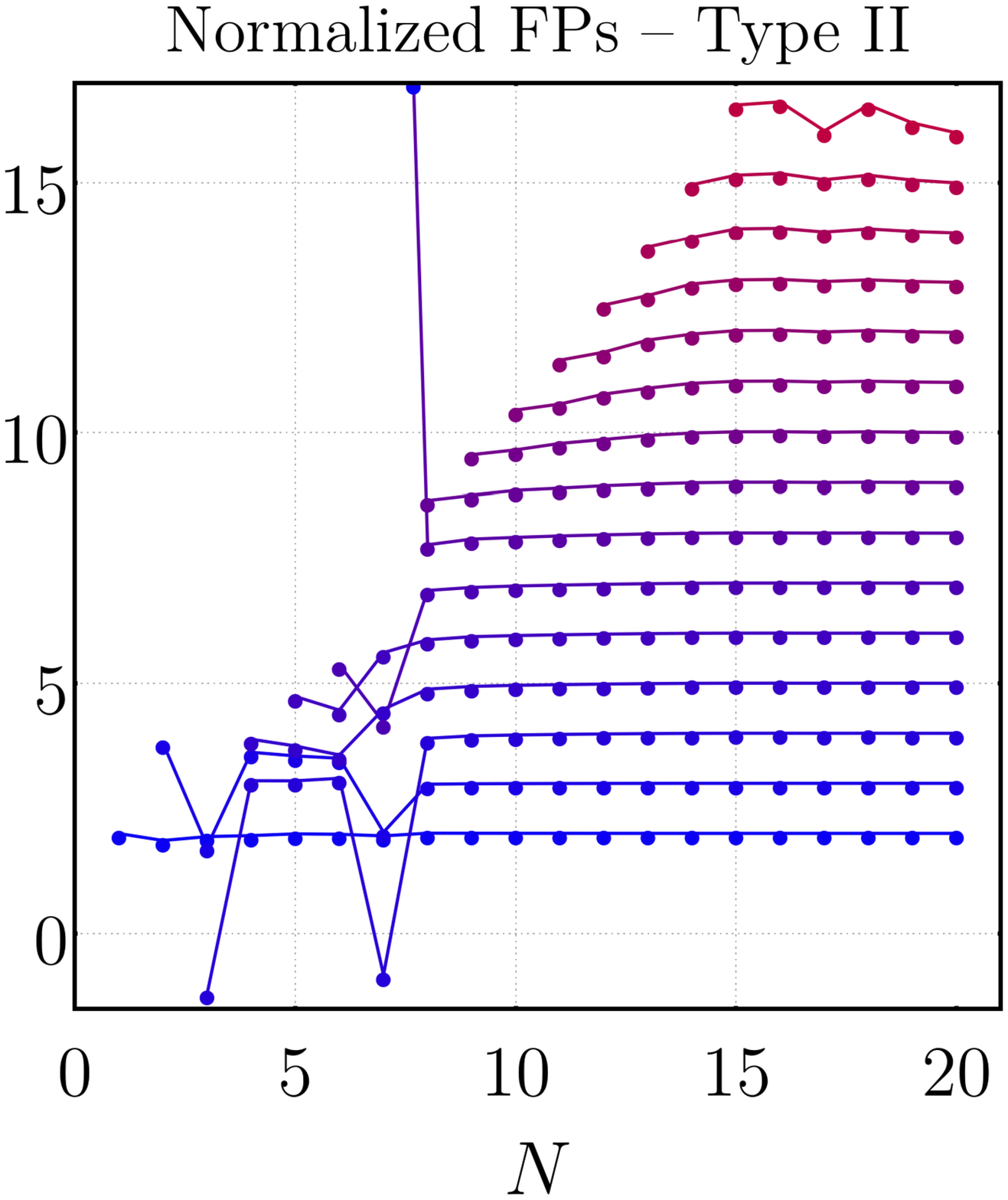}
    \caption{\footnotesize{Normalized fixed points in the $f(R)$-truncation. The convergence pattern is exhibited with the normalization $\lambda_1(N)=\frac{G^*(N)}{G^*(N_{\text{max}})}+1$ and $\lambda_n(N)=\frac{\alpha_n^*(N)}{\alpha^*_n(N_{\text{max}})}+n$ (for $n>1$). From bottom to top, we display $\lambda_{n}(N)$ for $n=1,\cdots,15$ in the $N_{\text{max}}=20$ truncation (see main text). The left panel shows the normalized fixed-points values associated with the Type I regularization scheme, while the right panel corresponds to results obtained via the Type II regularization scheme. Both schemes of calculation are closed with the RG-improved prescription.}}
  \label{fig:NormalizedFPsfR}
  \end{center}
\end{figure}

In order to provide a better visualization of the stabilization pattern against higher-order extensions for both regularization schemes, we consider in Fig.~\textbf{\ref{fig:NormalizedFPsfR}} a convenient normalization for the fixed-point couplings. Following \cite{Falls:2014tra, Falls:2017lst,Falls:2018ylp}, we define the set of normalized fixed points $\{\lambda_n\}_{n=1,\,\cdots,\, N}$ according to
\begin{equation}
\lambda_1(N)=\frac{G^*(N)}{G^*(N_{\text{max}})}+1 \qquad \text{and} \qquad \lambda_n(N)=\frac{\alpha_n^*(N)}{\alpha^*_n(N_{\text{max}})}+n,
\end{equation}
where $G^*(N)$ and $\alpha^*_n(N)$ represent the fixed-point values of the dimensionless couplings computed at order $N$. The couplings $\lambda_n$ are normalized in units of the fixed-point values computed at the largest approximation order, which in the present case is $N_{\text{max}}=20$, and are shifted by $n$. Fig.~\textbf{\ref{fig:NormalizedFPsfR}} gives evidence for the rapid apparent stabilization of the fixed points.

\begin{figure}[!t]
  \begin{center}
    \includegraphics[height=5.8cm]{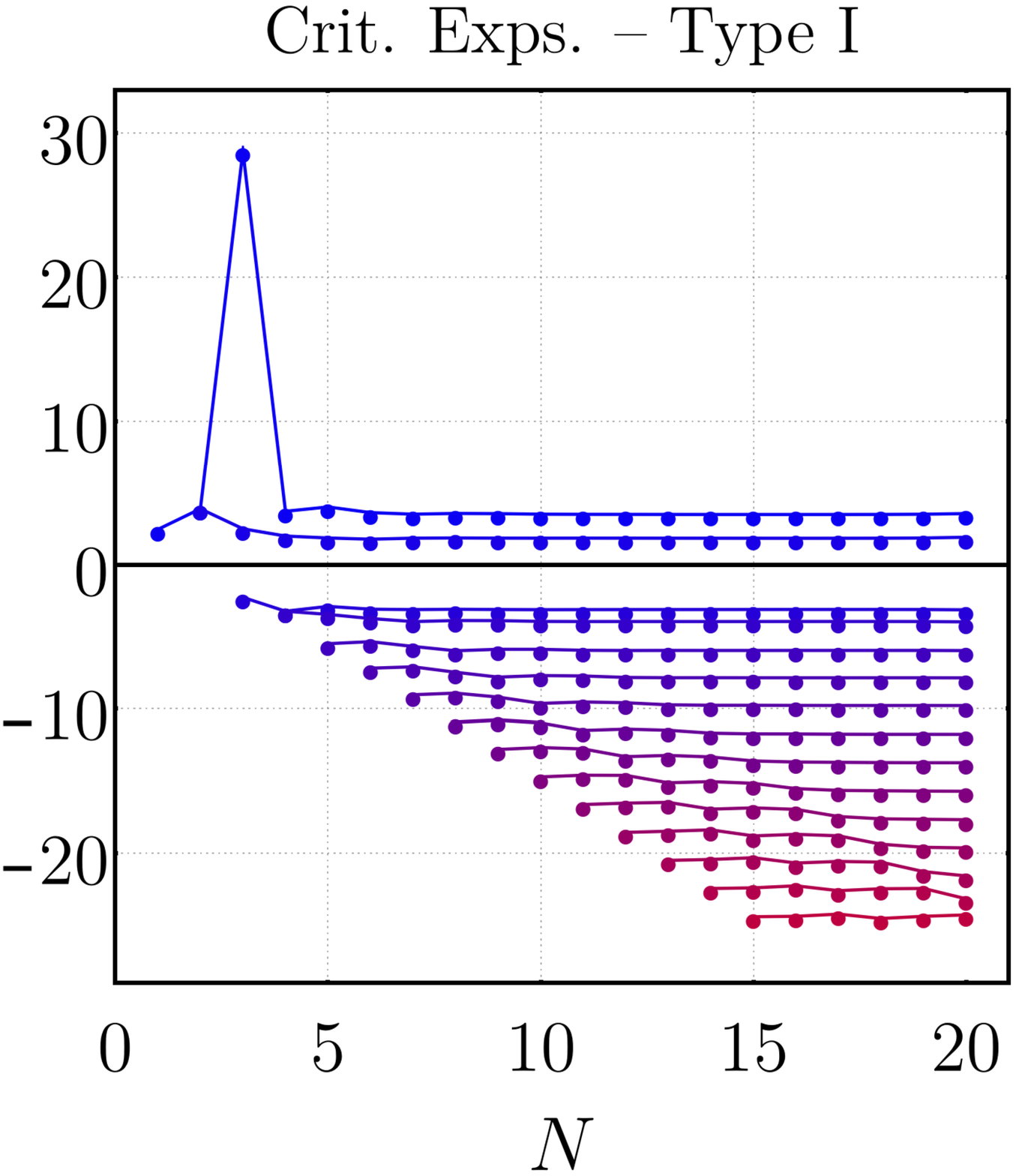}\quad
    \hspace{3em}
    \includegraphics[height=5.8cm]{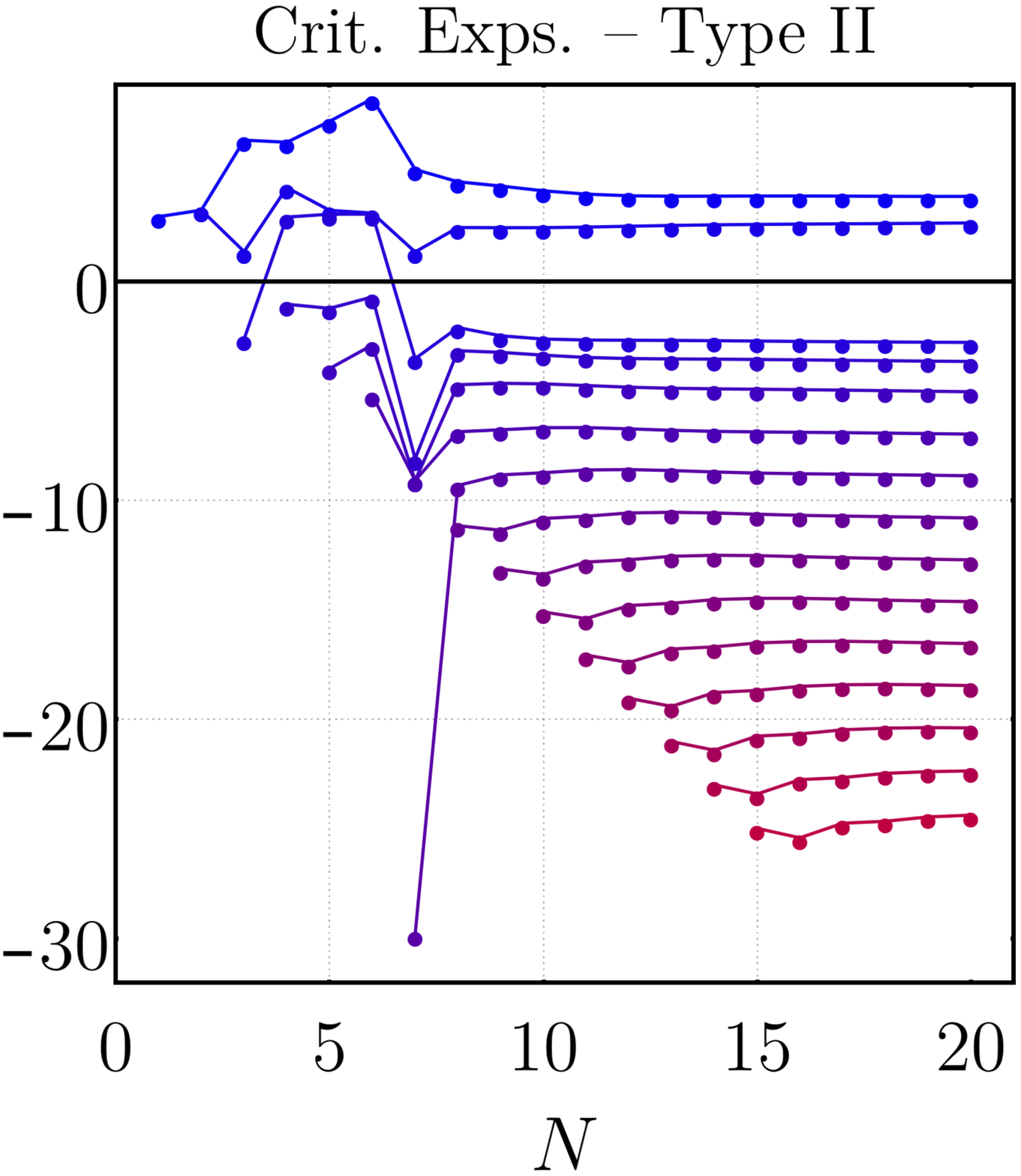}
    \caption{\footnotesize{Critical exponents associated with the fixed-point structure in the $f(R)$-approximation for the range $n=1,\cdots, 15$ in the $N_{\textmd{max}}=20$ truncation within the RG-improved closure. The left panel corresponds to results obtained under the Type I regularization, while the right panel displays the results obtained under the Type II regularization.}}
  \label{fig:fR&CritExps}
  \end{center}
\end{figure}

Fig.~\textbf{\ref{fig:fR&CritExps}} displays the corresponding critical exponents for both types of coarse-graining operators as functions of $N$ and gives evidence for a non-increasing number of relevant directions\footnote{We follow the convention that relevant directions are characterized by positive (real part of the) critical exponents which, in turn, are the eigenvalues of the stability matrix multiplied by minus one.}. This indicates that the dimensionality of the UV critical hypersurface does not grow up to the dimension of the truncated unimodular theory space, which is a crucial feature for the asymptotic safety program. As occurred for the fixed-point values, higher-order additional invariant operators do not seem to spoil the stabilization of the critical exponents. In particular, for the critical exponents computed within the Type I regularization, i.e., Bochner-Laplacian as a coarse-graining operator. Fig.~\textbf{\ref{fig:fR&CritExps}} (left) indicates that the number of relevant directions saturates at two (except obviously for $N=1$). For the Type II regularization, characterized by the Lichnerowicz-Laplacians (right), the analysis is a bit more subtle. In this case, we observe a small oscillation in the neighborhood of positive values for lower-order truncations ($N<6$). In spite of that, the inclusion of additional invariant operators drives the number of relevant directions to stabilize at two as well.

An interesting feature also displayed in Fig.~\textbf{\ref{fig:fR&CritExps}} is the near-canonical character of the critical exponents , i.e., a small deviation of the critical exponents in comparison with the canonical scaling of the operators appearing in our truncation. Indeed, the critical exponents computed within the $f(R)$-expansion behave like $\theta_n\sim\Delta_n$, where $\Delta_n=4-2n$ is the canonical scaling dimension of an invariant of the form $\bar{R}^n$. The two positive critical exponents appear as exceptions, since they deviate from the corresponding canonical scaling dimension by a greater gap. The near-canonical character of the critical exponents was already observed in a unimodular setting based on a polynomial expansion of $f(R)$ up to $\bar{R}^{10}$~\cite{Eichhorn:2015bna}. Additionally, it is worth mentioning that a near-canonical spectrum of critical exponents has been investigated in detail within standard ASQG~\cite{Falls:2013bv,Falls:2014tra,Falls:2017lst,Eichhorn:2018akn,Eichhorn:2018ydy,Falls:2018ylp,Kluth:2020bdv}. In particular, such a property suggests that power-counting can be a good guiding principle in the construction of truncations of the flowing effective action. 

As stated earlier, as an attempt to go beyond the RG-improvement prescription, the anomalous dimensions of the fluctuating metric and ghost fields may be independently computed through a simultaneous vertex and derivative expansion of the effective average action in the same fashion as discussed previously in the unimodular setting~\cite{deBrito:2020rwu} (see also \cite{Christiansen:2012rx,Codello:2013fpa}). This provides a second way of closing the system of RG equations by combining the background-field approximation for the couplings with independent anomalous dimensions for fluctuating fields in a hybrid approach, as in \cite{Eichhorn:2010tb,Codello:2013fpa,Dona:2013qba,Dona:2014pla}. Our setup for the generation of the  interaction proper vertices employs the same ansatz (\ref{workingtruncation}). In order to capture higher-curvature effects, the Lagrangian $f(R,R_{\mu\nu}^2)$ is decomposed into an Einstein-Hilbert term supplemented by quadratic-curvature invariants such that the gravitational sector is in the form
\begin{equation}\label{Quadratic_Action}
\Gamma_k^{\text{gravity}}[g_{\mu\nu}]=\frac{k^2}{16\pi G_{k}}\int_x\omega\,\big(-R+k^{-2}\alpha_{k,2}R^2+k^{-2}\rho_{k,2} R_{\mu\nu}R^{\mu\nu}\big)\,,
\end{equation}
with $\alpha_{k,2}$ and $\rho_{k,2}$ being the same dimensionless couplings as in (\ref{fR_Truncation}) and (\ref{FfunctionFZ}), respectively. In particular, for computational simplicity, curvature-squared contributions to the vertices are neglected. We emphasize that this is an additional approximation that should be refined in a future investigation. After expanding the gravitational action in powers of the fluctuation field $h_{\mu\nu}$, we set $\bar{g}_{\mu\nu}=\delta_{\mu\nu}$. This setup allows us to obtain the anomalous dimensions in the form $\eta_i=\eta_i(G_k,\alpha_{k,2},\rho_{k,2},\beta_{\alpha},\beta_{\rho},N_{\Psi})$. The explicit expressions are given in App.~\textbf{\ref{etas}} within a semi-perturbative approximation\footnote{The semi-perturbative approximation consists in setting to zero all the $\eta$'s that would arise from the RG-scale derivative on the regulator function. This amounts to neglecting the $\eta$'s on the right-hand side of the expressions for the anomalous dimensions \cite{Dona:2013qba,Dona:2015tnf,Eichhorn:2017lry,Eichhorn:2017sok,Eichhorn:2018nda,Eichhorn:2019yzm,deBrito:2020rwu}.} and, when inserted in the RG equations for the $f(R)$-truncation, the coupling $\rho_{k,2}$ and its beta function are set to zero, since the $f(R)$-approximation does not contain any $R_{\mu\nu}^2$-dependence. Similarly, when treating the system of RG equations for the FZ-truncation, the coupling $\alpha_{k,2}$ and its beta function are not considered.

\begin{figure}[!t]
  \begin{center}
    \includegraphics[height=5.8cm]{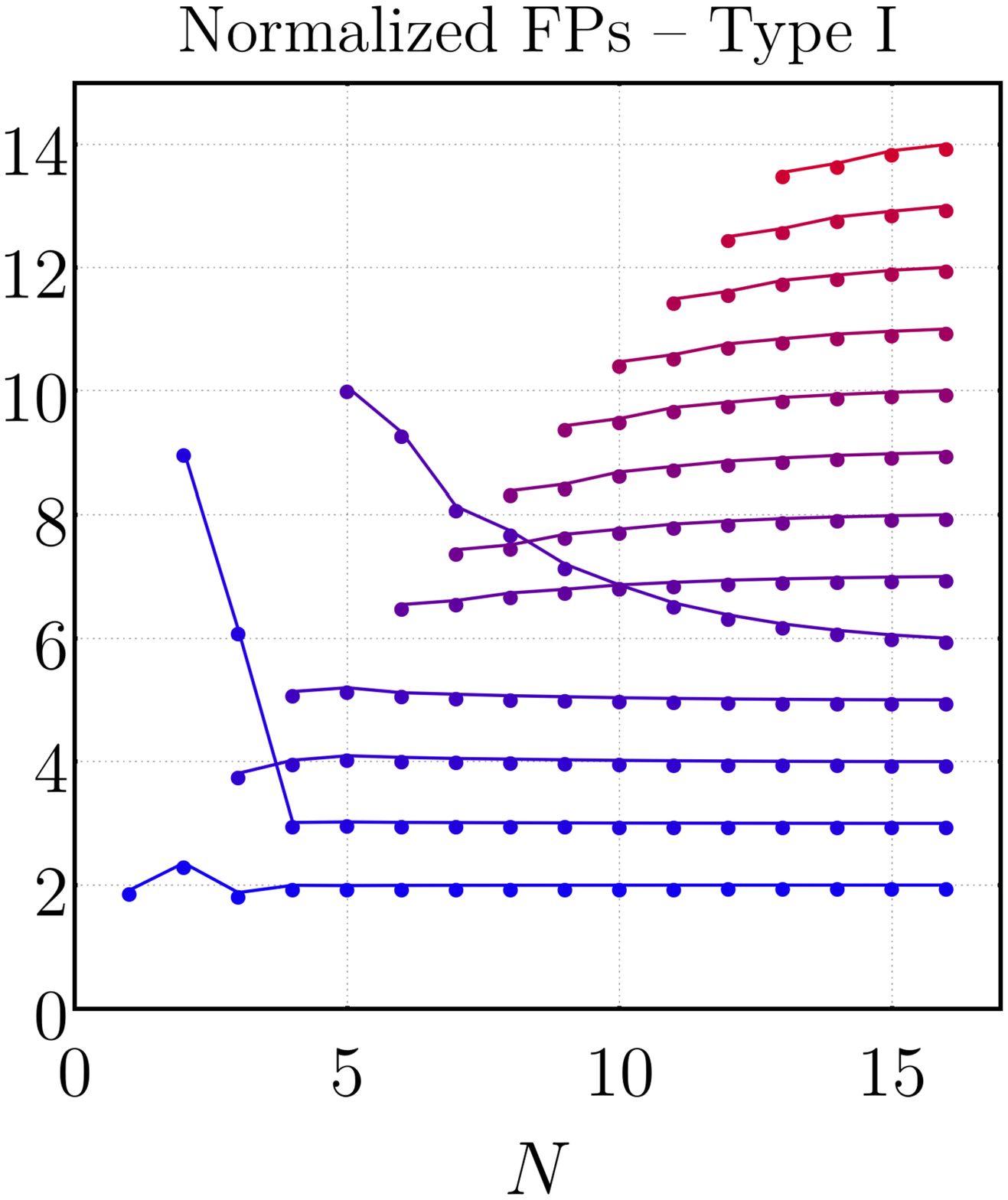} \quad
    \hspace{3em}
    \includegraphics[height=5.8cm]{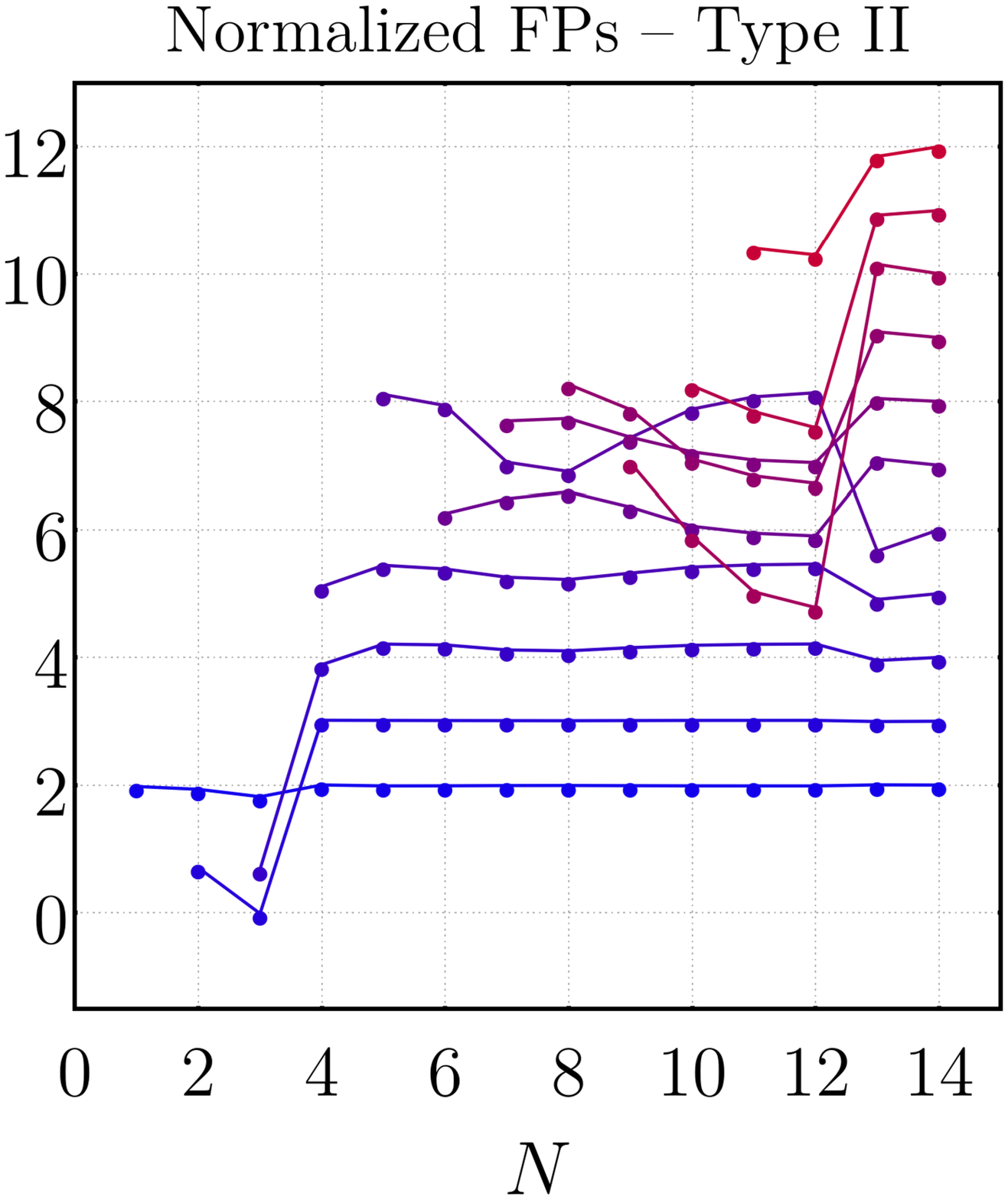}
   \caption{\footnotesize{Plots of the convergence pattern for the normalized fixed-point values of the couplings $\lambda_{n}(N)$  for the $f(R)$-approximation evaluated within the hybrid semi-perturbative closure. The left plot exhibits the convergence pattern for the the range $n=1,\cdots,13$ in the $N_{\text{max}}=16$ truncation under the Type I regularization, while the right plot displays the convergence pattern for the range $n=1,\cdots,11$ in the $N_{\text{max}}=14$ truncation under the Type II regularization. All couplings follow the same normalization convention as defined previoulsy. The truncations are smaller with respect to previous results and different for different coarse-graining operators due to numerical instabilities.}}
  \label{fig:NormalizedFPsfRSemi}
  \end{center}
\end{figure}

\begin{figure}[!t]
  \begin{center}
    \includegraphics[height=5.95cm]{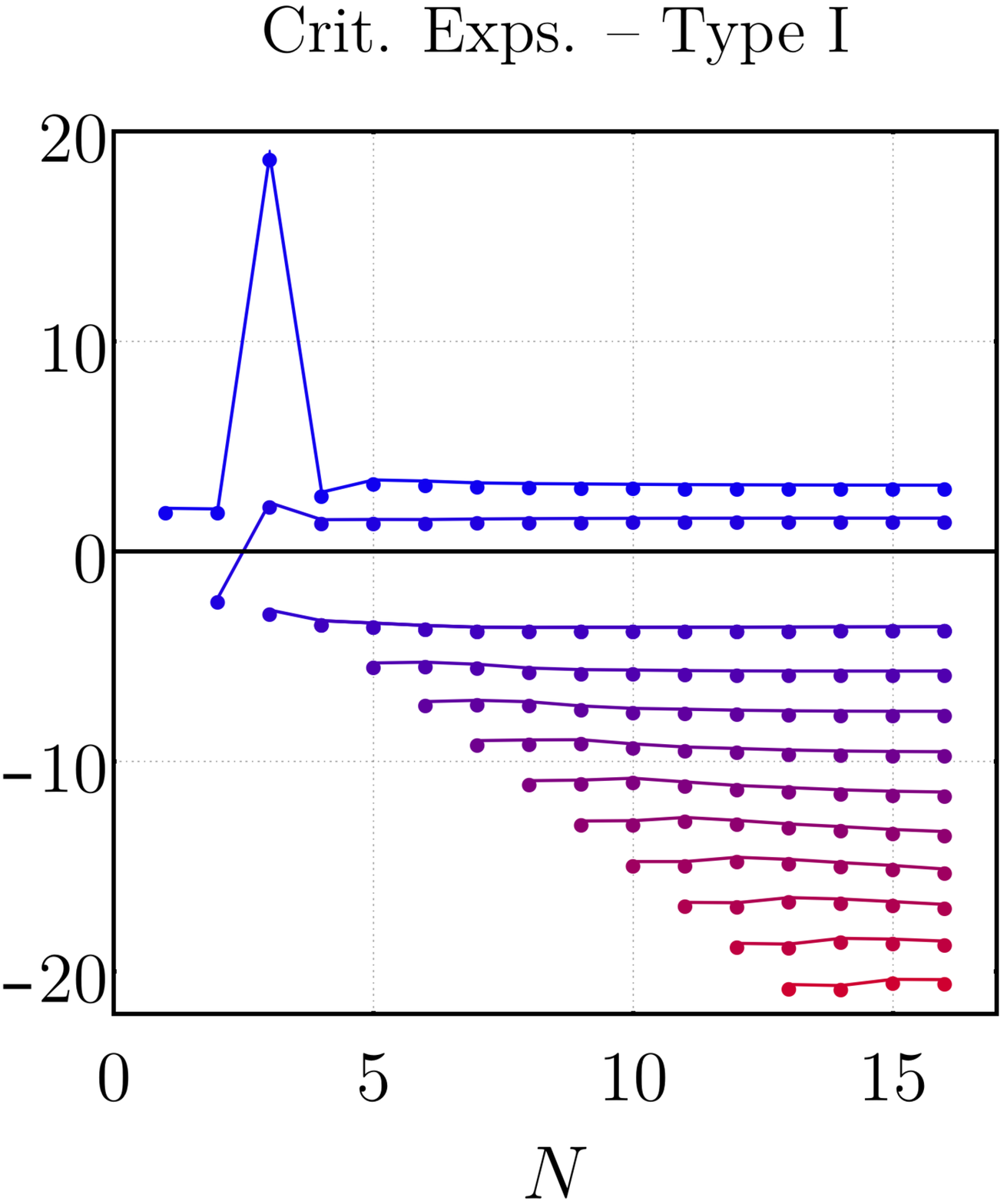} \quad
    \hspace{3em}
    \includegraphics[height=5.95cm]{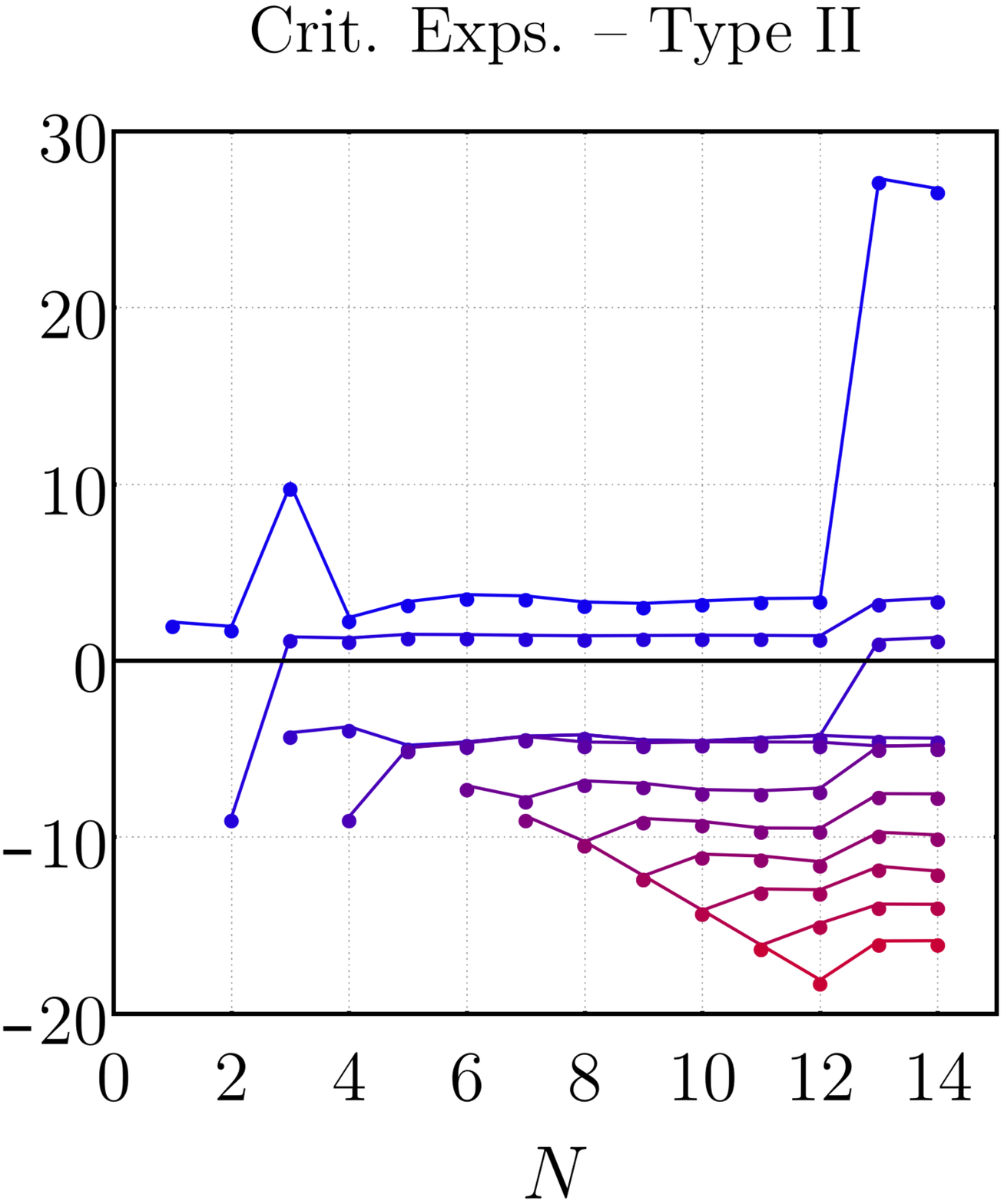}
   \caption{\footnotesize{Critical exponents associated with the fixed-point structure in the $f(R)$-approximation within the hybrid semi-perturbative closure. The left panel corresponds to results for the range $n=1,\cdots,13$ in the $N_{\text{max}}=16$ truncation under the Type I regularization. In particular, only the third and fourth set of critical exponents, under this regularization, are complex conjugate pairs and, consequently, the lines representing their real parts fall on top of each  other. The right plot exhibits the results for the range $n=1,\cdots,11$ in the $N_{\text{max}}=14$ truncation under the Type II regularization. Clearly, in the latter case, the truncation needs to be further extended in order to verify if apparent convergence is restored for the critical exponents.}}
  \label{fig:fR&CritExpsSemi}
  \end{center}
\end{figure}

To avoid a proliferation of similar plots, we refrain from showing the plots of convergence of individual fixed-point values and only exhibit results for the normalized fixed points and the critical exponents in Fig.~\textbf{\ref{fig:NormalizedFPsfRSemi}} and Fig.~\textbf{\ref{fig:fR&CritExpsSemi}}.


As argued in \cite{Meibohm:2015twa}, for a generic class of regulators proportional to the two-point functions, the imposition that the regulators must diverge in the UV leads to the constraint $\vec{\eta}\,\big|_*<2$. Since our regulators fall in this class, we have selected, within the hybrid semi-perturbative prescription, fixed-point values for $G^*$ and $\alpha_2^*$ which respect the bound\footnote{The constraint verified in this work is subjected to the approximations made here. In particular, for $\eta_{\sigma}$ the range of fixed-point values may be more restrict if one considers the effects of symmetry-breaking graviton mass terms (and full closure), as discussed in \cite{deBrito:2020rwu}.} $\vec{\eta}\,\big|_*<2$. The convergence pattern of the normalized fixed-point values $\lambda_n(N)$ for the $f(R)$-approximation are displayed in Fig.~\textbf{\ref{fig:NormalizedFPsfRSemi}} for both types of regularization schemes. Considering the non-linear character of the expressions for the anomalous dimensions given by (\ref{Eta_TT_Grav}), (\ref{Eta_Sigma_Grav}), (\ref{Eta_Gh}), as opposed to the RG-improved case, we managed to find suitable fixed-point solutions for the polynomial truncation up to $N_{\text{max}}=16$ for the Type I regularization and up to $N_{\text{max}}=14$ for Type II. As in the case of the RG-improved prescription, the Type I regularization leads to stable fixed-point solutions, apart from a late stabilization of the normalized coupling associated with the invariant $\bar{R}^5$. However, the Type II regularization only leads to a clear apparent stabilization for the first four lower-order operators and seems to be sensitive against the inclusion of higher-order invariants. This behavior is again evident in the plots of the critical exponents in Fig.~\textbf{\ref{fig:fR&CritExpsSemi}}. In order to tell if this behavior is a truncation artifact due to the independently-computed anomalous dimensions or simply reflects a limitation in our search method, an investigation of higher-order truncations would be needed. Interestingly though, the near-canonical character of the critical exponents is still manifest for both types of regularization schemes within this hybrid semi-perturbative approximation. This indicates that quantum fluctuations encoded in the anomalous dimensions provide a mild contribution to all invariant operators.

We move on to discuss the fixed-point structure of the polynomial FZ-truncation. The more complicated nature of this truncation naturally leads to larger expressions in contrast with the $f(R)$-truncation, thus demanding additional computational capacity. As a consequence, within the RG-improved prescription, we limit ourselves to explore the fixed-point equations within a truncation where the highest-order invariant operator corresponds to $R(R_{\mu\nu}R^{\mu\nu})^7$ (i.e., $N_{\text{max}}=15$). As in the $f(R)$ case, a numerical recursive solution of the fixed-point equations is implemented alongside a bootstrap search method.

\begin{figure}[!t]
  \begin{center}
    \includegraphics[height=5.0cm]{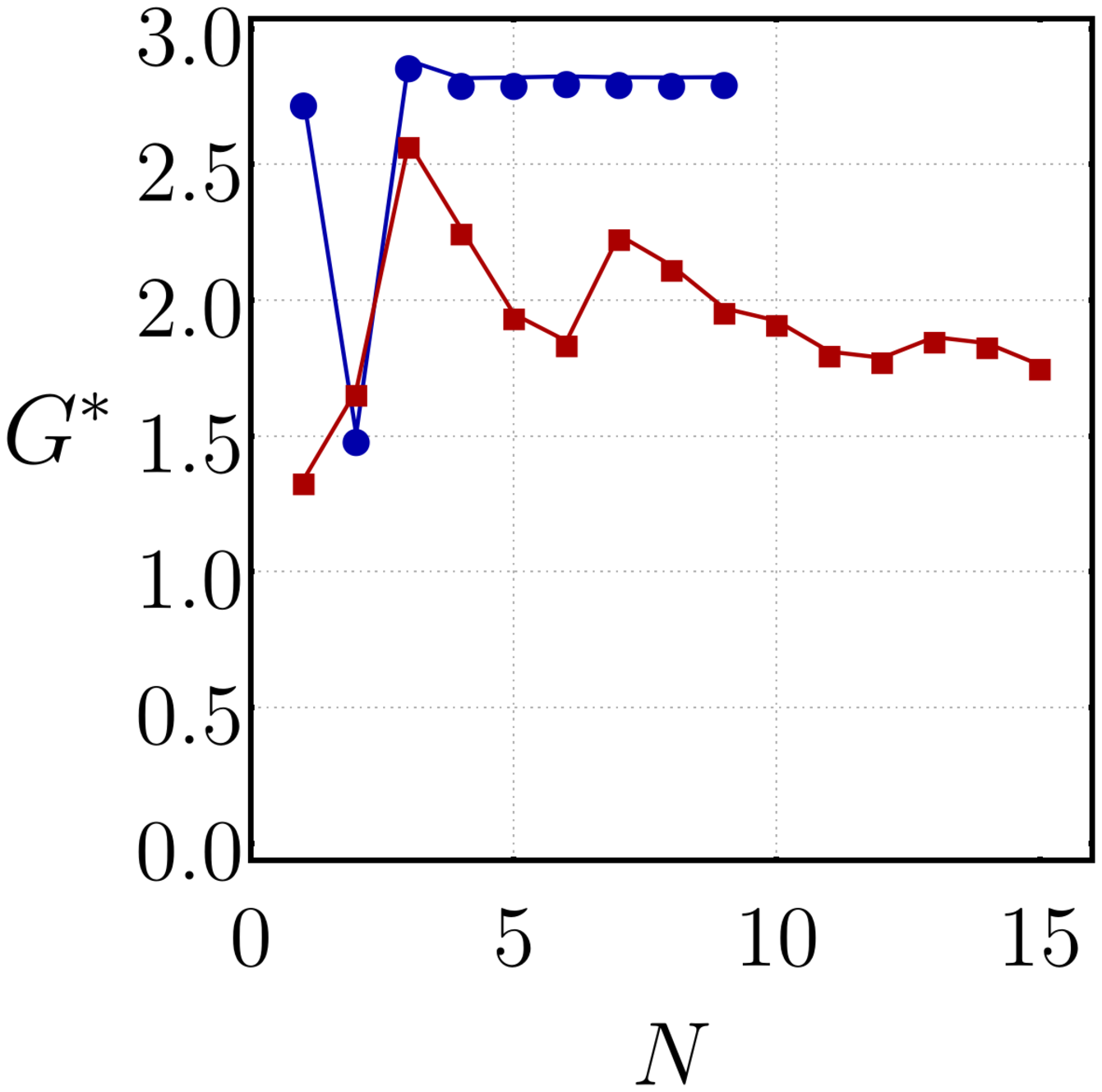} \quad
    \includegraphics[height=5.0cm]{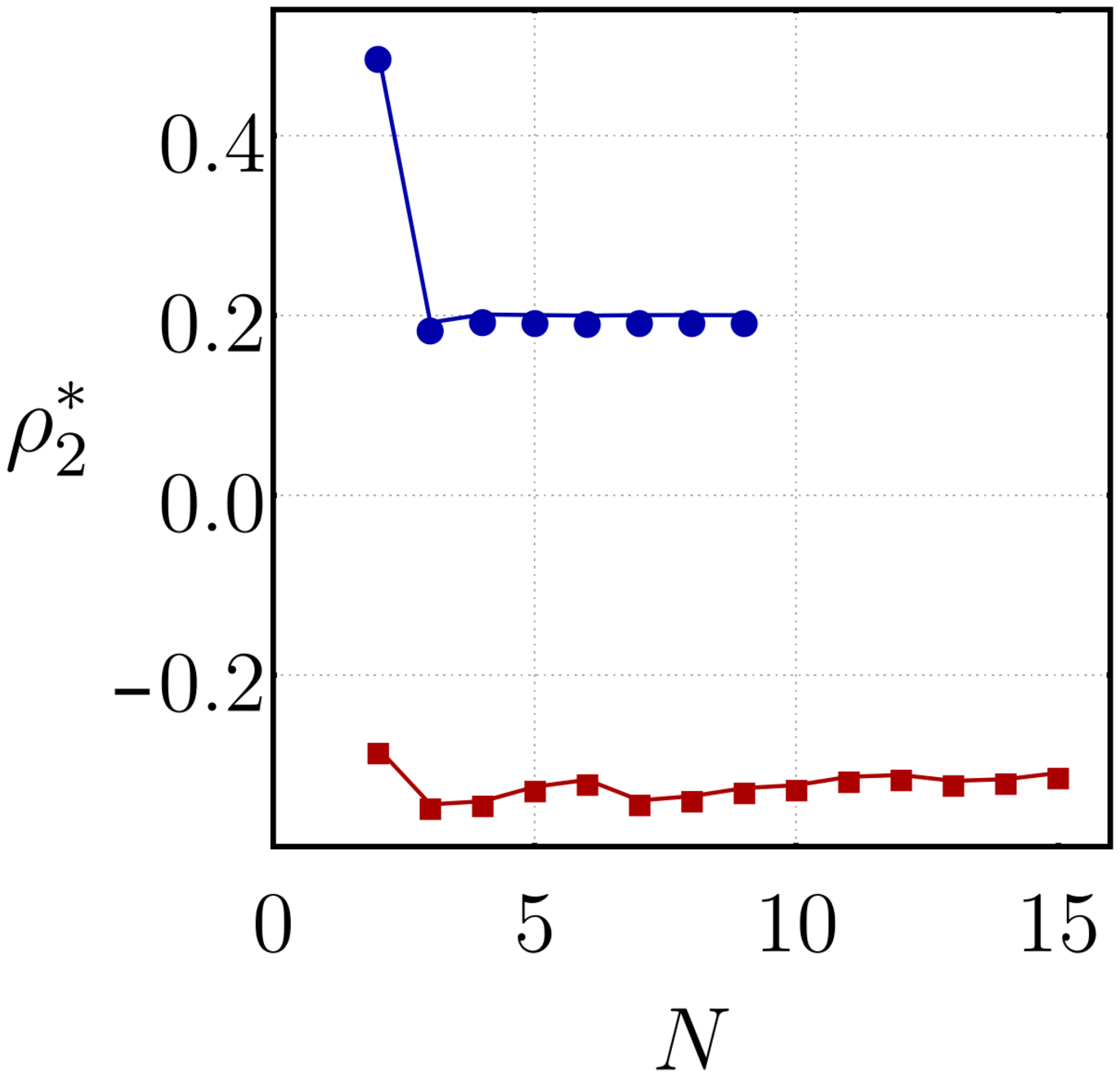} \quad
    \includegraphics[height=5.0cm]{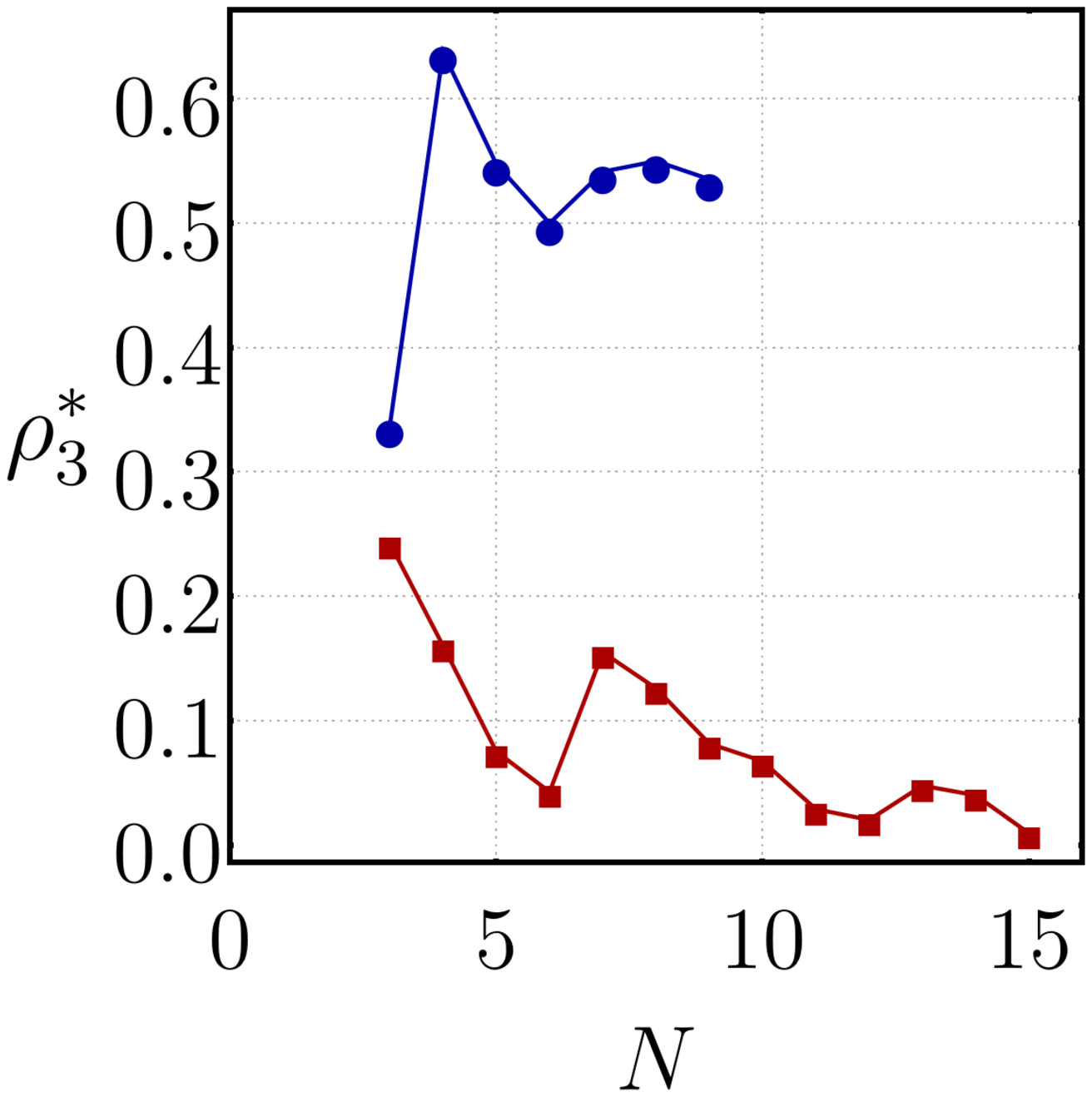} \quad
    \includegraphics[height=5.0cm]{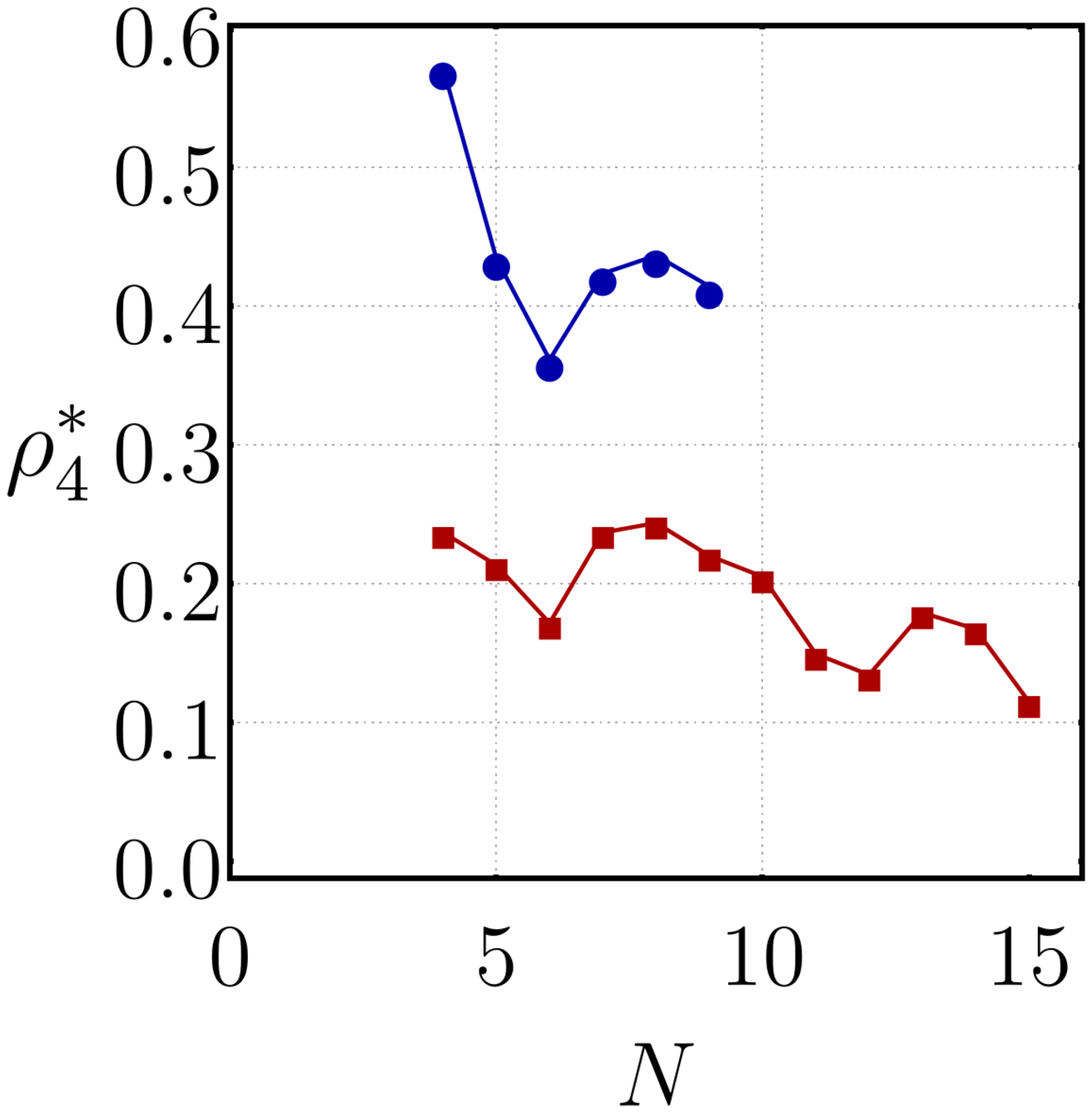} \quad
    \includegraphics[height=5.0cm]{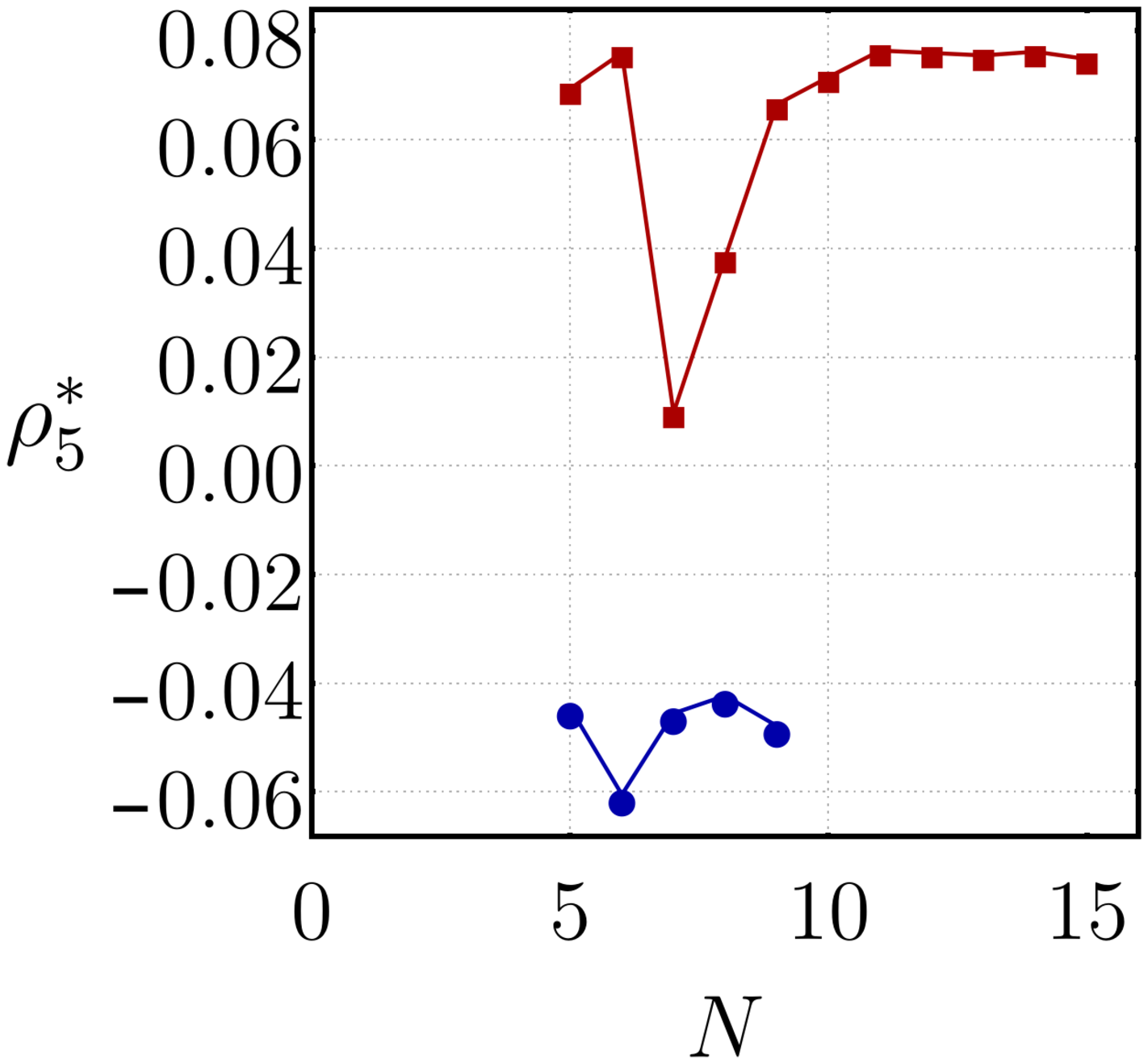} \quad
    \includegraphics[height=5.0cm]{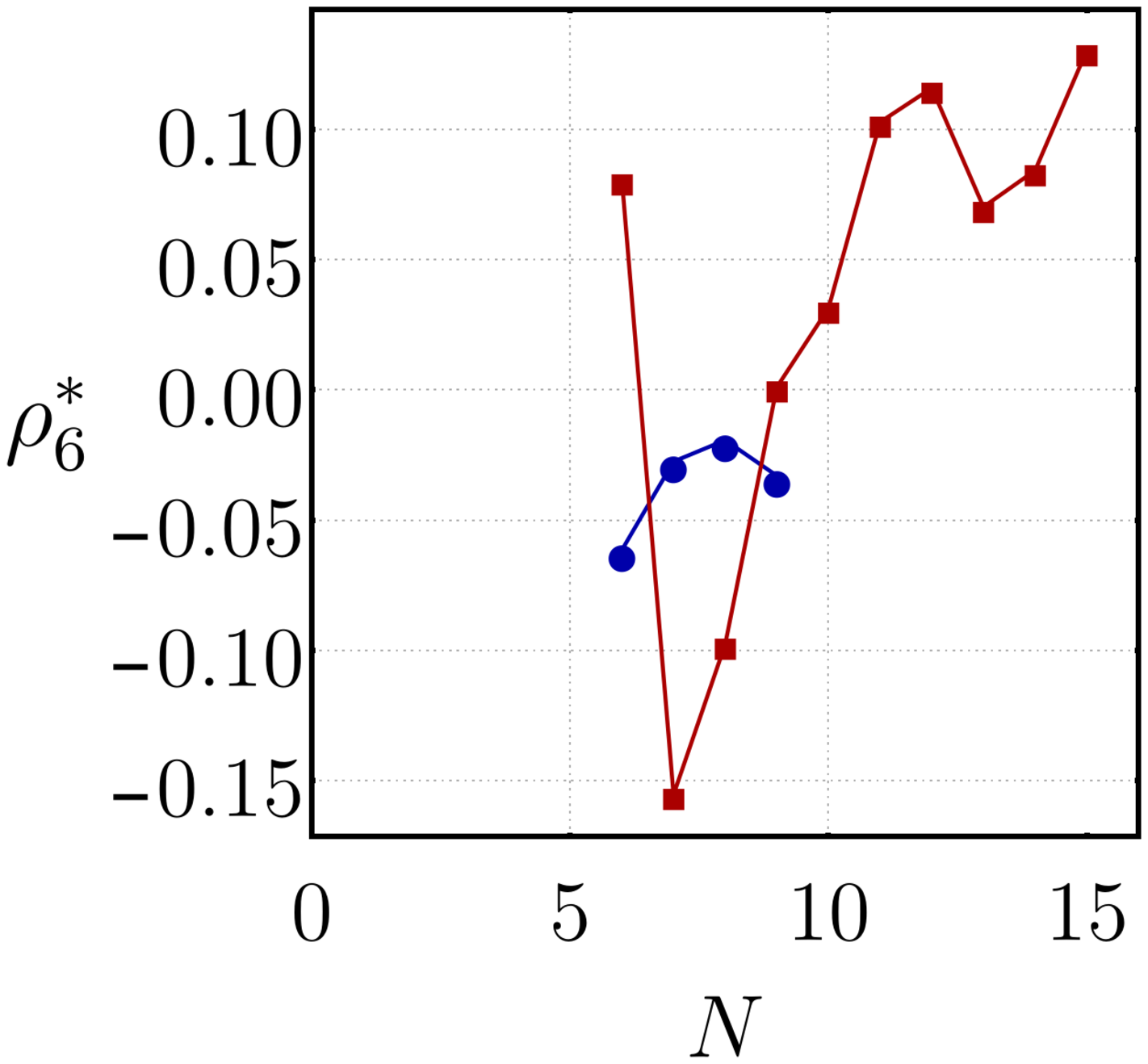}
    \caption{\footnotesize{
  Fixed-point values of the couplings $G_k$, $\rho_{k,2}$, $\rho_{k,3}$, $\rho_{k,4}$, $\rho_{k,5}$ and $\rho_{k,6}$ in the FZ-truncation. The blue circle indicates the Type I regularization (Bochner-Laplacian), whereas the red square indicates the Type II regularization (Lichnerowicz-Laplacians). All plots are computed within the RG-improved prescription.}}
  \label{fig:FPs_FZ_RGImprov}
  \end{center}
\end{figure}
\begin{figure}[t]
  \begin{center}
    \includegraphics[height=5.8cm]{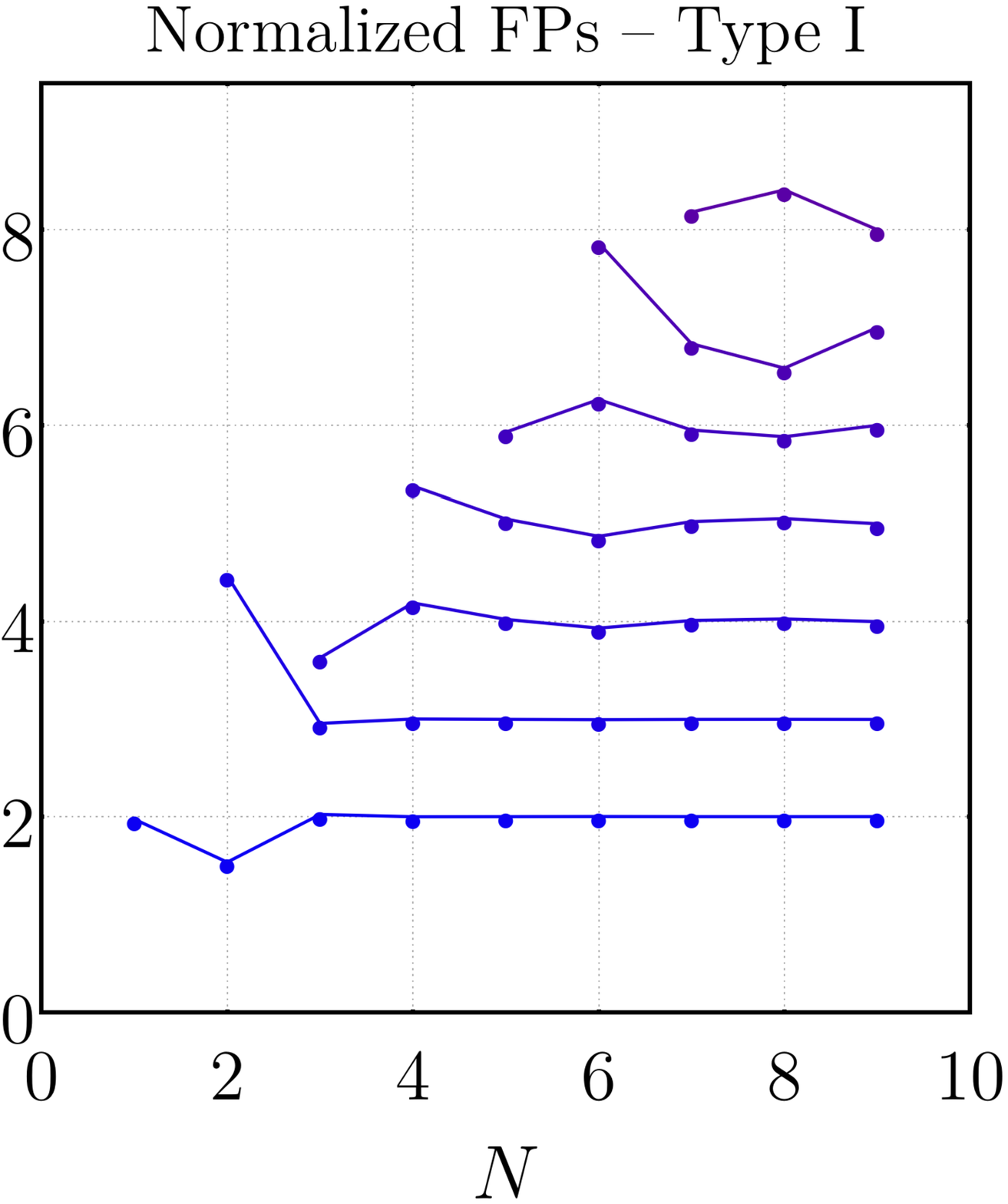} \quad
    \hspace{3em}
    \includegraphics[height=5.8cm]{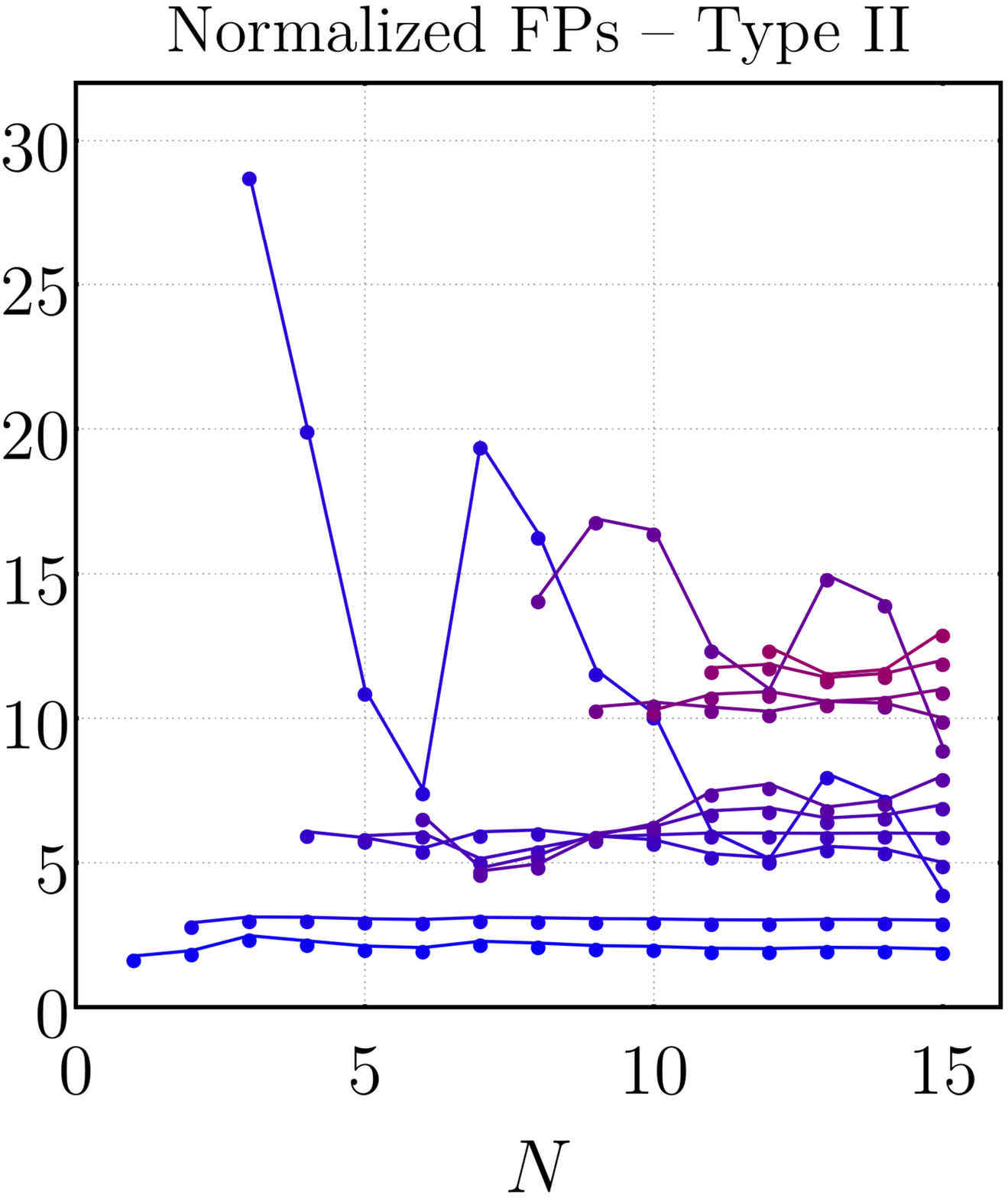}
    \caption{\footnotesize{Plots of the convergence pattern of the normalized fixed-point values of the couplings $\lambda_{n}(N)$ (now given in terms of $(G_k,\rho_{k,n})$) for the FZ-truncation evaluated within the RG-improved prescription. The left plot exhibits the convergence pattern for the range $n=1,\cdots,7$ in the $N_{\text{max}}=9$ truncation under the Type I regularization, while the right plot displays the convergence pattern for the range $n=1,\cdots,12$ in the $N_{\text{max}}=15$ truncation under the Type II regularization. All couplings follow the same normalization convention as in the $f(R)$ case.}}
  \label{fig:NormalizedFPsFZ}
  \end{center}
\end{figure}

In Fig.~\textbf{\ref{fig:FPs_FZ_RGImprov}}, we display our findings of the fixed-point values of the dimensionless couplings up to $\rho_{k,6}$ as functions of $N$ extracted from the FZ-truncation for both types of coarse-graining operators. We adopt the same convention for the plot markers as in the $f(R)$-truncation. Additionally, in Fig.~\textbf{\ref{fig:NormalizedFPsFZ}}, we display the convergence pattern of the normalized fixed-point values of the couplings $\lambda_n(N)$ defined in terms of ($G_k, \rho_{k,n}$). As one can notice from Figs.~\textbf{\ref{fig:FPs_FZ_RGImprov}} and~\textbf{\ref{fig:NormalizedFPsFZ}}, for the regularization employed by the Lichnerowicz-Laplacians (red squares in~\textbf{\ref{fig:FPs_FZ_RGImprov}} and right panel in~\textbf{\ref{fig:NormalizedFPsFZ}}), we managed to find suitable NGFP solutions for the polynomial truncation until $N_{\text{max}}=15$, exhibiting mild oscillations for higher-order operator invariants (with the exception of wild oscillations at the approximation orders $N=3$ and $N=8$). Contrarily, the regularization based on the Bochner-Laplacian (blue circles in~\textbf{\ref{fig:FPs_FZ_RGImprov}} and left panel in~\textbf{\ref{fig:NormalizedFPsFZ}}) leads to suitable, apparently stabler, NGFP solutions only up to $N_{\text{max}}=9$. This feature is attributed to a limitation in our numerical method implemented to generate fixed-point solutions. 

\begin{figure}[!t]
  \begin{center}
    \includegraphics[height=5.8cm]{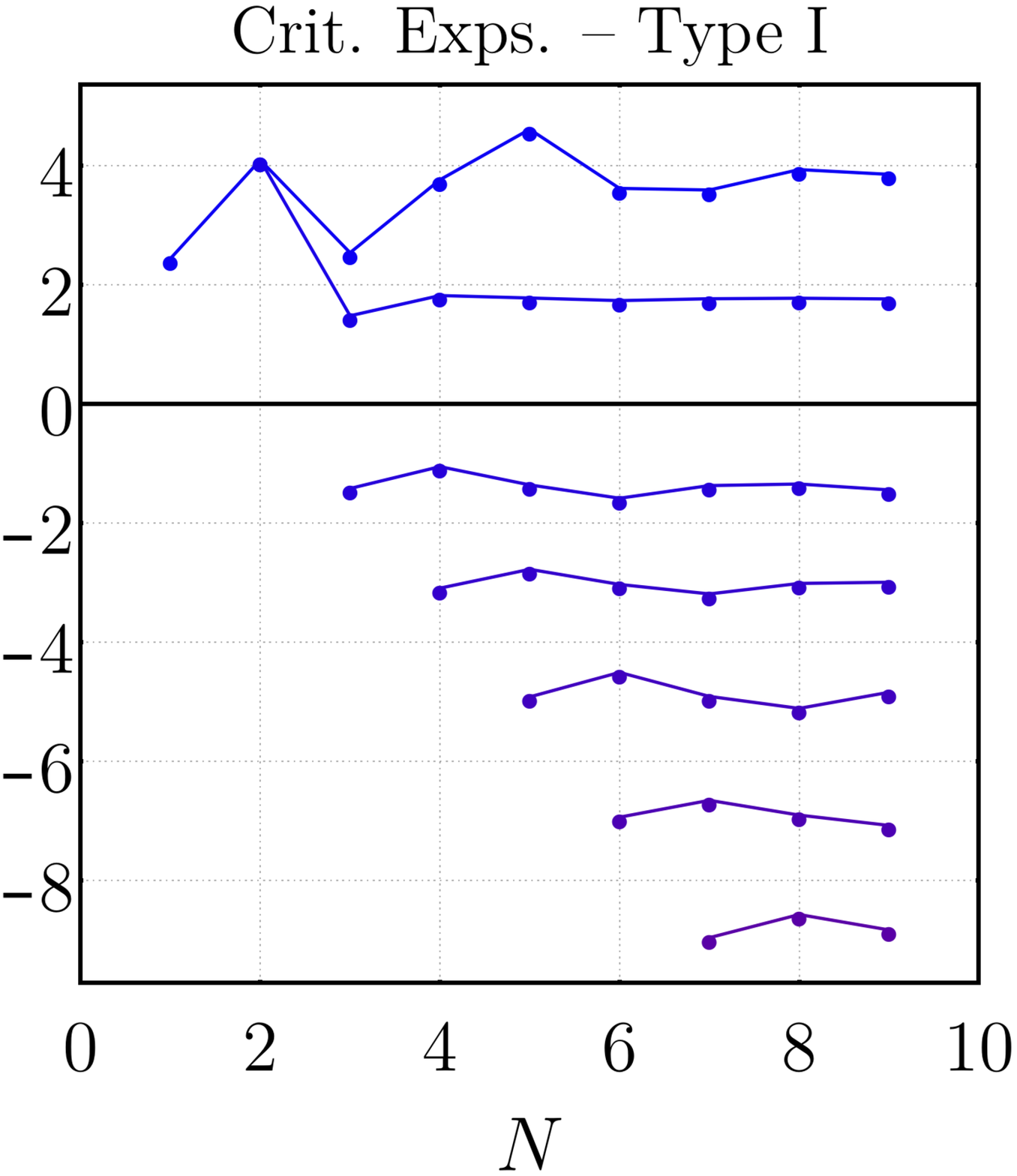} \quad
   \hspace{3em}
    \includegraphics[height=5.8cm]{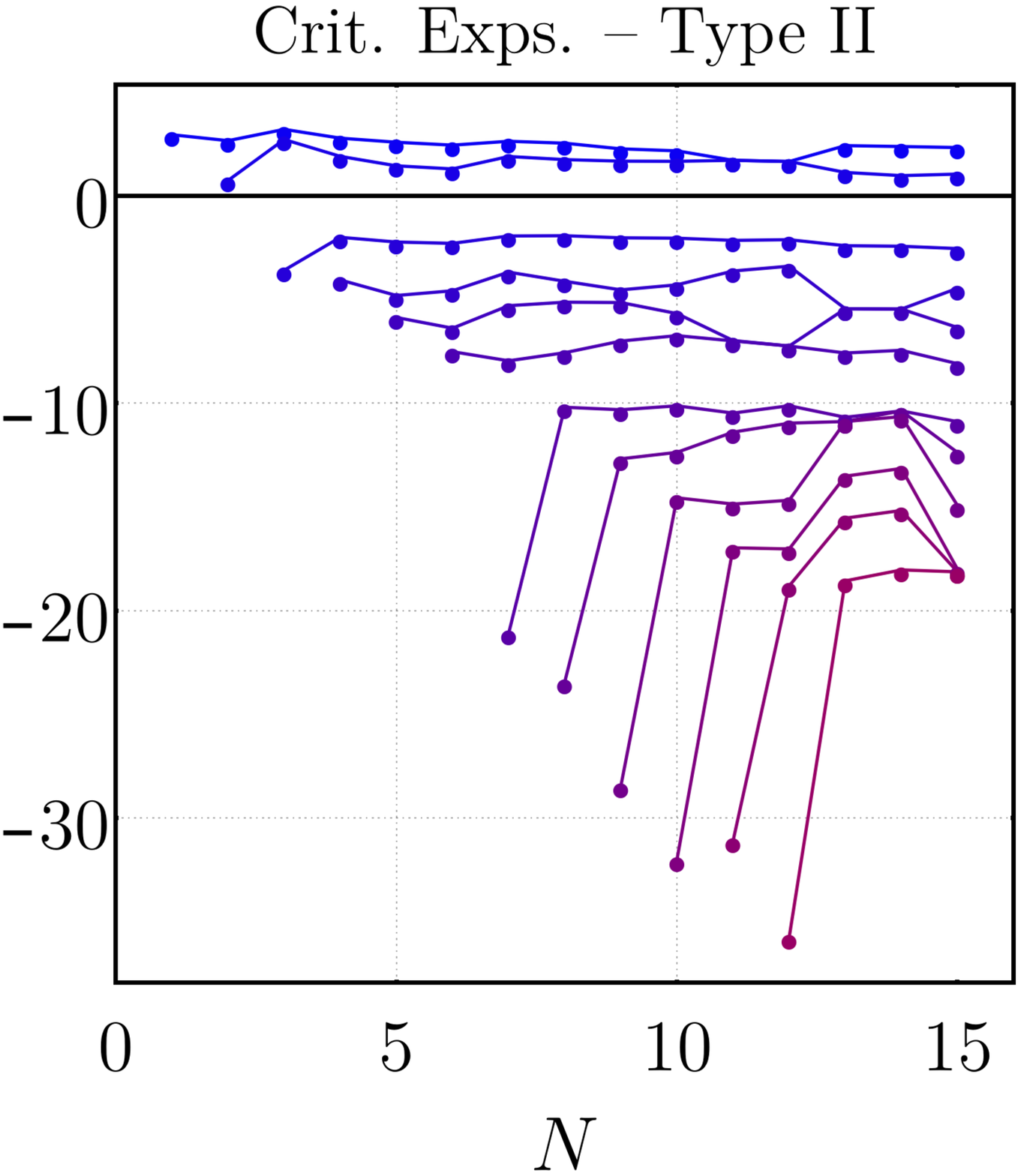}
  \caption{\footnotesize{Critical exponents associated with the fixed-point structure in the FZ-truncation within the RG-improved closure. The left panel corresponds to results for the range $n=1,\cdots,7$ in the $N_{\text{max}}=9$ truncation obtained under the Type I regularization, while the right plot displays the results for the range $n=1,\cdots,12$ in the $N_{\text{max}}=15$ truncation obtained under the Type II regularization.}}
  \label{fig:FZCritExps}
  \end{center}
\end{figure}

According to the critical exponents illustrated in Fig.~\textbf{\ref{fig:FZCritExps}}, our findings for the FZ-truncation still indicate that the UV critical hypersurface is characterized by two relevant directions for both types of coarse-graining operators. Despite the stabilization of the number of relevant directions, the numerical values for the critical exponents undergo the same unstable behavior as the fixed-point values depicted in Fig.~\textbf{\ref{fig:FPs_FZ_RGImprov}}. Albeit the difficulties in extending our analysis to truncations higher than $N=9$ for the Type I regularization scheme, the results shown in Fig.~\textbf{\ref{fig:FZCritExps}} (left) indicate that the critical exponents share the same near-canonical character as in the case of the $f(R)$-approximation. However, such a behavior is less apparent in the case of Type II coarse-graining operators. Here, as one can see from Fig.~\textbf{\ref{fig:FZCritExps}} (right), some critical exponents behave according to the near-canonical scaling. Nevertheless, for assorted choices of $N$ there are points which exhibit appreciable deviations from the canonical scaling of invariant operators within the truncation.

\begin{figure}[!t]
  \begin{center}
    \includegraphics[height=5.8cm]{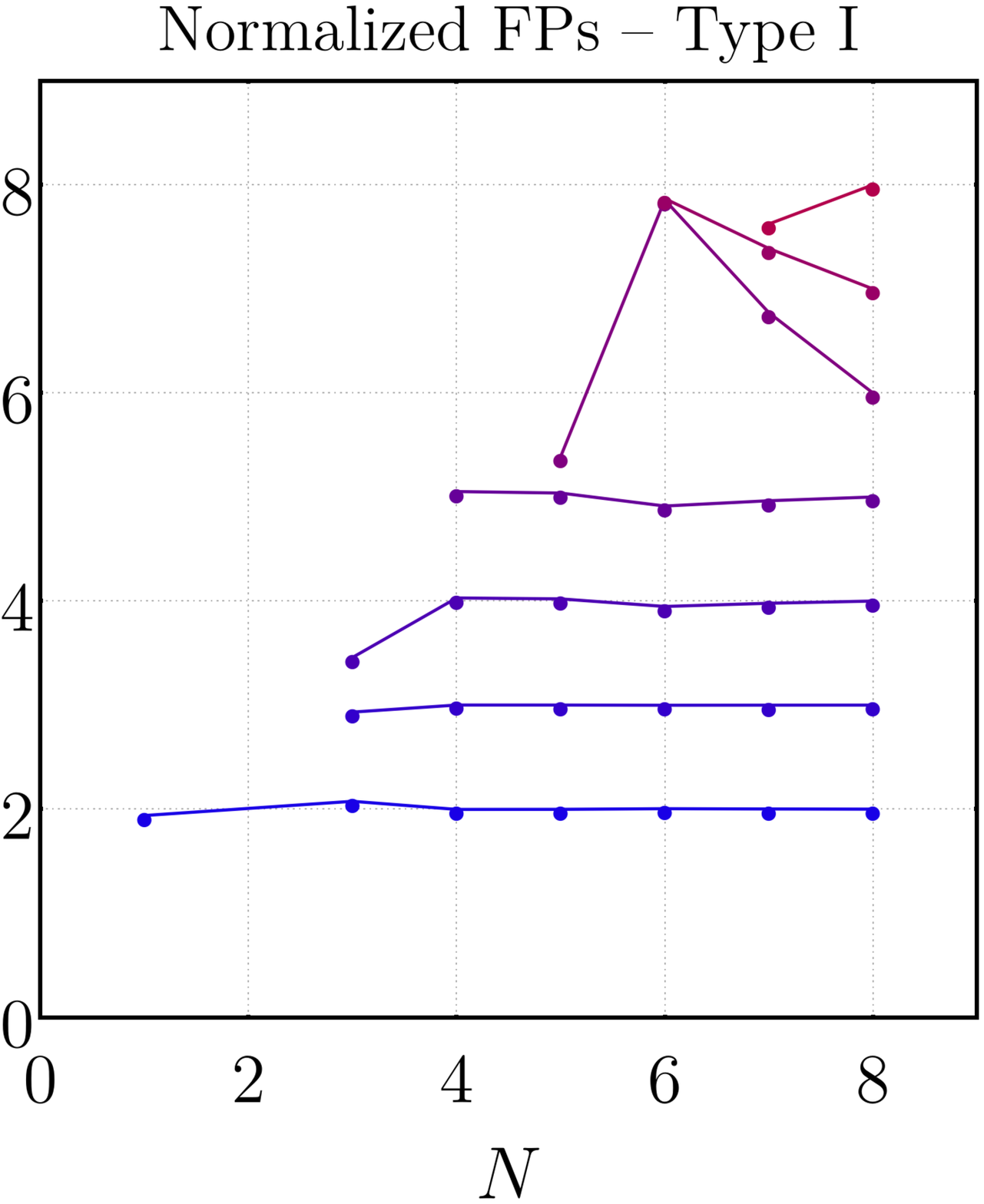} \quad
    \hspace{3em}
    \includegraphics[height=5.8cm]{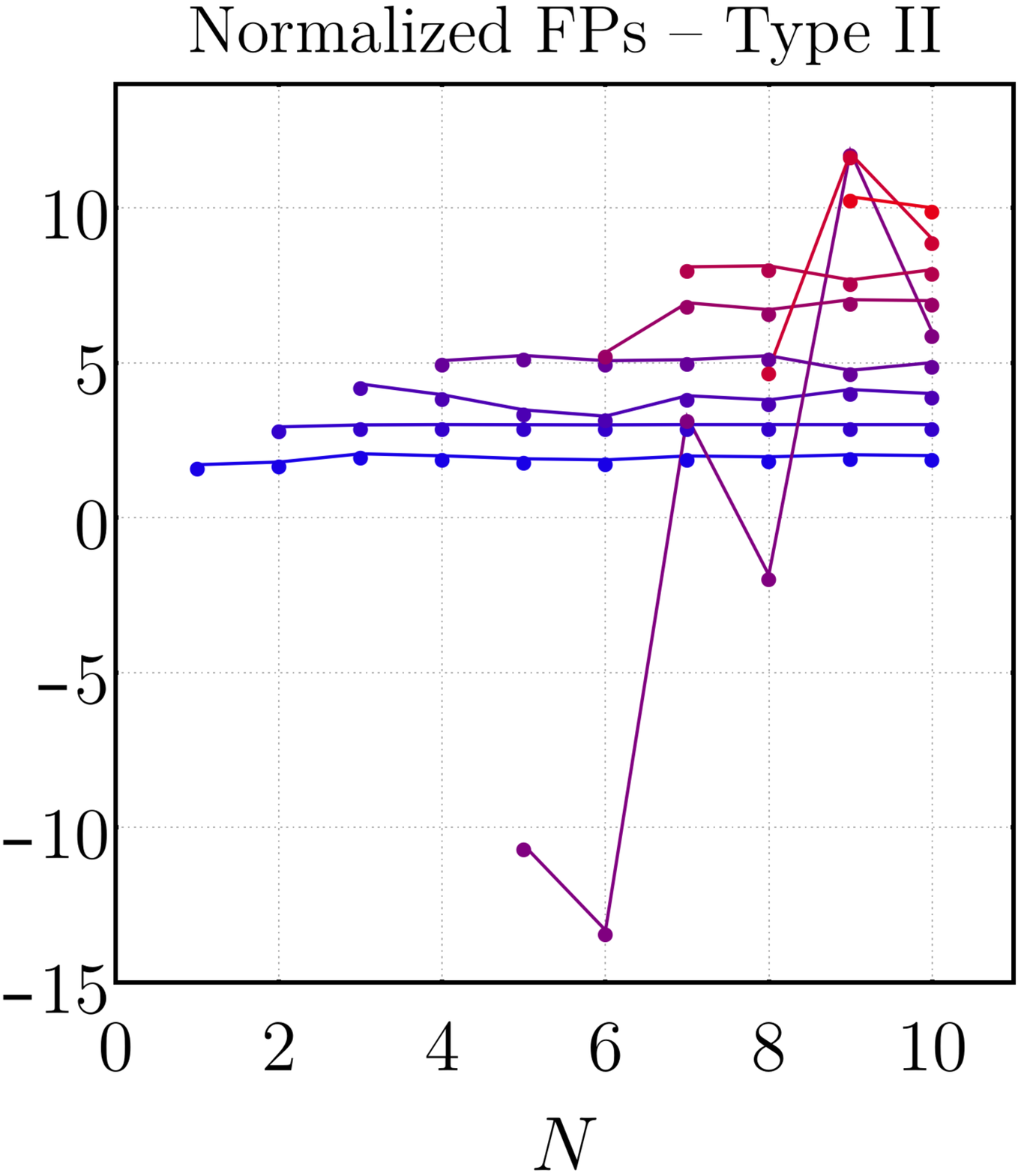}
    \caption{\footnotesize{Plots of the convergence pattern for the normalized fixed-point values of the couplings $\lambda_{n}(N)$ (given in terms of $(G_k,\rho_{k,n})$) for the FZ-truncation evaluated within the semi-perturbative prescription. The left plot exhibits the convergence pattern for the range $n=1,\cdots,7$ in the $N_{\text{max}}=8$ truncation under the Type I regularization, while the right plot displays the convergence pattern for the range $n=1,\cdots,9$ in the $N_{\text{max}}=10$ truncation under the Type II regularization. All couplings follow the same normalization convention as in the $f(R)$ case.}}
  \label{fig:NormalizedFPsFZSemi}
  \end{center}
\end{figure}

\begin{figure}[!t]
  \begin{center}
    \includegraphics[height=5.8cm]{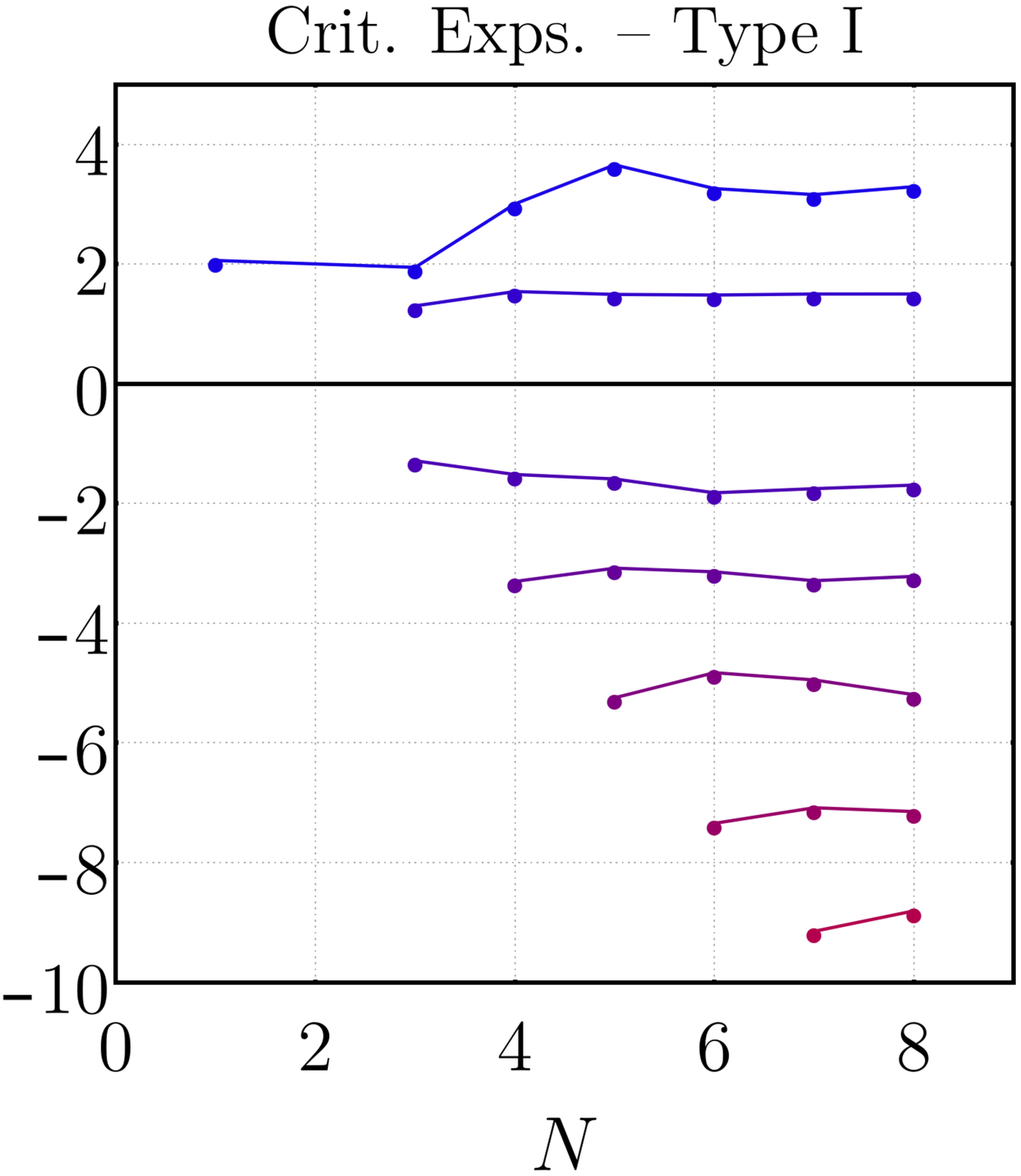} \quad
   \hspace{3em}
    \includegraphics[height=5.8cm]{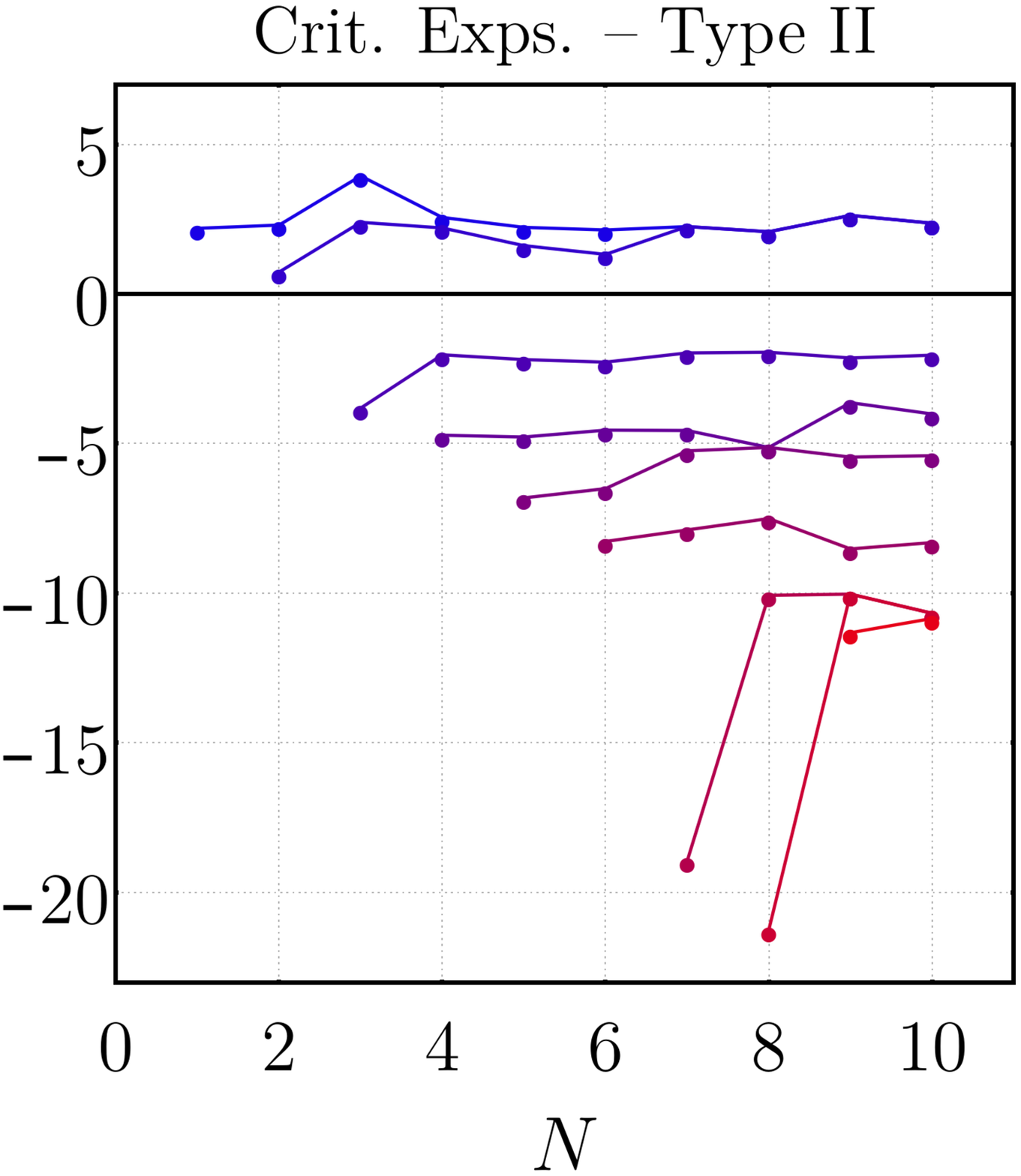}
  \caption{\footnotesize{Critical exponents associated with the fixed-point structure in the FZ-truncation within the semi-perturbative closure. The left panel corresponds to results for the range $n=1,\cdots,7$ in the $N_{\text{max}}=8$ truncation obtained under the Type I regularization, while the right plot displays the results for the range $n=1,\cdots,9$ in the $N_{\text{max}}=10$ truncation obtained under the Type II regularization.}}
  \label{fig:FZCritExpsSemi}
  \end{center}
\end{figure}

To conclude this section, we display in Figs. \textbf{\ref{fig:NormalizedFPsFZSemi}} and \textbf{\ref{fig:FZCritExpsSemi}} the results for the normalized fixed-point values and critical exponents in the FZ-truncation when the semi-perturbative prescription is adopted. For the regularization employed by the Lichnerowicz-Laplacians (right panel in Fig.~\textbf{\ref{fig:NormalizedFPsFZSemi}}), suitable NGFP solutions were found for polynomial truncation until $N_{\max}=10$, with improved stabilization of the fixed-point coordinates, apart from a severe oscillation at order $N=5$. These are better results in comparison with the previous case (given the simplicity of our truncation). Regarding the Bochner-Laplacian operator (left panel in Fig.~\textbf{\ref{fig:NormalizedFPsFZSemi}}), stable results were achieved only up to $N_{\max}=8$. A similar limitation was observed in the previous analysis. Moreover, at order $N=2$, we have disregarded the only would-be suitable NGFP solution for the pair $(G^*,\rho_2^*)$, since one of the two corresponding critical exponents is $\sim 110$ and may be regarded as a truncation artifact. Conclusive statements regarding the stability of the fixed point requires an extensive analysis of more sophisticated truncations.


As in the RG-improved case, the dimensionality of the UV critical hypersurface is still two for both regularization schemes. However, for the Type II case, the two positive critical exponents exhibit mild oscillations and, as opposed to the corresponding RG-improved result, the gap controlling their near-canonical scaling is severely reduced by the anomalous dimensions contributions when higher-order invariant operators are included. Nevertheless, in contrast with the RG-improved analysis, the critical exponents do not exhibit appreciable deviations from canonical scaling for several choices of the approximation order $N$.  

Notably, our findings suggest that in the unimodular version of the FZ-truncation the search for fixed-point candidates gets hampered by difficulties in extending the approximation order beyond $N=16$ (which is the case considering Lichnerowicz-Laplacians within the RG-improvement prescription) in comparison with the fixed-point analysis in the unimodular version of the $f(R)$-approximation. The possibility of extension gets more restricted when independent anomalous dimensions are adopted. As a consequence, the FZ-truncation in the unimodular setting generates less stable solutions than its $f(R)$ relative overall. This characteristic is opposed to considerations previously made in the standard ASQG setting. In particular, the systematic investigation carried out in~\cite{Falls:2017lst} reveals that the FZ-truncation presents a faster stabilization, including higher-order extensions, than the $f(R)$-approximation. Considering the approximations we have used, our findings reveals the opposite behavior in the unimodular version.

\subsection{Gravity-matter systems}\label{Impact_Matter}

Several works in standard ASQG provide strong hints for the existence of a NGFP in the RG flow within different truncations, ranging from the Einstein-Hilbert approximation to more sophisticated ones \cite{Souma:1999at,Reuter:2001ag,Lauscher:2002sq,Litim:2003vp,Codello:2006in,Codello:2007bd,Machado:2007ea,Codello:2008vh,Eichhorn:2009ah,Benedetti:2009rx,Benedetti:2009gn,Eichhorn:2010tb,Manrique:2010am,Manrique:2011jc,Christiansen:2012rx,Benedetti:2012dx,Demmel:2012ub,Dietz:2012ic,Dietz:2013sba,Demmel:2013myx,Falls:2013bv,Benedetti:2013jk,Ohta:2013uca,Codello:2013fpa,Demmel:2014sga,Demmel:2014hla,Falls:2014tra,Christiansen:2014raa,Falls:2014zba,Falls:2015qga,Christiansen:2015rva,Demmel:2015oqa,Ohta:2015efa,Ohta:2015fcu,Gonzalez-Martin:2017gza,Knorr:2017mhu,Christiansen:2017bsy,Gies:2015tca,Gies:2016con,Biemans:2016rvp,Denz:2016qks,Falls:2016msz,Falls:2017lst,Falls:2018ylp,deBrito:2018jxt,Knorr:2019atm,Burger:2019upn,Falls:2020qhj,Kluth:2020bdv}. On top of that, a growing number of investigations provide compelling evidence for the persistence of the NGFP against the introduction of a large class of matter fields, such as the field content corresponding to the SM of particle physics and  some beyond SM (bSM) extensions, see \cite{Dou:1997fg,Percacci:2002ie,Percacci:2003jz,Shaposhnikov:2009pv,Narain:2009fy,Zanusso:2009bs,Eichhorn:2011pc,Eichhorn:2012va,Dona:2012am,Dona:2013qba,Dona:2014pla,Labus:2015ska,Oda:2015sma,Meibohm:2015twa,Dona:2015tnf,Meibohm:2016mkp,Eichhorn:2016esv,Eichhorn:2016vvy,Biemans:2017zca,Hamada:2017rvn,Christiansen:2017qca,Eichhorn:2017eht,Eichhorn:2017egq,Eichhorn:2017sok,Christiansen:2017cxa,Eichhorn:2017als,Christiansen:2017gtg,Christiansen:2017qca,Eichhorn:2017ylw,Eichhorn:2017lry,Alkofer:2018fxj,Alkofer:2018baq,Eichhorn:2018akn,Eichhorn:2018ydy,Eichhorn:2018nda,Pawlowski:2018ixd,Eichhorn:2018whv,deBrito:2019epw,Wetterich:2019zdo,Wetterich:2019rsn,Wetterich:2019rsn,Reichert:2019car,Burger:2019upn,Domenech:2020yjf,Daas:2020dyo,Eichhorn:2020sbo,deBrito:2020dta}. 
In this section, we explore the impact of matter degrees of freedom on the interacting gravitational fixed-point structure in the unimodular setting for both $f(R)$ and $F(R_{\mu\nu}^2)+RZ(R_{\mu\nu}^2)$ polynomial truncations. By varying the number of matter fields, we can probe the compatibility of non-trivial fixed-point solutions in the unimodular theory space coupled to the field content corresponding to the SM of particle physics as well as to some bSM extensions.

Following the same strategy employed in the pure gravity case, a numerical recursive solution and a bootstrap search method were adopted for the selection of suitable fixed-point candidates. For both $f(R)$- and FZ-truncations we have limited our search to fixed-point solutions within polynomial approximations ranging from $N=1$ to $N=10$. The results reported in the case of gravity-matter systems are restricted to the RG-improved treatment for the anomalous dimensions, i.e., using a prescription that relates $\eta_\TT$ and $\eta_\sigma$ to the beta function of the Newton coupling, while setting all other anomalous dimensions to zero. The other prescription considered in the pure gravity, with anomalous dimensions computed via derivative expansion, will not be reported here. The reason is related to the existence of certain bounds on the anomalous dimensions, as it was pointed out in \cite{Meibohm:2015twa}, appearing as a consistency requirement for an appropriate behavior of the FRG regulator at $k\to \infty$. For gravity-matter systems we have found fixed-point values that violate such bounds and we can argue that these results are not self-consistent.

In Table \textbf{\ref{fig:Table_FPsMatter}} we exhibit a summary of the results concerning the stability of NGFPs for specific matter contents for both $f(R)$ and $F(R_{\mu\nu}^2)+RZ(R_{\mu\nu}^2)$ approximations. In this case, we just report the main qualitative features, i.e., in which cases we find evidence for fixed-point solutions and the corresponding number of relevant directions. In all cases, we have investigated polynomial truncations including operators up to $\mathcal{O}(R^{10})$.

\begin{table}
	\begin{center}
		\resizebox{13cm}{!}{%
			\vspace{0pt}
			\begin{tabular}{|c | c | c | c | c |c | c |c|}
				\toprule
				\multicolumn{8}{|c|}{Stability of NGFP for some specific matter models}\\
				\midrule
				Model &  \multicolumn{3}{c|}{Matter content}  & \multicolumn{2}{c|}{Type I }   & \multicolumn{2}{c|}{Type II } \\
				\midrule
				& $N_{\phi}$ & $N_A$ & $N_{\psi}$ &  $f(R)$ & $F(R_{\mu\nu}^2)+R Z(R_{\mu\nu}^2)$  & $f(R)$ & $F(R_{\mu\nu}^2)+R Z(R_{\mu\nu}^2)$  \\
				\midrule
				SM & 4 & 12 & 45/2 & \cmark $(2)$  &  \cmark $(2)$  & \cmark $(2)$    & \cmark $(2)$ \\
				\midrule
				SM + $3\nu_{\text{R}}$ & 4 & 12 & 24 & \cmark $(2)$  &   \cmark $(3)$ & \cmark $(2)$   &\cmark $(2)$  \\
				\midrule
				MSSM & 49 & 12 & 61/2  & \xmark  & \xmark  & \xmark  &\xmark  \\
				\midrule
				SU(5) GUT & 124 & 24 & 24  & \xmark    & \xmark & \xmark  &\xmark  \\
				\midrule
				SO(10) GUT & 97 & 45 & 24   & \cmark$^*(2)$    & \cmark$^*(3)$    & \cmark$^*(2)$  &\cmark$^*(2)$   \\
				\bottomrule
			\end{tabular}%
		}
		\caption{\footnotesize{Collection of the results on the stability of NGFPs arising from the matter content of the Standard Model and some of its commonly studied extensions for both $f(R)$ and $F(R_{\mu\nu}^2)+RZ(R_{\mu\nu}^2)$ polynomial projections in the unimodular setting. The RG-improved closure is adopted. The symbols go as follows: checkmarks $\,$\cmark$\,$ indicate the underlying setup possesses a suitable NGFP which converges for increasing order of approximation $N$. The number between parenthesis indicates the number of relevant directions observed. An asterisk simply indicates that there is no NGFP at the level of the Einstein-Hilbert truncation ($N=1$), converging afterwards towards a suitable NGFP. Finally, an $\,$\xmark$\,$ means that there is no NGFP at all orders of approximation, except for one appearance at only one power of curvature.}}
		\label{fig:Table_FPsMatter}
	\end{center}
\end{table}

The minimal requirement for a phenomenological viable fixed-point solution is its compatibility with the matter content of the SM, i.e., $N_{\phi}=4$, $N_A=12$ and $N_{\psi}=45/2$. As we can observe in Table \ref{fig:Table_FPsMatter}, our result points towards the existence of this fixed point for both truncations under investigation and for both types of regulators employed in the coarse-graining procedure. The fixed-point solution corresponding to the SM matter content exhibits similarities with the results observed for pure gravity. In both truncations and regularization schemes, we have found evidence for two-dimensional UV critical surfaces. Furthermore, the numerical values for the fixed-point solutions, as well as for the critical exponents, seem to stabilize for truncations characterized by $N\gtrsim 6$. The exception is the FZ-approximation with Type I regulator, which presents a mild deviation from the ``convergence'' pattern at $N=9$ and $N=10$.
	
To complement our analysis, we also have considered the matter content associated with some bSM scenarios. The first extension, which is motivated by the necessity to accommodate neutrino masses, corresponds to the choice $(N_{\phi}=4,N_A=12,N_{\psi}=24)$, i.e., including $3/2$ additional Dirac fermions (or 3 Weyl fermions), accounting for 3 right-handed neutrinos. In this case, our results also point towards the existence of UV fixed-point solutions. The main difference in comparison with the SM matter content is the appearance of an extra relevant direction in the FZ-truncation with Type I regularization. For the other approximations/schemes, our results indicate two relevant directions.
	
It is also interesting to consider matter content corresponding to bSM scenarios characterized by larger symmetry groups, e.g., supersymmetric  models and grand unified theories (GUTs). In Table \ref{fig:Table_FPsMatter}, we report our findings for matter content associated with the minimally supersymmetric Standard Model (MSSM), SU(5) and SO(10) GUTs. Among these options, only the SO(10) GUT ($N_\phi=97$, $N_A=45$ and $N_\psi = 24$) exhibits suitable fixed-point solutions. In this case, most of the schemes under investigation leads to UV fixed-points characterized by two relevant directions. The exception, once again, is the FZ-truncation with Type I regulator, where we have found three relevant directions. It is also interesting to emphasize that, in the case of matter content corresponding to SO(10) GUT, the fixed-point solutions do not appear at the level of the lowest truncation, i.e., $N=1$.
	
For the matter content corresponding to the MSSM and SU(5) GUT, we do not find evidence for the existence of suitable fixed-point solutions within the aforementioned truncations. Our findings show a qualitative agreement with the results for non-unimodular settings reported in \cite{Dona:2013qba,Alkofer:2018fxj}. As it was pointed out in \cite{Dona:2013qba}, the absence of suitable UV fixed-points for gravity-matters systems with field content corresponding to the MSSM and SU(5) GUT can be explained by the inclusion of too many scalars and fermions, without being compensated by the inclusion of extra vector fields. It is important to emphasize that this explanation is restricted to the calculations based on the background-field approximation. It is worth mentioning that results from the fluctuation approach, see \cite{Pawlowski:2020qer}, for ASQG indicate that the inclusion of too many scalars pushes the scalar anomalous dimension to a regime that violates certain regulator bounds \cite{Meibohm:2015twa}.

\section{Conclusions}\label{Conclusions}

In this work, the renormalization group flow of unimodular quantum gravity was analyzed. This was motivated by the possibility of such a quantum theory to be asymptotically safe and, thus, well defined up to arbitrarily short distances. We have explored larger theory spaces with respect to previous analyses, by considering truncations which involve the tensorial structure of Ricci-tensor invariants and anomalous dimensions which are computed from the running of the two-point function of gravitons and Faddeev-Popov ghosts. Moreover, in the background approximation, we have used the background-dependent correction to the flow equation  discussed in \cite{deBrito:2020rwu}. Such improvements enabled us to confront previous results \cite{Eichhorn:2013xr,Eichhorn:2015bna} with truncations enlargements and, apart from quantitative differences which follows from truncation-induced effects, we have found evidence for the persistence of the fixed point.

Technically, we have also tested how the underlying fixed-point structure is affected by different choices of the endomorphism parameter in the regulator function. In particular, we discussed results obtained for Bochner and Lichnerowicz coarse-graining operators. As expected, different choices of such operators, in the background approximation, directly affect the projection onto curvature invariants in the flow equation and can lead to substantial different qualitative results such as the number of relevant directions (see, e.g., \cite{Bridle:2013sra,deBrito:2018jxt}). For this discrete choice of the endomorphism parameter, we have observed stable qualitative results both in the $f(R)$ and FZ truncations, where the fixed point features two relevant directions. Nevertheless, different classes of truncations lead to different computational subtleties and we have verified that in this setup, the $f(R)$ truncation has better (apparent) convergence properties. More efficient methods must be employed for the FZ-truncation in order to probe whether the fixed point structure stabilize for larger truncations. In any case, it is remarkable that by changing the endomorphism parameter and the anomalous dimension prescription in each class leads to a fixed point which features the same qualitative features, leading to the expectation that this is a consequence of the near-perturbative nature of the fixed point (which is reflected in the near-canonical scaling).

Finally, we have considered the interaction of unimodular quantum gravity with matter degrees of freedom. Intuitively, matter fluctuations will affect the running of the gravitational couplings and since we aim at describing a realistic theory of quantum gravity, the fixed point must exist in the presence of matter fields. As a first approximation, we have included scalars, spinors and vectors without self-interactions coupled to the unimodular gravitational field. As discussed, the matter content of the SM and of some of its extensions does not destroy the fixed point, leading to evidence for the existence of a complete theory of quantum gravity and matter. However, as pointed out, for some extensions of the SM, the matter content is ``too big" and destroys the fixed point, i.e., they act against scale invariance in the UV. Hence, it is a concrete realization that even for truncated theory spaces, one might indeed find systems which do not feature a fixed point.

The present work suggests several different ways of improving the truncations of unimodular quantum gravity and matter systems. In particular, a promising and necessary direction is the consideration of approximations that go beyond the background one. In this work we have performed a purely background approximation and a hybrid one. However, it is necessary to investigate momentum-dependent correlation functions in unimodular quantum gravity and compare the results with our present findings. This is work in progress. 

Lastly, there is the discussion about the equivalence of unimodular quantum gravity and standard quantum gravity. Conceptually, from the point of view of asymptotically safe quantum gravity, this is an important puzzle to be resolved. Different symmetry groups define, in principle, different theory spaces and hence, a different set of essential couplings that should reach a non-trivial fixed point. In the unimodular setting, there is no room for a cosmological constant as an essential coupling while in standard gravity, it is usually treated as an essential coupling and it is required to reach a fixed point in the asymptotic safety program. However, it is far from clear if this necessarily leads to incompatible pictures. In standard gravity, the cosmological constant corresponds to a relevant direction and, thus, a free parameter that should be fixed by ``experiments". In unimodular gravity, the cosmological constant appears as an integration constant which is also fixed by initial conditions. In the end, it remains to be understood if such theories share the same observables or not.

\section*{Acknowledgments}

The authors are grateful to R. Alkofer, A. Eichhorn, J. Pawlowski and R. Percacci for discussions about unimodular gravity in the last months/years. The authors also acknowledge M. Schiffer for helping with some computational issues. GPB is supported by VILLUM FONDEN under grant number 29405. GPB is also grateful to CNPq no.~142049/2016-6 for the financial support during part of this work. The work of AFV is supported by CNPq under the grant no. 140968/2020-2. ADP acknowledges CNPq under the grant PQ-2 (309781/2019-1), FAPERJ under the “Jovem Cientista do Nosso Estado” program (E26/202.800/2019), and NWO under the VENI Grant (VI.Veni.192.109) for financial support.

\appendix

\section{The decomposed Hessian}\label{hessianelements}

In this Appendix we report the Hessians employed in the computation of the beta functions of the gravitational couplings. Expanding the gravitational part of the flowing action (\ref{workingtruncation}) up to second order in the fluctuation field $h$ leads to the expressions
\begin{subequations}
\begin{align}
\Gamma_{\text{TT}}^{(2)}&=\mathcal{Z}_{k,\text{TT}}\bigg[f_k^{(0,1)}\bigg(\Delta_2+(\gamma_2-1)\bar{R}\bigg)-f_k^{(1,0)}\bigg]\bigg(\Delta_2+\frac{2\gamma_2-1}{2}\bar{R}\bigg)\\
\Gamma_{\xi\xi}^{(2)}&=\frac{2\mathcal{Z}_{k,\xi}}{\alpha}\bigg(\Delta_1+\frac{2\gamma_1-1}{2}\bar{R}\bigg)^2\\
\Gamma_{\sigma\sigma}^{(2)}&=\frac{9\mathcal{Z}_{k,\sigma}}{8}\bigg[\mathbf{P}_k\bigg(\Delta_0+\frac{3\gamma_0-1}{3}\bar{R}\bigg)+\mathbf{Q}_k\bigg]\bigg(\Delta_0+\frac{3\gamma_0-1}{3}\bar{R}\bigg)(\Delta_0+\gamma_0\bar{R})^2\\
\Gamma_{C\bar{C}}^{(2)}&=\sqrt{2}\mathcal{Z}_{k,C}\bigg(\Delta_1+\frac{2\gamma_1-1}{2}\bar{R}\bigg),
\end{align}
\end{subequations}
where we have defined
\begin{subequations}
	\begin{align}
	\mathbf{P}_k &= 
	f_k^{(2,0)} + \frac{1}{4} \bar{R}^2\,f_k^{(0,2)} + 4 \,\bar{R} \,f_k^{(1,1)} + \frac{2}{3} f_k^{(0,1)} \,,\\
	\mathbf{Q}_k &= 
	\frac{1}{3}f_k^{(1,0)} + \frac{2}{9} \bar{R} \, f_k^{(0,1)} \,.
	\end{align}
\end{subequations}
Furthermore, as defined in the main text, we define the coarse-graining operator for each spin-$s$ sector as $\Delta_s=\Delta_{\L s}-\gamma_s\bar{R}$, where the endomorphism parameters are introduced such that the choice $\gamma_0 = \gamma_{\frac{1}{2}} = \gamma_1 = \gamma_2=0$ implements the Lichnerowicz-Laplacians and $\gamma_0=0$, $\gamma_{\frac{1}{2}}=1/4$, $\gamma_1=1/4$ and $\gamma_2=2/3$ provide the Bochner-Laplacian.
As the matter action is already second order in the fields, its Hessian elements are given by	
\begin{subequations}
\begin{align}
\Gamma_{\phi\phi}^{(2)}&=\Delta_{\L0}\\
\Gamma_{A^{\text{T}}A^{\text{T}}}^{(2)}&=\Delta_{\L1}\\
\Gamma_{A^{\text{L}}A^{\text{L}}}^{(2)}&=\frac{1}{\zeta}\Delta_{\L0}\\
\Gamma_{\psi\bar{\psi}}^{(2)}&=i\slashed{\nabla}\\
\Gamma_{c\bar{c}}^{(2)}&=\Delta_{\L0}.
\end{align}
\end{subequations}

\section{Heat kernel evaluation and trace technology}\label{heatkernel}

We follow the standard heat kernel techniques to compute the functional traces needed throughout this work. We restrict the calculations below to $d=4$. On general grounds, a functional trace can be expanded in terms of  heat kernel coefficients~\cite{Buchbinder:1992rb,Avramidi:2000bm,Percacci:2017fkn}, namely
\begin{align}\label{Heat_kernel_expansion}
\Tr_{(s)}\big[ W(\Delta_s)  \big] = 
\frac{1}{16\pi^2}\sum_{n=0}^\infty \int_x \sqrt{\bar{g}} \,\, Q_{2-n}[W] \,\, \textmd{tr}\big[ \textbf{b}_{2n}(\Delta_s) \big],
\end{align}
with $Q_n$-functional defined (for arbitrary real $n$) according to
\begin{align}
Q_n[W] = \frac{(-1)^k}{\Gamma(n+k)} \int_0^\infty dz \, z^{n+k-1} \,\frac{d^k W(z)}{dz^k} ,
\end{align}
where $k$ denotes some (arbitrary) positive integer satisfying the following restriction $n+k>0$. Moreover, $\textmd{tr}\big[ \textbf{b}_{2n}(\Delta_s) \big]$ denotes the trace of the (non-integrated) heat kernel coefficient $\textbf{b}_{2n}(\Delta_s)$ associated with the coarse-graining operator $\Delta_s$. When the background is evaluated over a sphere $S^4$, we can express
\begin{align}
\textmd{tr}\big[ \textbf{b}_{2n}(\Delta_s) \big] = c_s \, \bar{R}^n \,,
\end{align}
where $c_s$ denotes a numerical coefficient depending on the choice of the coarse-graining operator. In Tables \ref{Table_Heat-Kernel_Coeffs-Bochner} and \ref{Table_Heat-Kernel_Coeffs-Lichnerowicz}, we report the relevant $c_s$-coefficients for the analysis presented in this work.

\begin{table}[!htb]
	\begin{center}
		\begin{tabular}{|c|ccccccc|}
			\hline\hline \xrowht[()]{10pt}
			$\,\, s \,\,\,\,$ & $n=0$ & $n=2$                 & $n=4$                       & $n=6$                           & $n=8$                             & $n=10$                             & $n=12$                                \\
			\hline \xrowht[()]{20pt}
			0   & $1$   & $\frac{5}{6}$  & $\frac{749}{2160}$ & $\frac{26141}{272160}$ & $\frac{130117}{6531840}$ & $\frac{203161}{61585920}$ & $\frac{925711}{2037934080}$  \\ \xrowht[()]{20pt}
			1   & $3$   & $\frac{3}{2}$  & $\frac{259}{720}$  & $\frac{4931}{90720}$   & $\frac{1373}{241920}$    & $\frac{8527}{20528640}$   & $\frac{261865}{13450364928}$ \\ \xrowht[()]{20pt}
			2   & $5$   & $-\frac{5}{6}$ & $-\frac{1}{432}$    & $\frac{311}{54432}$    & $\frac{109}{1306368}$    & $-\frac{317}{12317184}$   & $-\frac{6631}{4483454976}$  \\
			\hline\hline
		\end{tabular}
		\caption{\footnotesize $c_s$-coefficients associated with the Bochner-Laplacian as the coarse-graining operator. All the coefficients were computed within the 4-sphere background.}
		\label{Table_Heat-Kernel_Coeffs-Bochner}
	\end{center}
\end{table}
\begin{table}[!htb]
	\begin{center}
		\begin{tabular}{|c|ccccccc|}
			\hline\hline \xrowht[()]{10pt}
			$\,\, s \,\,\,\,$ & $n=0$ & $n=2$                 & $n=4$                       & $n=6$                           & $n=8$                             & $n=10$                             & $n=12$                                \\
			\hline \xrowht[()]{20pt}
			0   & $1$   & $\frac{1}{6}$   & $\frac{29}{2 160}$ & $\frac{37}{54 432}$      & $\frac{149}{6 531 840}$     & $\frac{179}{431 101 440}$     & $-\frac{1387}{201 755 473 920}$     \\ \xrowht[()]{20pt}
		   1/2  & $4$   & $-\frac{1}{3}$   & $-\frac{11}{2160}$   & $\frac{31}{544 320}$ & $\frac{41}{26 127 360}$      & $\frac{31}{492 687 360}$     & $\frac{10331}{3 228 087 582 720}$          \\ \xrowht[()]{20pt}			 
			1   & $3$   & $-\frac{1}{2}$  & $\frac{19}{720}$   & $-\frac{5}{18 144}$      & $-\frac{11}{2 177 280}$     & $-\frac{19}{143 700 480}$     & $-\frac{347}{67 251 824 640}$       \\ \xrowht[()]{20pt}
			2   & $5$   & $-\frac{25}{6}$ & $\frac{719}{432}$  & $-\frac{23 125}{54 432}$ & $\frac{101 981}{1 306 368}$ & $-\frac{952 135}{86 220 288}$ & $\frac{50 728 409}{40 351 094 784}$ \\
			\hline\hline
		\end{tabular}
		\caption{\footnotesize $c_s$-coefficients associated with the Lichnerowicz-Laplacian as the coarse-graining operator. All the coefficients were computed within the 4-sphere background.}
		\label{Table_Heat-Kernel_Coeffs-Lichnerowicz}
	\end{center}
\end{table}

For the Litim's cutoff (\ref{litimcutoff}), the $Q_n$-functionals can be computed analytically even for a general function of the form $f_k(\bar{R},\bar{R}^{2}_{\mu\nu})$. Due to specific properties of this choice of profile function, only a finite number of $Q_n$-functionals (with negative $n$) lead to non-vanishing results. As a consequence, the heat kernel expansion in (\ref{Heat_kernel_expansion}) involve only a finite number of terms.

Since the FRG equation written in the York basis features traces over differential constrained fields, spurious eigenvalues of the coarse-graining operator must be properly removed. In the main text this was indicated with the inclusion of an appropriate number of primes in some functional traces. These ``primed'' traces can be computed according to~\cite{Lauscher:2001ya,Machado:2007ea,Codello:2008vh}
\begin{align}
\Tr^{\prime \cdots \prime}_{(s)}\big[ W(\Delta_s)  \big] = \Tr_{(s)}\big[ W(\Delta_s)  \big] - \sum_{l \in M_s} D_l(s) W(\lambda_l(s)),
\end{align}
where $M_s=\{s,s+1,\cdots,m-1+s\}$ with $m$ denoting the number of spurious modes (``primes''). Moreover, $\lambda_l(s)$ denotes the $l$-th eigenvalue of the ``interpolating'' background Laplacian $\Delta_s$ defined on the 4-sphere and $D_l(s)$ represents the degree of degeneracy associated with $\lambda_l(s)$. For the calculation performed in this work, the relevant expressions when $s = 0$ or $1$ are given by
\begin{subequations}
	\begin{align}
	\lambda_l(s) = 
	\frac{(l+3)\,l-s}{12} \,\bar{R}  
	- \gamma_0 \,\delta_{0,s}\,\bar{R} +
	\left( \frac{1}{4} - \gamma_1\right)\,\delta_{1,s}\,\bar{R}\,,\qquad\,
	\end{align}
	\begin{align}
	\quad
	D_l(s) = \frac{(2l+3) \,(l+2)!}{6\,l!}\delta_{0,s} + 
	\frac{l\,(l+3)\,(2l+3)}{2} \delta_{1,s} \,.
	\end{align}
\end{subequations}

\section{UG and background-field approximation}

In the background-field approximation, the unimodular background effective action splits in a transverse diffeomorphism-invariant part and a gauge-fixing part in the form
\begin{equation}
\Gamma_k[\bar{g},\varphi,\Psi]=\bar{\Gamma}_k[g,\Psi]+\hat{\Gamma}_k[\bar{g},\varphi],
\end{equation}
with $\hat{\Gamma}_k[\bar{g},\varphi]\approx S_{k,\textrm{gf}}[\bar{g},h]+S_{k,\textrm{gh}}[\bar{g},\textrm{ghosts}]$, where $\varphi=(h,\bar{C},C,\bar{c},c)$ is the fluctuating multiplet comprising the metric fluctuation and the ghosts associated with \textit{TDiff} and U(1) gauge symmetries. The multiplet $\Psi=(\phi,A,\psi)$ collects the matter part. The covariant approach provides a local invariance associated with the use of the background-field method, namely a split transformation of the background metric and the fluctuation field which renders the full metric invariant, i.e.,
\begin{equation}
g_{\mu\nu}(\bar{g},h)\,\,\mapsto\,\, g_{\mu\nu}(\bar{g}+\delta_{\textrm{split}} \bar{g},h+\delta_{\textrm{split}} h)=g_{\mu\nu}(\bar{g},h).
\end{equation}
This split transformation is tantamount to guaranteeing background independence (see \cite{Eichhorn:2018yfc,Bonanno:2020bil,Pawlowski:2020qer} for a discussion in the context of the FRG approach). For the linear split of the metric, $g_{\mu\nu}=\bar{g}_{\mu\nu}+h_{\mu\nu}$, the split transformation is given by $\delta_{\textrm{split}}\bar{g}_{\mu\nu}=-\chi_{\mu\nu}$ and $\delta_{\textrm{split}} h_{\mu\nu}=\chi_{\mu\nu}$, with $\chi_{\mu\nu}=\chi_{\mu\nu}(x)$ being a local transformation parameter. For the non-linear exponential parametrization, more convenient for unimodular gravity, the explicit form of $\delta_{\textrm{split}} h_{\mu\nu}$ is not straightforward. In this case, we denote $\delta_{\textrm{split}} h_{\mu\nu}=\mathscr{N}^{\alpha\beta}_{\mu\nu}[\bar{g},h]\chi_{\alpha\beta}$, and its explicit form can be determined by an iterative procedure (see \cite{Eichhorn:2017sok,deBrito:2020rwu}). In the following, we
discuss on the corresponding functional identity associated with the split symmetry.

The invariance of $\bar{\Gamma}_k$ under local split transformations\footnote{Actually, the full effective action explicitly breaks the split symmetry since the gauge-fixing and regulator terms are not shift-symmetric. Hence, the Ward identity associated with shift symmetry is actually deformed by those explicit-breaking sources. Nevertheless, ultimately, one is interested in integrating down to $k=0$ which eliminates the spurious breaking coming from the regulator. As for the gauge-fixing, it arises as a BRST-exact term and, therefore, can be handled in conjunction with the Slavnov-Taylor identity. Since we are working in the background approximation where fluctuations are turned off at the level of the flow equation after the computation of the Hessian, we simply ignore those symmetry-breaking contributions.}, i.e., $\delta_{\textrm{split}}\bar{\Gamma}_k[g,\Psi]=0 $, leads to the following functional identity written in schematic form:
\begin{equation}\label{funcid}
\frac{\delta \bar{\Gamma}_k}{\delta \bar{g}}-\mathscr{N}[\bar{g},h]\circ \frac{\delta \bar{\Gamma}_k}{\delta h}=0,
\end{equation}
where the $\circ$ notation indicates a space-time integration and a contraction of all indices and it is understood that $h(\bar{g},g)$. By acting with $(\delta/\delta\bar{g}+\delta/\delta h)$ on (\ref{funcid}) and noting that, at $h=0$, we have $\mathscr{N}[\bar{g},0]=\mathds{1}$ and $\delta_{\bar{g}}\mathscr{N}[\bar{g},h]|_{h=0}=0$, the functional identity (\ref{funcid}) turns into
\begin{equation}
\frac{\delta^2\bar{\Gamma}_k}{\delta \bar{g}^2}\bigg|_{h=0}-\frac{\delta^2\bar{\Gamma}_k}{\delta h^2}\bigg|_{h=0}=\bigg[\frac{\delta\mathscr{N}}{\delta h}\circ \frac{\delta\bar{\Gamma}_k}{\delta\bar{g}}\bigg]_{h=0}.
\label{NI1}
\end{equation}
By requiring that the background is on-shell, i.e., $\bar{g}=\bar{g}_{\text{EoM}}$, the right-hand side of \eqref{NI1} vanishes, and the background and the fluctuation two-point correlation functions agree in the exponential parametrization:
\begin{equation}\label{funcid2}
\frac{\delta^2\bar{\Gamma}_k}{\delta \bar{g}^2}\bigg|_{h=0}=\frac{\delta^2\bar{\Gamma}_k}{\delta h^2}\bigg|_{h=0}.
\end{equation}
The projection adopted for the anomalous dimensions consists of taking two derivatives w.r.t to quantum fields of the flow equation. Therefore, applying a scale derivative on both sides of (\ref{funcid2}) provides an identification of the anomalous dimensions of the graviton modes to be proportional to the background anomalous dimensions. Although the background-field approximation is still useful, Eq.~(\ref{funcid}) is not preserved along the flow as it does not correspond to the modified Nielsen identity, $\text{mNI}=0$, namely, \cite{Safari:2015dva,Eichhorn:2018akn,Eichhorn:2018yfc,Bonanno:2020bil,Pawlowski:2020qer},
\begin{equation}
\text{mNI}=\frac{\delta \Gamma_k}{\delta \bar{g}}-\mathscr{N}[\bar{g},h]\circ \frac{\delta \Gamma_k}{\delta h}-\expval{\bigg[ \frac{\delta }{\delta \bar{g}}- \mathscr{N}[\bar{g},h]\circ \frac{\delta }{\delta h}\bigg]\hat{\Gamma}_k}+\Xi_k[\bar{g},\varphi],
\end{equation}
where $\Xi_k[\bar{g},\varphi]$ is a regulator-dependent contribution. Therefore, for our purposes and for technical reasons, we have chosen to go one step further and also consider a hybrid closure of the system of beta functions by improving the background-field approximation with anomalous dimensions computed in an independent way via the vertex expansion employing a flat background metric.

\section{Anomalous dimensions}\label{etas}
Here we report the (non-vanishing) contributions to the anomalous dimensions of the graviton modes and ghost fields. The expressions are presented within the semi-perturbative approximation where all anomalous dimensions contributions on the RHS of the equations are set to zero. We omit the $k$-dependence for simplicity. For the graviton modes, the gravitational contribution to the anomalous dimension of $\eta_{\textrm{TT}}$ and $\eta_{\sigma}$ yield the following results
\begin{subequations}
\begin{align}\label{Eta_TT_Grav}
 \eta_{\textrm{TT}}\big|_{\text{grav}}&= \frac{G}{432\pi}\bigg[54+\frac{90}{(1+\rho)^4}-\frac{4(53\beta_{\rho}-69)}{(1+\rho)^3}-\frac{1656-155\beta_{\rho}+\alpha(290\beta_{\rho}-3252)}{(1-2\alpha)(1+\rho)^2}\nn\\
& +\frac{20(50+2\alpha(87\alpha-95)-\beta_{\alpha})}{(1-2\alpha)^2(1+\rho)}+\frac{18}{(1-6\alpha-2\rho)^4}+\frac{8(12\beta_{\alpha}+4\beta_{\rho}-3)}{(1-6\alpha-2\rho)^3}\nn\\
& +\frac{76(3\beta_{\alpha}+\beta_{\rho})-8\alpha(15+87\beta_{\alpha}+29\beta_{\rho})}{(1-2\alpha)(1-6\alpha-2\rho)^2}-\frac{8(11+2\alpha(87\alpha-47)+5\beta_{\alpha})}{(1-2\alpha)^2(1-6\alpha-2\rho)}\bigg],
\end{align}
\begin{align}\label{Eta_Sigma_Grav}
 \eta_{\sigma}\big|_{\text{grav}}&=\frac{G}{432\pi}\bigg[-504-\frac{720}{(1+\rho)^4}+\frac{8 (291+23 \beta_{\rho})}{(1+\rho)^3}-\frac{22 \alpha  (5 \beta_{\rho} -138)-47 \beta_{\rho} +1494}{(1-2 \alpha) (1+\rho)^2}\nn\\
&+\frac{2 \left(660 \alpha ^2-532 \alpha +8 \beta_{\alpha} +113\right)}{(1-2 \alpha )^2
   (1+\rho )}-\frac{144}{(1-6\alpha-2\rho)^4}-\frac{8(42\beta_{\alpha}+14\beta_{\rho}-51)}{(1-6\alpha-2\rho)^3}\nn\\
   & -\frac{4 (\alpha  (66 \beta_{\alpha} +22 \beta_{\rho} -42)-57 \beta_{\alpha} -19 \beta_{\rho}
   +9)}{(1-2 \alpha) (1-6 \alpha -2 \rho)^2}-\frac{4 (4 \alpha  (33 \alpha -65)-8 \beta_{\alpha} +85)}{(1-2 \alpha )^2 (1-6 \alpha -2
   \rho)}\bigg].
\end{align}
\end{subequations}

The ghost anomalous dimension is given by
\begin{equation}\label{Eta_Gh}
\eta_c=\frac{G}{270\pi}\bigg[\frac{5(3\beta_{\rho}-4)}{(1+\rho)^2}+\frac{20}{1+\rho}-\frac{4(7+9\beta_{\alpha}+3\beta_{\rho})}{(1-6\alpha-2\rho)^2}+\frac{148}{1-6\alpha-2\rho}\bigg].
\end{equation}

\bibliographystyle{apsrev4-2}
\frenchspacing
\bibliography{refs}
\end{document}